\documentclass[iop, revtex4, numberedappendix]{emulateapj}

\usepackage[backref,breaklinks,colorlinks,citecolor=blue]{hyperref}
\usepackage{epstopdf}
\usepackage{amssymb}
\usepackage{amsmath}
\usepackage[pdftex]{lscape}
\usepackage[all]{hypcap}       
\usepackage{color}
\usepackage{ulem}

\newcommand{\pasa}[1]{Publications of the Astronomical Society of Australia}

\shorttitle{Rest frame Properties of $z\sim2$ Star-forming Galaxies}
\shortauthors{Tacchella et al.}

\begin{document}

\title{The SINS/zC-SINF Survey of $z\sim2$ Galaxy Kinematics: Rest frame Morphology, Structure, and Colors from Near-infrared Hubble Space Telescope Imaging\altaffilmark{\dag}}

\author{S. Tacchella\altaffilmark{1},
P. Lang\altaffilmark{2},
C. M. Carollo\altaffilmark{1},
N. M. F\"orster Schreiber\altaffilmark{2},
A. Renzini\altaffilmark{3,4},
A. E. Shapley\altaffilmark{5}
S. Wuyts\altaffilmark{2},
G. Cresci\altaffilmark{6}, 
R. Genzel\altaffilmark{2,7,8},
S. J. Lilly\altaffilmark{1},
C. Mancini\altaffilmark{3},
S. F. Newman\altaffilmark{7},
L. J. Tacconi\altaffilmark{2},
G. Zamorani\altaffilmark{9},
R. I. Davies\altaffilmark{2}, 
J. Kurk\altaffilmark{2}, \&
L. Pozzetti\altaffilmark{9}
}
\altaffiltext{1}{Department of Physics, Institute for Astronomy, ETH Zurich, CH-8093 Zurich, Switzerland}
\email{sandro.tacchella@phys.ethz.ch} 
\altaffiltext{2}{Max-Planck-Institut f\"ur extraterrestrische Physik (MPE), Giessenbachstr. 1, D-85748 Garching, Germany}
\altaffiltext{3}{INAF Osservatorio Astronomico di Padova, Vicolo dell'Osservatorio 5, I-35122 Padova, Italy}
\altaffiltext{4}{Department of Physics and Astronomy Galileo Galilei, Universita degli Studi di Padova, via Marzolo 8, I-35131 Padova, Italy}
\altaffiltext{5}{Department of Physics and Astronomy, University of California, Los Angeles, CA 90095-1547, USA}
\altaffiltext{6}{Istituto Nazionale di Astrofisica -- Osservatorio Astronomico di Arcetri, Largo Enrico Fermi 5, I-50125 Firenze, Italy}
\altaffiltext{7}{Department of Astronomy, Campbell Hall, University of California, Berkeley, CA 94720, USA}
\altaffiltext{8}{Department of Physics, Le Conte Hall, University of California, Berkeley, CA 94720, USA}
\altaffiltext{9}{INAF Osservatorio Astronomico di Bologna, Via Ranzani 1, I-40127 Bologna, Italy}

\altaffiltext{\dag}{
Based on observations made with the NASA/ESA {\em Hubble Space Telescope\/}, obtained at the Space Telescope Science Institute, which is operated by the Association of Universities for Research in Astronomy, Inc., under NASA contract NAS 5$-$26555 (programs \# GO9822, \# GO10092, \# GO10924, \# GO11694, \# GO12578, \# GO12060, \# GO12061, \# GO12062, \# GO12063, \# GO12064, \# GO12440, \# GO12442, \# GO12443, \# GO12444, \# GO12445). Also based on observations obtained at the {\em Very Large Telescope\/} of the European Southern Observatory, Paranal, Chile (ESO Programme IDs 075.A-0466, 076.A-0527, 079.A-0341, 080.A-0330, 080.A-0339, 080.A-0635, 081.A-0672, 183.A-0781, 087.A-0081, 088.A-0202, 088.A-0209, 091.A-0126).}

\slugcomment{{\sc ApJ accepted:} \today}

\begin{abstract}
We present the analysis of Hubble Space Telescope (HST) $J$- and $H$-band imaging for 29 galaxies on the star-forming main-sequence at $z\sim2$, which have adaptive optics Very Large Telescope SINFONI integral field spectroscopy from our SINS/zC-SINF program. The SINFONI H$\alpha$ data resolve the ongoing star formation and the ionized gas kinematics on scales of $1-2$ kpc; the NIR images trace the galaxies' rest frame optical morphologies and distributions of stellar mass in old stellar populations at a similar resolution. The global light profiles of most galaxies show disk-like properties well described by a single S\'{e}rsic profile with $n\sim1$, with only $\sim15\%$ requiring a high $n>3$ S\'{e}rsic index, all more massive than $10^{10}M_\odot$. In bulge+disk fits, about $40\%$ of galaxies have a measurable bulge component in the light profiles, with $\sim15\%$ showing a substantial bulge-to-total ratio $B/T\ga0.3$. This is a lower limit to the frequency of $z\sim2$ massive galaxies with a developed bulge component in stellar mass because it could be hidden by dust and/or outshined by a thick actively star-forming disk component. The galaxies' rest-optical half-light radii range between $1$ and $7$ kpc, with a median of 2.1 kpc, and lie slightly above the size-mass relation at these epochs reported in the literature. This is attributed to differences in sample selection and definitions of size and/or mass measurements. The $(u-g)_{rest}$ color gradient and scatter within individual $z\sim2$ massive galaxies with $\ga10^{11}M_\odot$ are as high as in $z=0$ low-mass, late-type galaxies and are consistent with the high star formation rates of massive $z\sim2$ galaxies being sustained at large galactocentric distances. 
\end{abstract}


\section{Introduction} \label{sec:intro}

The peak of cosmic star formation rate (SFR) at $z\sim1-3$ is thought to be the epoch of the major buildup of the massive spheroids that dominate the stellar mass budget in today's Hubble sequence \citep[e.g.,][]{bell03, baldry04, thomas05, cimatti06, leitner12, behroozi13b, ilbert13, muzzin13}. Precisely when, where, and how the dominant bulges emerge inside the disks is still unknown. The morphology of typical galaxies at $z\sim2$ is undoubtedly very different from the one of local galaxies of similar mass, in particular at rest frame ultraviolet wavelengths, where the high-redshift galaxies are highly star-forming and much clumpier than their local counterparts \citep[e.g.][]{abraham96, elmegreen05, lotz06, law07, wuyts11}. The advent of the WFC3 on the Hubble Space Telescope (HST) has, however, enabled the study of the rest frame optical morphologies of these high-$z$ galaxies, and revealed that often, although not always, star-forming galaxies (SFGs) at $z\sim2$ have a more regular appearance at the longer relative to the shorter wavelengths \citep[e.g.,][]{toft07, elmegreen09, cameron11a, forster-schreiber11b, wuyts12, law12a}.

Information on galaxy kinematics at $z\sim2$ has also become recently available through integral field spectroscopy (IFS) of optical emission lines such as H$\alpha$, tracing the ionized gas \citep[e.g.,][]{genzel06, forster-schreiber06b, genzel08, forster-schreiber09, epinat09, epinat12, law09, cresci09, jones10, gnerucci10, mancini11, law12c, swinbank12, glazebrook13, newman13a}. Adaptive optics (AO) has been crucial in increasing the resolution to the $1-2$ kpc scales at $z\sim2$, and in enabling a much better disentanglement, at least for the largest galaxies, of rotationally supported disks from galaxy mergers. A fair summary is that while the precise occurrence of disk-like kinematics is still debated, it appears that at least $\sim50\%$ of $\sim10^9-10^{11}~M_{\odot}$ SFGs on the $z\sim2$ `main-sequence' (i.e., SFR vs. galaxy stellar mass) plane are rotationally supported structures with typical rotation velocities of $100-300~\mathrm{km~s}^{-1}$ \citep[e.g.,][]{glazebrook13}. With typical velocity dispersions in H$\alpha$ of $50-100~\mathrm{km~s}^{-1}$, they are, however, different from $z=0$ disks.

Progress in understanding the development at early epochs of the massive spheroids (and thus of the Hubble sequence) can be made by studying, simultaneously, the spatially resolved distributions of stellar mass and star formation rate in high-$z$ galaxies with different morphologies and structural versus kinematic properties. Thus, as part of our SINS/zC-SINF of spatially resolved H$\alpha$ kinematics and star formation properties of $z\sim2$ galaxies with SINFONI at the Very Large Telescope \citep[VLT; e.g., ][]{forster-schreiber09, mancini11} and building on our pilot near-infrared (NIR) imaging with NICMOS on HST \citep{forster-schreiber11a, forster-schreiber11b}, we carried out a major program of sensitive, high-resolution, NIR imaging in the $J$ and $H$ bands with HST.  We targeted in particular sources with AO-assisted SINFONI observations of the rest frame H$\alpha$ (and [NII]) emission, reaching a resolution of $\sim1-2$ kpc (presented in a complementary paper by F\"{o}rster Schreiber et al. (in preparation, hereafter FS15)). The $J-H$ color at the $z\sim2$ redshifts of the SINS/zC-SINF galaxies straddles across the Balmer/4000$\mathrm{\AA}$-break; the NIR color images thus well map the resolved stellar mass distributions inside the galaxies. 

In this paper we present a detailed analysis of the new WFC3 NIR images for a total sample of 29 SINS/zC-SINF galaxies. These measurements provided key information for our studies of the nature of dispersion-dominated galaxies \citep{newman13a}, of powerful active galactic nuclei (AGN) driven nuclear outflows \citep{forster-schreiber14}, and of constraints on gravitational quenching \citep{genzel14a}.  In a forthcoming paper, we will compare, on 1 - 2 kpc scales, the SFR distributions with the stellar mass distributions inside galaxies, in order to constrain the timescales and properties of the quenching mechanism that shuts down star formation in massive $z\sim2$ main-sequence galaxies \citep{tacchella15b}.

The paper is organized as follows. We start with the description of the target selection (Section~\ref{sec:GalaxySample}). We then describe the reduction of the newly obtained HST WFC3/IR imaging data and the ancillary VLT SINFONI/AO data used in this work (Section~\ref{sec:ObsData} and \ref{subsec:Sinfoni}). In Section~\ref{sec:LinkingSINFHST} we use the combination of rest-optical light distributions and H$\alpha$ kinematics to classify our sample into four classes of regular disks, irregular disks, mergers, and unresolved galaxies. The following Section~\ref{sec:GalacticShapes} presents in detail the structural measurements performed by modeling the $J$ and $H$ light distribution. Specifically, we model the light profiles in the $J$- and $H$-band and use the models to quantify galaxy sizes and degree of light concentration (through S\'ersic index), and we determine the bulge+disk decomposition properties of all galaxies for which such a decomposition is reliable. The results of this modeling are presented in Section~\ref{sec:results}. In Section~\ref{sec:color} we discuss the color properties within the galaxies based on color profiles, as well as the full 2D color maps. To give a better view of the global properties of our sample, in Section~\ref{sec:discussion} we briefly explore how it compares with the $z=0$ galaxy population and with other $z\sim2$ samples on the S\'{e}rsic index versus stellar mass and size versus stellar mass planes; here we also compare the resolved color properties of the $z\sim2$ galaxies with corresponding measurements for $z=0$ Hubble sequence galaxies. The paper is summarized in Section~\ref{sec:Summary}.

Throughout this paper, we adopt $WMAP9$ cosmology: $H_0=69.3~\mathrm{km~s^{-1}~Mpc^{-1}}$, $\Omega_{\Lambda,0}=0.71$, and $\Omega_{m,0}=0.29$ \citep{hinshaw13}. For this cosmology, $1\arcsec$ corresponds to $\approx8.4~\mathrm{kpc}$ at $z=2.2$. Magnitudes are given in the AB photometric system. All sizes and radii presented in this paper are circularized, i.e. $r=r_a\sqrt{(b/a)}$.

\section{The Galaxy Sample: Typical star-forming galaxies at $z\sim2$}\label{sec:GalaxySample}

In this section we briefly present the selection of the 29 galaxies in our sample and show that they form a representative sample of the main-sequence population of SFGs at $z\sim2$. 

\subsection{Target Selection} \label{subsec:Sample}

A detailed discussion of the selection of the SINS/zC-SINF AO sample can be found in FS15. In short, the galaxies were drawn from the optical spectroscopic surveys of parent samples selected photometrically based on: 

\begin{itemize}
\item s$BzK$ or `BX/BM' photometric criteria from the spectroscopic zCOSMOS-DEEP survey collected with VIMOS at the VLT (`zC' objects, \citealt{lilly07, lilly09}; 18 targets);
\item $U_\mathrm{n}GR$ optical colors (`BX' objects, \citealt{steidel04}; 8 targets);
\item Spitzer/IRAC 4.5$\micron$ fluxes (Galaxy Mass Assembly ultra-deep Spectroscopic Survey or “GMASS”, \citealt{cimatti08,kurk13} 3 targets);
\item combination of $K$-band and $BzK$ color criteria (survey by \citealt{kong06} of the `Deep-3a' field; 3 targets); and
\item $K$-band magnitudes (`K20' survey, e.g. \citealt{cimatti02}; Gemini Deep Deep
Survey or `GDDS', \citealt{abraham04}; 3 targets).
\end{itemize}

The specific selection criteria for the SINFONI instrument were then the observability of the H$\alpha$ emission line and a minimum expected H$\alpha$ line flux (corresponding roughly to a minimum SFR of $\sim10~M_{\odot}$/yr; \citealt{forster-schreiber09, mancini11}). We considered only sources with redshifted H$\alpha$-line falling in either the SINFONI $H$- or $K$-band, within spectral regions of high atmospheric transmission, and at least $400~\mathrm{km}~\mathrm{s}^{-1}$ away from OH airglow lines. These criteria constrain the target redshifts either in the range $z\approx1.3-1.7$ or $z\approx2-2.5$, respectively. 

Our SINS/zC-SINF AO sample consists of 35 galaxies. From these, 29 galaxies have HST $J$- and $H$-band imaging and therefore make up the sample analyzed here. The whole sample is summarized in Table~\ref{tbl:aux_data_stel}, which lists the H$\alpha$ redshift, $K$-band magnitude, and main stellar properties. All but one galaxy lies in the redshift range $z\approx2-2.5$ (GK-2540 lies at $z_{\mathrm{H}\alpha}=1.6146$).

The stellar mass $M_{\star}$\footnotemark[1]\footnotetext[1]{Note that we adopt for the definition of galaxy stellar mass the integral of the SFR, i.e., we do not subtract the mass `returned' to the gas through stellar evolution processes.}, age, visual extinction ($A_V$), and absolute and specific SFR (sSFR) are obtained from stellar population synthesis modeling of the optical to NIR broadband spectral energy distributions (SEDs) supplemented with mid-IR $3-8\micron$ photometry when available. All details of the SED modeling are presented in \citet{forster-schreiber09, forster-schreiber11a, forster-schreiber11b} and \citet{mancini11}. Briefly, we adopt the best-fit results obtained with the \citet{bruzual03} code, a \citet{chabrier03} initial mass function (IMF), solar metallicity, the \citet{calzetti00} reddening law, and either constant or exponentially declining SFRs. The uncertainties on the derived quantities are dominated by model assumptions: the uncertainty on the stellar mass is a factor of $\sim2-3$, on the SFRs and stellar ages even larger. Changing the star formation histories shifts the stellar mass less than the SFR, but both systematically (see Figures 2 and 3 of \citealt{mancini11}). For observed SEDs coverage from optical out to at least NIR wavelengths, the adoption of similar evolutionary tracks, and similar star formation histories return a robust relative ranking of galaxies in these properties.

\begin{figure}
\includegraphics[width=\linewidth]{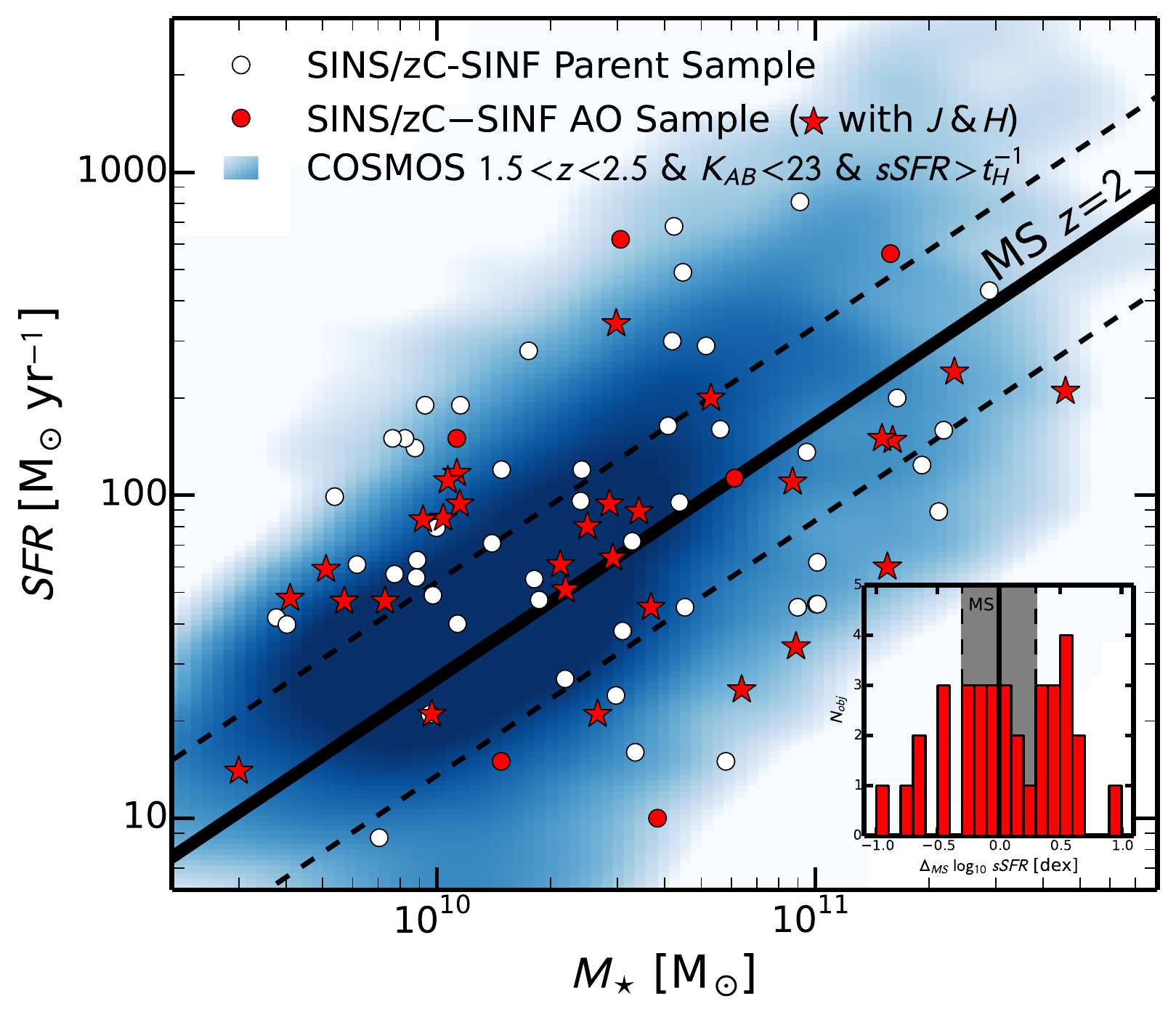} 
\caption{Distribution of the galaxy sample in the stellar mass - SFR ($M_{\star}$-SFR) plane. The red markers (circles and stars) show the complete SINS/zC-SINF AO sample, while the sample indicated with red stars represents the 29 galaxies in our sample with additionally $J$- and $H$-band imaging. The SINS/zC-SINF AO sample is a subsample of the SINS/zC-SINF H$\alpha$ sample (SINFONI in seeing-limited mode) shown with white circles \citep{forster-schreiber09, mancini11}. The solid black line indicates the `main-sequence' of SFGs at $z\sim2$ \citep{rodighiero11}, while the scatter of $0.4$ dex is shown by black dashed lines. We compare our sample to $1.5<z<2.5$ SFGs in the COSMOS field at $K_{s,AB}<23.0$ mag and with sSFR$>t_H^{-1}(z)$. These galaxies are shown with the blue color map, where the darker blue indicated more galaxies in a given bin of stellar mass and SFR. The inset shows the distributions of the offsets in specific SFR from the main-sequence (in logarithmic units) of the comparison SFG sample, indicating that the SINS/zC-SINF AO sample lies, on average, $0.13$ dex above the main-sequence, i.e. lying well within the $0.4$ dex scatter.} 
\label{fig:main_sequence}
\end{figure}

\subsection{Location on the $M_{\star}$-SFR Plane} \label{subsec:Sample_MS}

As mentioned before, our galaxies were selected based on a combination of criteria: firstly, the required optical redshift means in practice an optical magnitude cutoff (to ensure sufficient signal-to-noise ratio [S/N]; in the case of zCOSMOS: $B<25$); secondly, the requirement of minimum H$\alpha$ flux (or SFR) likely emphasizes younger, less obscured, and more actively star-forming systems \citep{forster-schreiber09,mancini11}. Our sample is therefore neither a mass nor sSFR complete sample. One expects our galaxies to lie above the main-sequence and toward bluer colors than the bulk of the $z\sim2$ population. 

However, our galaxies follow the trend of the main-sequence well, as visible in the $M_{\star}$-SFR plane in Figure~\ref{fig:main_sequence}. The red points show the 35 SINS/zC-SINF AO galaxies, a subsample of the 110 galaxies of the SINS/zC-SINF H$\alpha$ sample (shown with white points). Our AO sample represents well the underlying larger parent non-AO sample. The red stars mark the 29 galaxies that are in our sample on which we will focus in this paper. The SFRs are taken from the SED fits described above. Our targets probe two orders of magnitude in stellar mass ($\approx3\times10^9-4\times10^{11}~\mathrm{M_{\sun}}$), SFR ($\approx10-300~\mathrm{M_{\sun}~yr^{-1}}$), and sSFR ($\approx0.4-11.7~\mathrm{Gyr^{-1}}$) and cover the kinematic diversity of massive $z\sim2$ SFGs. Figure~\ref{fig:main_sequence} compares our targets with a sample of $\sim6000$ galaxies of the COSMOS sample ($2.0<z<2.5$ SFGs with $K_{s,AB}<23.0$ mag and with sSFR$>t_H^{-1}(z)$ at the respective redshift of each object; SFR from UV+IR estimates). Our targets probe the main-sequence of SFGs at $z\sim2$ \citep{daddi07, rodighiero11, whitaker12}, preferentially lying only slightly at higher sSFRs (on average 0.13 dex above the main-sequence, which itself has a scatter of $Í\sim0.4$ dex) and bluer rest frame optical $(U-V)_{rest}$ colors (on average 0.1 AB mag) compared to the underlying population of massive SFGs, which have them self a scatter of $(U-V)_{rest}\approx0.3$ AB mag (see FS15 for a more extended discussion). This validates that our galaxies are a representative sample of $\ga10^{10}~M_{\odot}$ main-sequence galaxies at $z\sim2$.

\section{WFC3/IR Imaging Data: Observations and Reduction}\label{sec:ObsData}

Our goal of mapping out the distribution of stars dominating the bulk of the stellar mass was driving the choice of filters: the HST $J$- and $H$-bands bracket optimally the age-sensitive Balmer/4000$\mathrm{\AA}$-break, enabling us to measure stellar masses very accurately on resolved $1-2$ kpc scales. Table~\ref{tbl:hst_data_overview} gives an overview of the available HST imaging data for all galaxies in our sample. We describe the observations of the newly obtained data in Section~\ref{subsec:Obs}. For several targets, we were able to supplement our data with archival HST data (Section~\ref{subsec:HST}). Section~\ref{subsec:DataRed} highlights the main points of the data reduction, and Sections \ref{subsec:PSF} and \ref{subsec:Noise} discuss the point spread function (PSF) and image noise properties.

\subsection{New WFC3/IR Observations} \label{subsec:Obs}

As part of \# GO12578, we carried out WFC3/IR observations between March 2012 and November 2012 with the WFC3/IR camera on board HST and using the F160W ($H$) and F110W ($J$) filters. The targets for this program were the 17 (21) for which HST $H$-band ($J$-band) imaging was not available from other HST imaging campaigns. The $H$-band, centered at 1536.9 nm (width = 268.3 nm), probes the longest wavelengths at which the HST thermal emission is unimportant, taking full advantage of the lack of sky background that limits the sensitivity of ground-based NIR observations. The $J$-band is centered at 1153.4 nm and has a width of 443.0 nm. Simulations based on population synthesis models indicate that the adjacent nonoverlapping $J$ and $H$ filter pair is best in terms of combined speed (integration time) and accuracy of $M_{\star}/L$ estimates for the typical brightnesses and redshift range ($2.0<z<2.5$) of our targets. The $H$-band filter samples approximately the SDSS g-band in the rest frame at the median $z=2.2$ of our targets, and the $J-H$ color minimizes extinction effects, brackets optimally the age-sensitive Balmer/4000$\mathrm{\AA}$-break, and the wide bandpasses maximize flux sensitivity. 

Each target was observed for two orbits in $H$ and one orbit in $J$, with each orbit split into four exposures with a subpixel dither pattern to ensure good sampling of the PSF and minimize the impact of hot / cold bad pixels and other such artifacts (e.g., cosmic rays, satellite trails). The individual exposure time was 653 s, giving a total on-source integration of 5224 s for the $H$-band and 2612 s for the $J$-band respectively. We used a square four-point dither pattern with a dither box size that is driven by the two constraints of dithering over the science target while simultaneously ensuring that nearby bright objects (most notably the bright AO stars) do not overlap with any of the target location on the detector.

\subsection{Archival HST Data}\label{subsec:HST}

For the other 8 (12) galaxies with missing $H$-band ($J$-band) imaging, we used publicly available data. Eight galaxies lie in CANDELS fields \citep{grogin11, koekemoer11}, and therefore have public available imaging data in the following bands: ACS/WFC F606W ($V$), ACS/WFC F814W ($I$), WFC3/IR F105W ($Y$), WFC3/IR F125W ($J$), and WFC3/IR F160W ($H$). These data are treated in the same way as our newly obtained imaging data and resampled to a pixel scale of $0\farcs05$; the same pixel scale as the rest of our imaging data.  The differences between the two filter bands $J$ (F110W) and $J$ (F125W) are very minor, and the Balmer/4000$\mathrm{\AA}$-break is equally well traced with both filters. Furthermore, six targets have already deep HST $H$ data: three galaxies have four-orbit integration from our NIC2 Pilot program \citep[PI Shapley, ][]{forster-schreiber11a}, and one has three-orbit WFC3/IR $H$ data from Cycle 17 program \# GO11694 (PI Law), which we reduced in the same way as our new data. In addition, 18 galaxies lie in the zCOSMOS field and therefore have HST/ACS F814W ($I$) imaging.

\subsection{WFC3/IR Data Reduction} \label{subsec:DataRed}

\subsubsection{WFC3/IR Detector Calibration}

We have used the standard procedure to convert the raw data to a set of final, flat-fielded, flux-calibrated images. We used the Pyraf/STSDAS task \texttt{calwf3} to construct the bad pixel array (data quality array) and to do the bias and dark current subtraction for each readout. Once all the separate MULTIACCUM read-outs are calibrated, the up-the-ramp slope fitting is done with \texttt{calwf3}. In this step, we have not applied the cosmic-ray rejection since we reject the cosmic rays with \texttt{MultiDrizzle} in a later step.

The available flat-field files in the STScI archive pipeline do not fully correct for all the flat-field features present in the data. There are IR `blobs' that have appeared. We averaged over $\sim30$ exposures in both bands and visually identified the areas of low sensitivity (see also \citealt{koekemoer11}). A static mask was then defined that masks all these areas of low sensitivity. This static mask is then used as an input file for the \texttt{MultiDrizzle} task later on.

In addition, several percent of exposures are affected by the passage of satellites across the field of view during the exposure. All exposures that are affected by this are identified visually, and the pixels are flagged in the data quality array.

\subsubsection{Relative Astrometry and Distortion Corrections}\label{subsub:astrometry}

Initial astrometric uncertainties were of order of $\sim0\farcs3-0\farcs5$ owing to guide star acquisition and reacquisition from one orbit to the next. Additionally, there are small contributions from other sources, such as spacecraft positioning error, optical offsets introduced by filter change, and changes in the optical path length to the detectors due to `breathing'. These contributions amount to $\sim0.2-0.3$ pixels \citep{koekemoer11}. 

The astrometric accuracy further depends on the degree to which the detector geometric distortions are calibrated. We have used the latest distortion solutions, from March 22 2012. 

To solve for and remove the residual uncertainties in the spacecraft dither offsets between all the exposures in each orbit, we used the Pyraf/STSDAS task \texttt{Tweakshifts}, which provides an automated interface for computing residual shifts between input exposures being combined using \texttt{MultiDrizzle}. The shifts computed by \texttt{Tweakshifts} correspond to pointing differences after applying the WCS information from the input image's headers. The utilized method for computing offsets between images consists of identifying sources in each image, matching them up and fitting the matched sets of positions. The sources are identified by using \texttt{SExtractor} software, version 2.8.6 \citep{bertin96}, selecting only point sources. The final output of \texttt{Tweakshifts} is a shift file, which contains the final shift, rotation, and scale change for each input image relative to the reference image.

\begin{figure*} \begin{center} \leavevmode
\includegraphics[width=\textwidth]{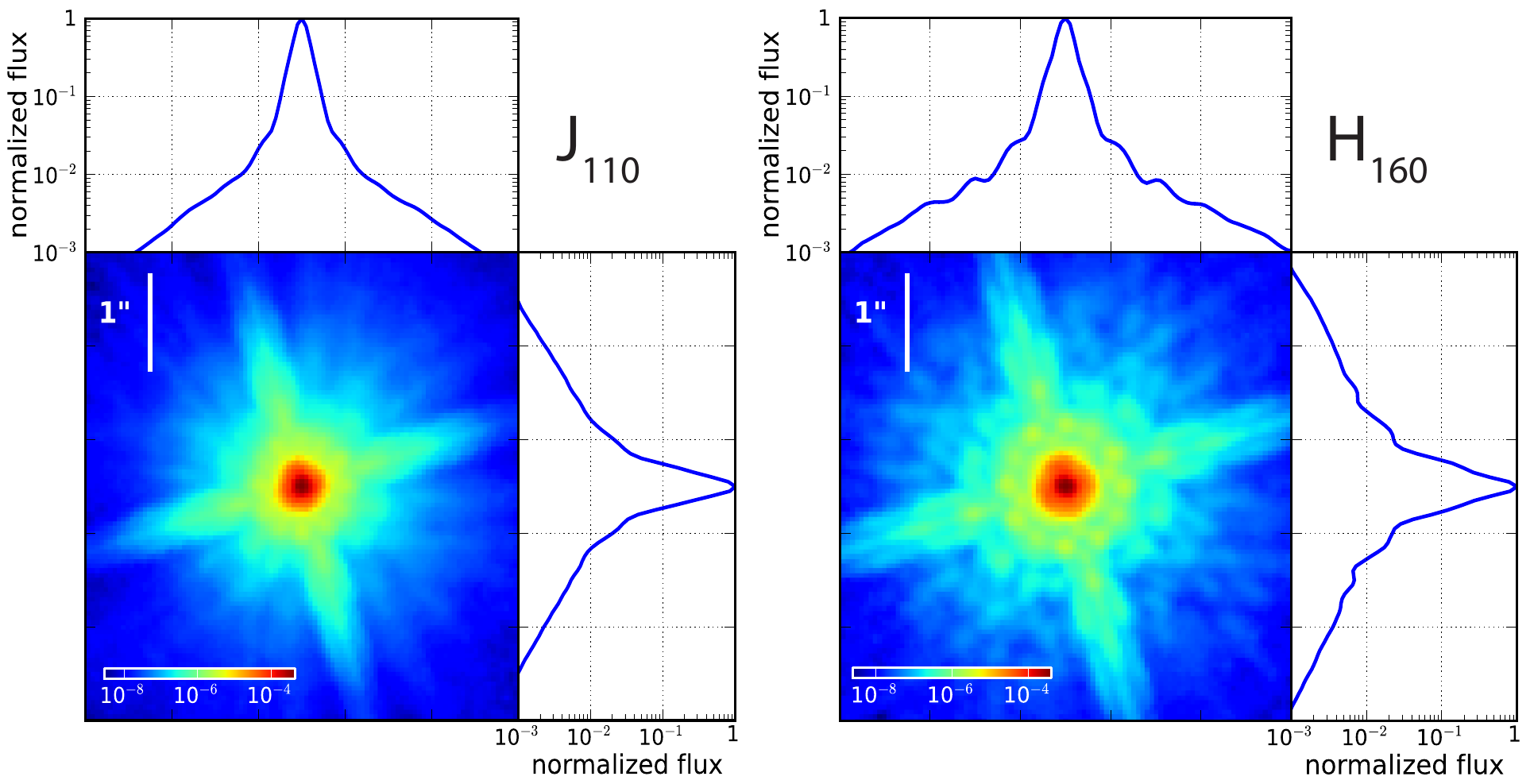} 
\caption{$J$- (left) and $H$-band (right) PSF constructed by stacking six well exposed and non-saturated stars. The FWHM of the $J$- and $H$-band PSF amount to $0\farcs16$ and $0\farcs17$, respectively. } 
\label{fig:PSF}
\end{center}\end{figure*}

\begin{figure}
\includegraphics[width=\linewidth]{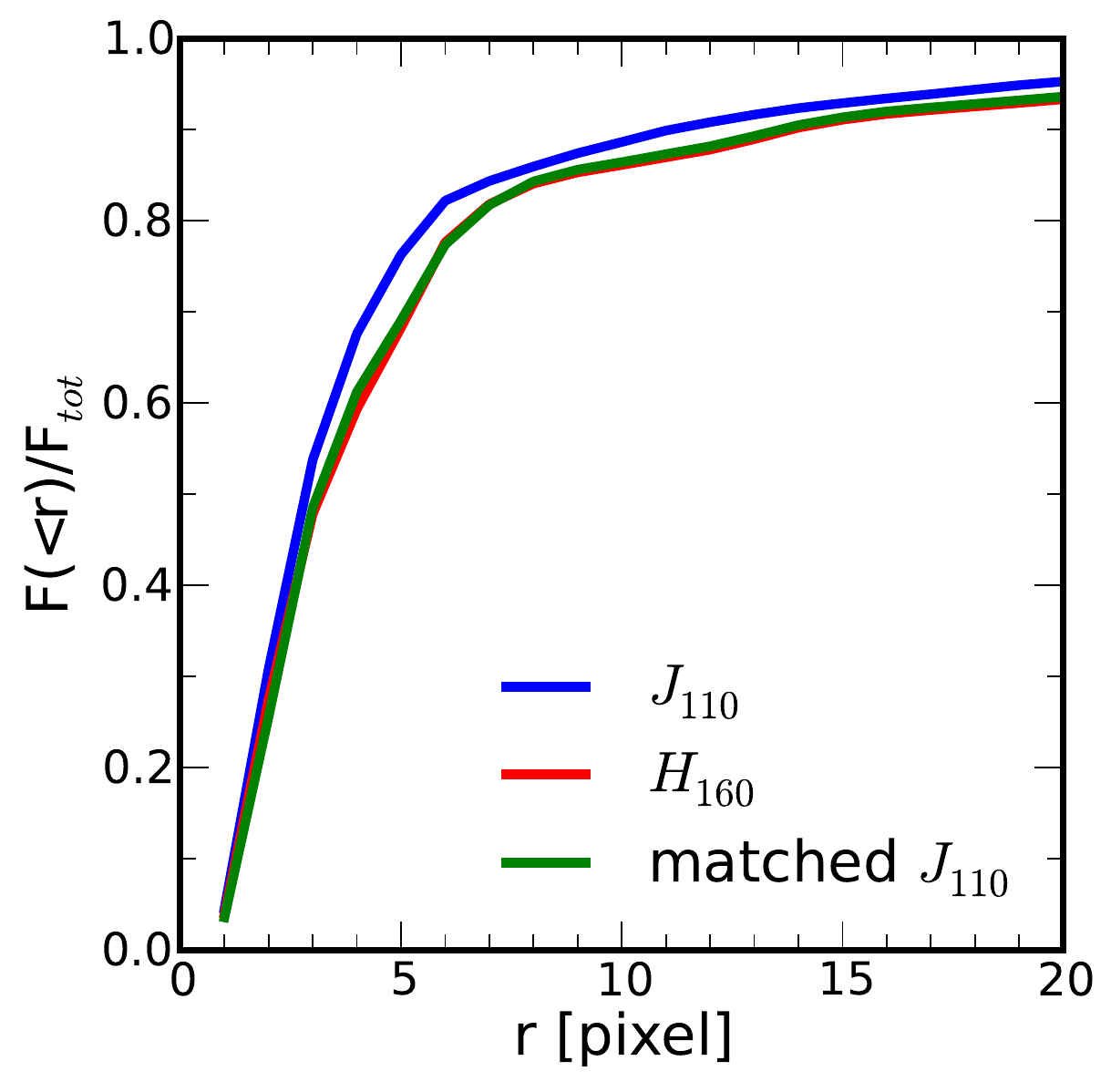} 
\caption{Normalized flux within a given radius as a function of radius (in pixels). The blue and red lines represent the PSF of the $J$- and the $H$-band, respectively. As expected, within a given radius, the PSF of $J$-band contains more light (i.e., is more concentrated) than the PSF of $H$-band. The green line shows the PSF of $J$-band after PSF matching.} 
\label{fig:PSF_Match}
\end{figure}

For sources in the COSMOS field, we have used the ACS $I$-band images as reference images, i.e. aligning all exposures for a galaxy (eight $H$ and four $J$ exposures) relative to it, since the COSMOS $I$-band images have good absolute astrometry. For all other sources, we have used the first $H$-band exposure as a reference image. The whole alignment process has been done in an iterative manner, i.e. using the output shift file of \texttt{Tweakshifts} as input for the next alignment step. After several iterations, the astrometric uncertainty is $\la0.20$ pixels.

\subsubsection{Combining the Individual Images with MultiDrizzle}

We use \texttt{MultiDrizzle} to detect cosmic rays and to dither the different exposures to one final image. Additionally to the science exposure with the updated data quality arrays, we use the static mask (IR `blobs' of low sensitivity) and the reference image with the shift file as input for \texttt{MultiDrizzle}. 

In \texttt{MultiDrizzle} task, we used the standard cosmic-ray rejection routine with standard parameters, i.e., a first pass going through all the pixels in the image and using $S=1.2$ and $S/N=3.5$, followed by a second pass in a 1 pixel wide region around each of the pixels flagged in the first pass, but using more stringent criteria of $S=0.7$ and $S/N=3.0$. This ensures that fainter pixels around cosmic rays are also flagged.

We choose an output pixel scale of $0.05\arcsec$ pixel$^{-1}$ for the final WFC/IR mosaics to match the pixel scale of the SINFONI/AO data. This pixel scale provides an adequate sampling of the PSF. Finally, we set pixfrac (defines how much the input pixels are reduced in linear size before being mapped onto the output grid) to 0.8, which was found from experimentation to give the best trade-off between gain in resolution and introduction of rms noise in the final images.

\begin{figure*} \begin{center} \leavevmode
\includegraphics[width=\textwidth]{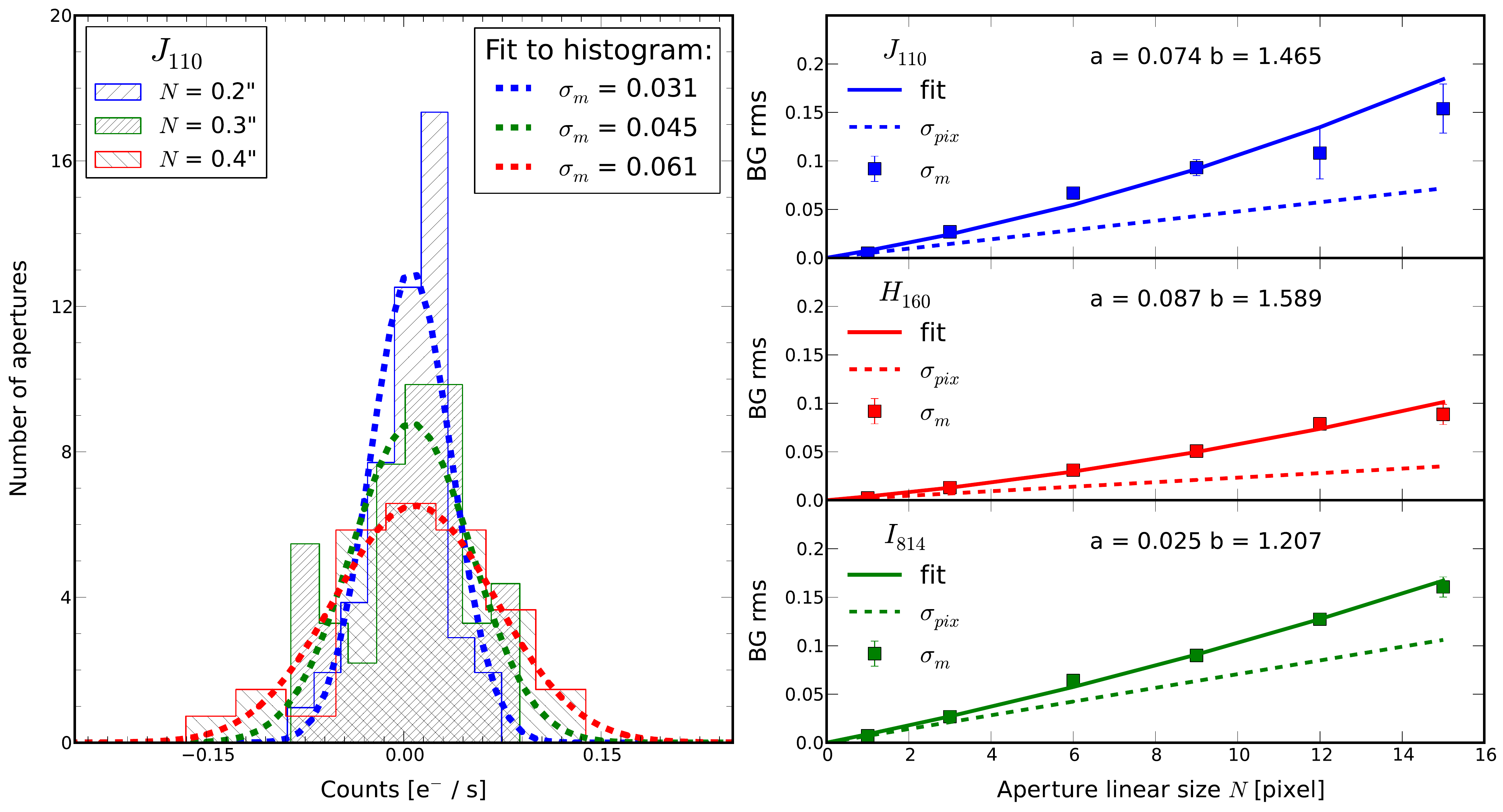} 
\caption{Background rms derived from the distribution of fluxes within empty apertures. Left: distribution of empty aperture fluxes within a $0.2''$, $0.3''$, and $0.4''$ aperture diameter on the $J$-band image of ZC400569. The noise properties are qualitatively the same for the other filter band, as well as other objects. The distribution is well described by a Gaussian with an increasing width for increasing apertures size. Right: Background rms vs. aperture size for the WFC3/IR $J$ (top), WFC3/IR $H$ (middle), and ACS $I$-band. Solid lines represent the function given in Equation~\ref{eq:sigma} fitted to the observed rms noise values (values for the parameter $a$ and $b$ are given). Dashed lines indicate a linear extrapolation of the pixel-to-pixel rms. Correlation between pixels introduces a stronger than linear scaling with aperture size.} 
\label{fig:noise}
\end{center}\end{figure*}

\subsubsection{Residual Background Subtraction}

The combined drizzled images showed residual background features on scales of a few tens of pixels from sources such as sky background from earthshine, zodiacal light, and low-level thermal background emission. These features were successfully modeled and subtracted from the combined drizzled images using background maps generated with the \texttt{SExtractor} software. Since our science goals require detection of low surface brightness emission of the galaxies to large radii, we took particular care in making these background maps. We masked out sufficiently large areas around each object and then ran \texttt{SExtractor} with a background mesh size of 50 pixels and filter size of four meshes. This combination provided the best match to the spatial frequency of the residual features, without compromising the faint extended emission from the galaxies.

\subsection{PSF Investigation} \label{subsec:PSF}

We measure the PSF in each band from six well-exposed and nonsaturated stars. We align all stars with the IRAF task \texttt{IMALIGN} and normalize them to have the same integrated brightness. We then combine them with the IRAF task \texttt{IMCOMBINE}, taking the median of all the stars. The PSFs for the $J$- and $H$-band are shown in Figure~\ref{fig:PSF}. We have constructed for each target galaxy its own PSF, since each exposure was taken at different times and with different telescope orientations. The variation from PSF to PSF for the different targets is $\la1\%$ in the central parts, while in the outskirts it is $\sim5\%$. The FWHM is $0\farcs16$, $0\farcs17$, and $0\farcs10$ for $J$-, $H$-, and $I$-band, respectively.

With the PSF for each band in hand, we use the package IRAF \texttt{PSFMATCH} to calculate a smoothing kernel to convolve all images to the resolution of the $H$-band image, since the $H$-band PSF is the largest one. We test the effectiveness of PSF matching by comparing the fractional encircled energy of each PSF before and after the procedure, shown in Figure~\ref{fig:PSF_Match}. The PSFs in all bands have identical profiles, especially within the central region, where the gradient is steepest.

\subsection{Noise Properties and Limiting Depths} \label{subsec:Noise}

The background noise properties of the raw data are well described by the rms of the signal measured in each pixel, since both the Poisson and readout noise should be uncorrelated. For such uncorrelated Gaussian noise, the effective background rms for an aperture of area $A$ is simply the pixel-to-pixel rms $\bar{\sigma}$ scaled by the linear size $N=A^{1/2}$ of the aperture, $\sigma(N)=N\bar{\sigma}$. Instrumental features, the data reduction, and PSF-matching have added significant systematics and correlated noise in our final data. Therefore, the simple linear scaling of the background rms $\sigma(N)$ would lead to underestimated flux uncertainties. To investigate the noise properties and the limiting depths, we used aperture photometry on empty parts of the image to quantify the rms of background pixels within the considered aperture size \citep{labbe03,forster-schreiber06a}. We randomly placed $\sim300$ nonoverlapping apertures at a minimum distance of $\sim5\arcsec$ from the nearest segmentation pixels in a \texttt{SExtractor} segmentation map. For a given aperture size, the distribution of empty aperture fluxes is well fitted by a Gaussian, as illustrated in Figure~\ref{fig:noise}. We model the background rms as a function of aperture size with a polynomial of the form

\begin{equation}
\sigma(N)=\frac{N\bar{\sigma}(a+bN)}{\sqrt{w}}
\label{eq:sigma}
\end{equation}

where the term $1/\sqrt{w}$ accounts for the spatial variations in the noise level related to the exposure time and is taken from the weight maps (it will be absorbed in the coefficients $a$ and $b$). The coefficient $b$  represents the correlated noise contribution that becomes increasingly important on larger scales. 

The noise behavior is qualitatively the same for all the targets and for all the filter bands. Figure~\ref{fig:noise} shows the result for one of our targets for the $J$-, $H$- and $I$-band, with the background rms measurements for the various apertures and the best-fit polynomial models (Equation \ref{eq:sigma}) compared to the expected linear relationship for uncorrelated Gaussian noise. The values for $J$-band are $a=0.09\pm0.03$ and $b=1.4\pm0.2$, for $H$-band $a=0.07\pm0.02$ and $b=1.7\pm0.2$ and for $I$-band $a=0.02\pm0.01$ and $b=1.2\pm0.1$. 

From the background rms, we can directly determine the limiting depths (sensitivity) for our images. The total limiting AB magnitudes ($5\sigma$) for point sources are $27.3\pm0.5$, $26.6\pm0.4$, and $26.4\pm0.1$ for $J$, $H$ and $I$, respectively. For a larger source of $1~\mathrm{arcsec}^2$, the $3\sigma$ limiting AB magnitudes are $25.9\pm0.5$, $25.2\pm0.4$, and $24.9\pm0.1$ for $J$, $H$ and $I$, respectively.

\section{VLT SINFONI/AO IFS Data} \label{subsec:Sinfoni}

The uniqueness of the galaxy sample presented here is the availability of SINFONI/AO observations, which give H$\alpha$ maps and kinematics resolved on $\approx1-2$ kpc scales. A detailed description of the data is given in FS15. Briefly, the SINFONI data were collected between April 2005 and November 2013. For the AO observations, the intermediate SINFONI pixel scale of 50 mas is used to achieve the full gain in resolution afforded by AO. The integration times range from about 2.0 hr (ZC410123) to 23.0 hr (D3A-15504), with an average of 7.9 hr and a median of 6.0 hr. Depending on the redshift of the sources, we used the $K$-band ($z>2$) or $H$-band ($z<2$) grating to map the main emission lines of interest (H$\alpha$ and [NII]$\lambda\lambda6548,6584$ doublet). With these choices of grating and pixel scale, the nominal spectral resolution is R $\sim$ 2900 and 4500 in the $H$- and $K$-band, respectively. 

The effective angular resolution was estimated by fitting a two-dimensional Moffat profile to the final PSF image associated with each galaxy. The Moffat profile fitted the profile best because beside the bright central AO part of the PSF, there is a significant wider halo reflecting the uncorrected part of the natural seeing. The major-axis FWHMs range from $0\farcs13$ to $0\farcs36$, with mean and median of $0\farcs20$ and $0\farcs19$, respectively. 

In this paper we use the extracted H$\alpha$ emission line flux and kinematic maps, such as the H$\alpha$ velocity and dispersion maps (see also \citealt{newman13a,genzel14a}).


\begin{figure*} \begin{center} \leavevmode
\includegraphics[width=\textwidth]{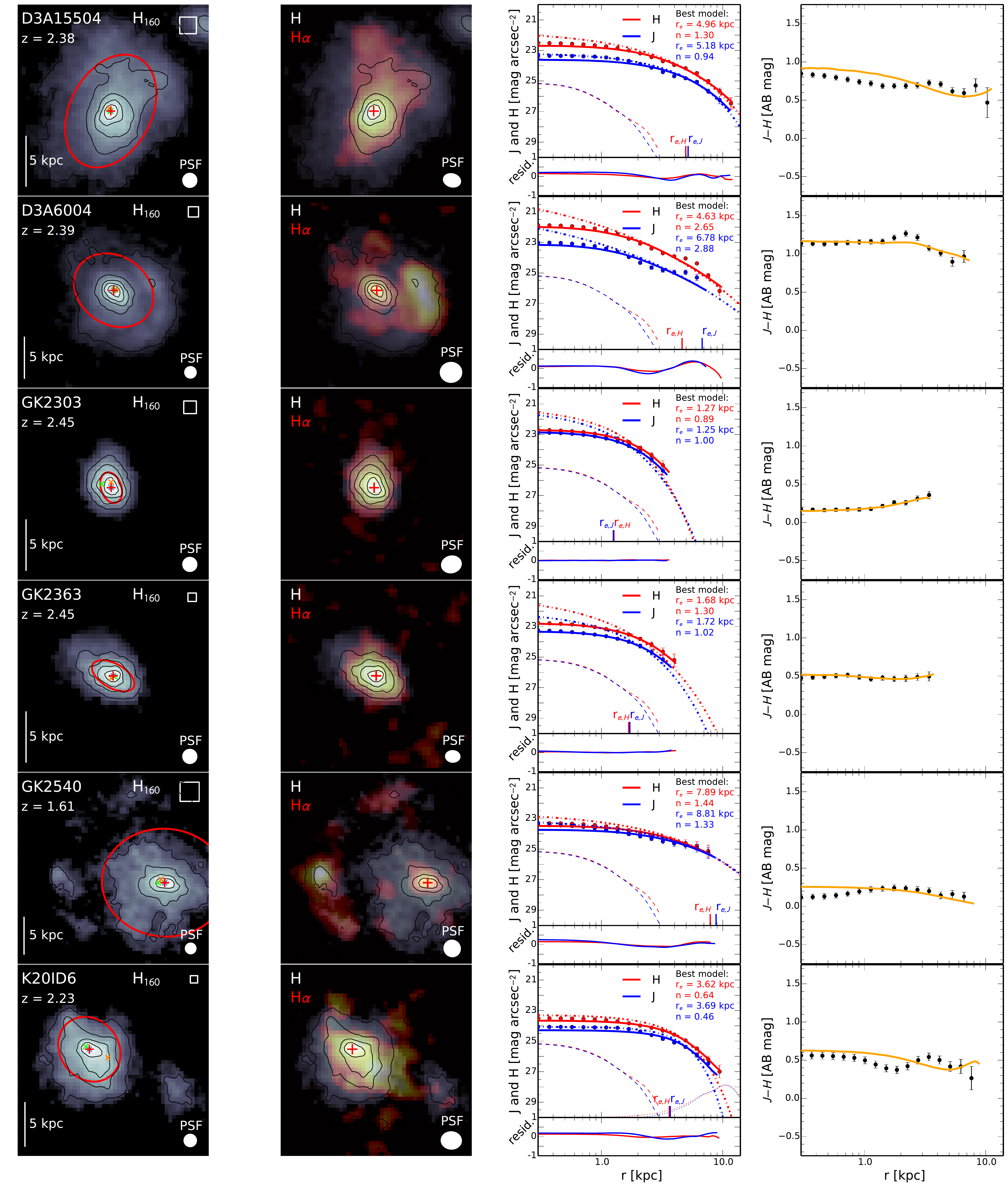} 
\caption{Overview of the data for each galaxy. For a given galaxy (row), the first column shows the HST $H$-band image with its contours, PSF, and a line indicating 5 kpc. The red ellipse represents the best \texttt{GALFIT} parameters $r_e$, $b/a$, and P.A.. The box in the upper right corner shows the area in which the center is varied to estimate the uncertainties on the best-fitted values. The fiducial center is the largest red plus sign `+', while the center of light, mass, and the dynamical center are shown with a cross `x' in blue, green, and orange, respectively. The second column shows a superposition of the $H$-band image (same scaling and contours as in the first column) and the SINFONI/AO H$\alpha$ emission line map, shown in red-yellow color scaling. The third column shows the $J$-band and $H$-band surface brightness profiles in blue and red, respectively. The dot-dashed lines show the \texttt{GALFIT} model, while the solid lines are the PSF-convolved profiles. The dashed lines represent the PSF, and the arrows on the right mark the surface brightness limit. For some galaxies, additional features are modeled and are shown here as thin dotted lines. On the bottom, the normalized residual profiles are shown. The last (fourth) column shows the PSF-convolved $J-H$ color profiles in comparison with the data from the color maps (black dots with error bars; see Section~\ref{sec:color}). 
}
\label{fig:data_overview}
\end{center}\end{figure*}

\begin{figure*} \begin{center} \leavevmode
\includegraphics[width=\textwidth]{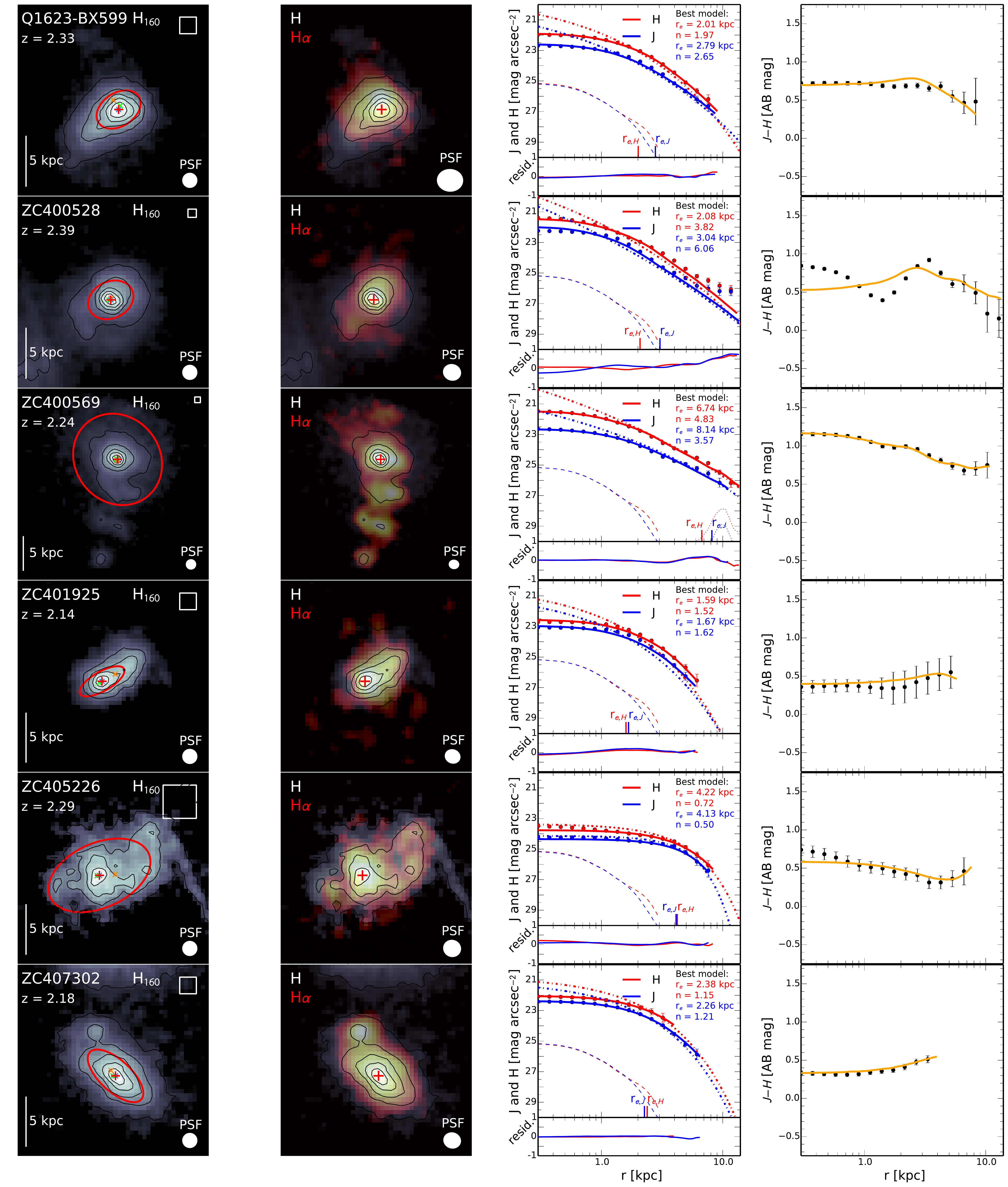} 
\caption{Same as Figure~\ref{fig:data_overview}.}
\end{center}\end{figure*}

\begin{figure*} \begin{center} \leavevmode
\includegraphics[width=\textwidth]{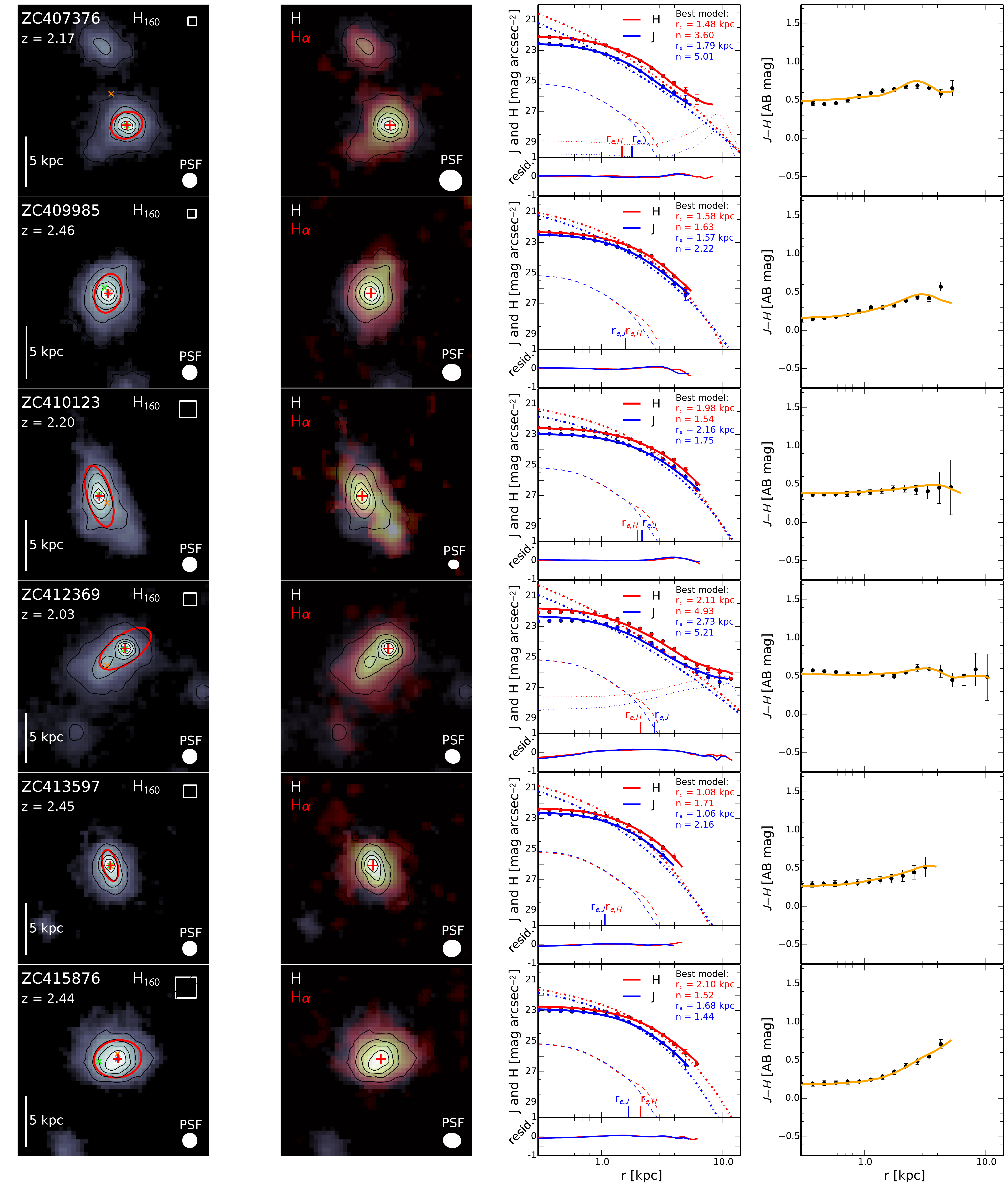} 
\caption{Same as Figure~\ref{fig:data_overview}.}
\end{center}\end{figure*}

\begin{figure*} \begin{center} \leavevmode
\includegraphics[width=\textwidth]{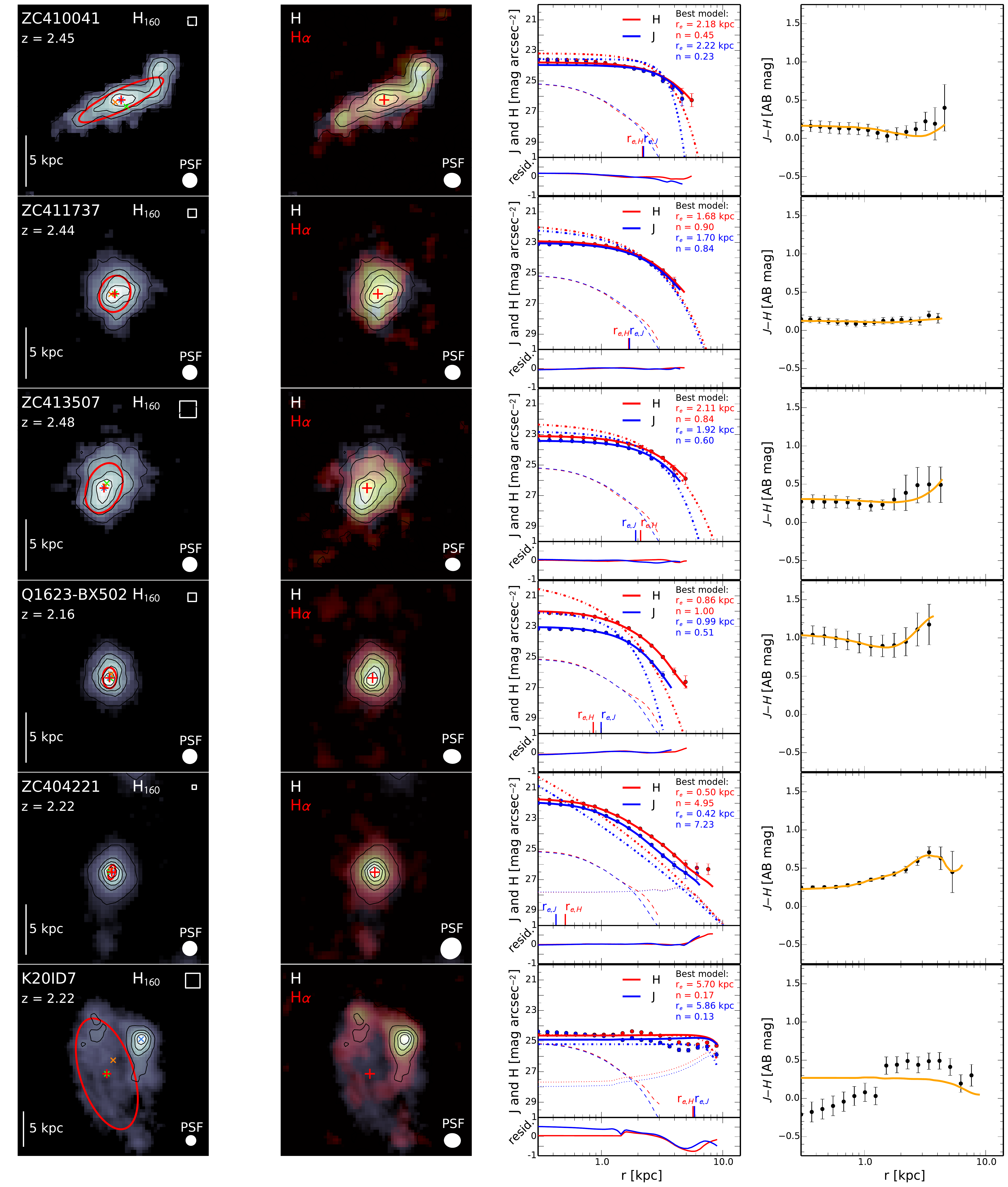} 
\caption{Same as Figure~\ref{fig:data_overview}. Galaxies Q1623-BX502 and ZC404221 are very compact, and their fits have to be taken with care. In addition, K20-ID7 shows a bright clump in its outskirts that had to be fitted as well. The fit is clearly less reliable.}
\end{center}\end{figure*}

\begin{figure*} \begin{center} \leavevmode
\includegraphics[width=\textwidth]{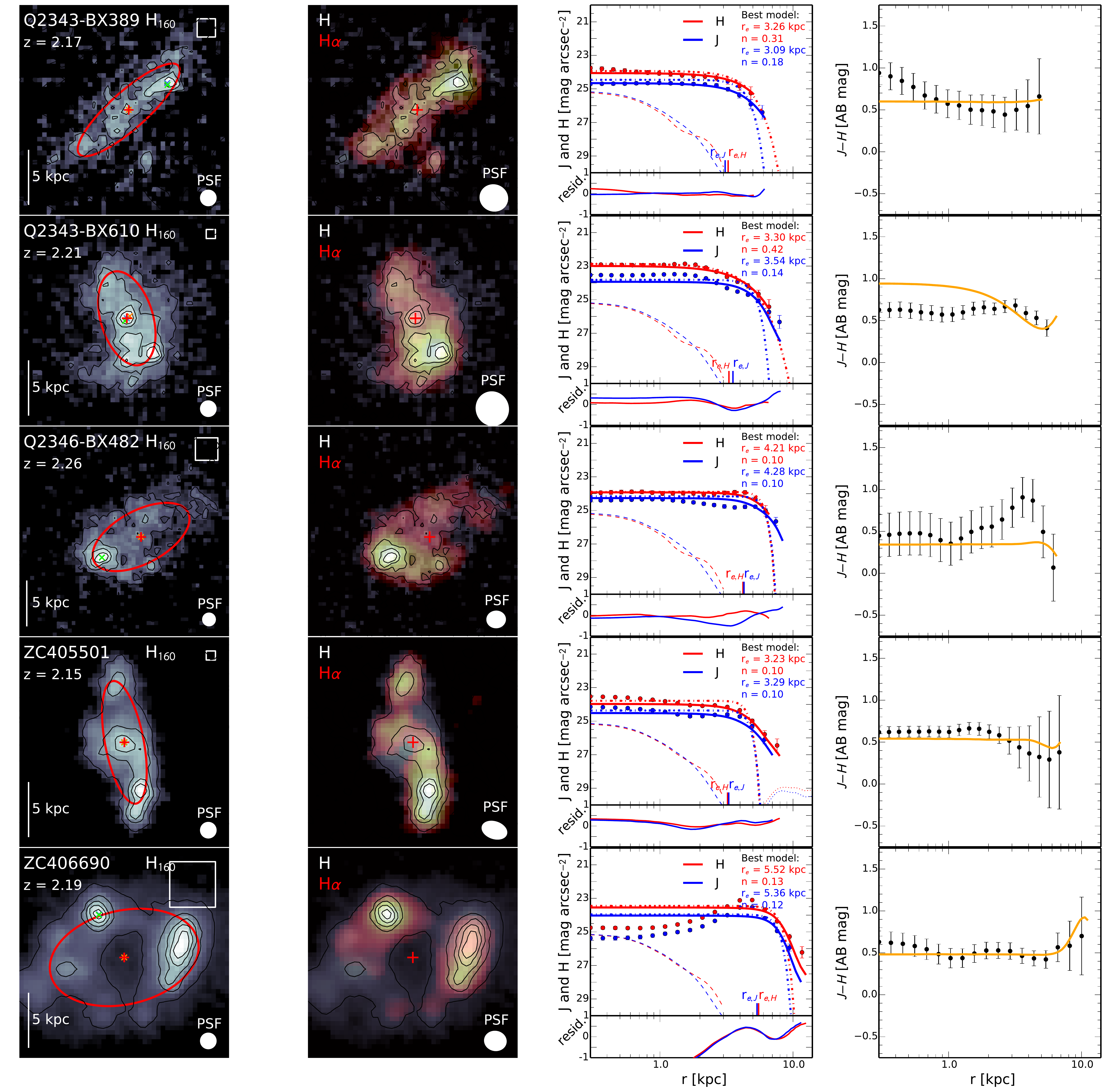} 
\caption{Same as Figure~\ref{fig:data_overview}. All these galaxies (Q2343-BX389, Q2343-BX610, Q2346-BX482, ZC405501, and ZC406690) have bright clumps in their outskirts, and their fits are therefore less reliable.}
\label{fig:data_overview_end}
\end{center}\end{figure*}


\section{Comparison of the Stellar Continuum Properties with the H$\alpha$ Kinematic Properties }\label{sec:LinkingSINFHST}
 
For each galaxy in our sample, a comprehensive description of the features detected in the NIR data is given in Appendix~\ref{app:galaxy_by_galaxy}. The availability of both rest frame optical surface brightness and resolved H$\alpha$ kinematic maps gives us the possibility of investigating the connection between the rest-optical morphological appearance of each galaxy and its kinematic state. Specifically, we present a comprehensive morphological classification, which also takes into account the global galactic kinematic properties.
 
Bright, kpc-sized clumps are common in SFGs at $z\sim2$. They have been first identified in rest frame UV images from HST \citep[e.g.][]{cowie95, van-den-bergh96, giavalisco96, elmegreen04a, elmegreen04b, elmegreen05, conselice04, lotz04, law07}. Recently, there have been an increasing number of surveys targeting the H$\alpha$ emission of high-$z$ galaxies with AO-resolution IFS \citep[e.g.][]{mannucci09, contini12, swinbank12, wright09, law09, law12a, wisnioski11, wisnioski12}. Since H$\alpha$ traces the ionized gas, i.e., the star-forming regions, it does not necessarily follow the underlying distribution of stellar mass. In addition, the ionized gas may also reflect stellar winds, i.e., strong outflows (prominent example is ZC406690, see \citealt{genzel11} and \citealt{newman12b}). However, we note that these outflows are generally detected as broad ($\mathrm{FWHM} > 400-500$ km/s) underlying line emission.  The fits applied to extract the H$\alpha$ flux and kinematic maps used here are primarily sensitive to the narrow emission component, and thus the maps trace the distribution of star formation and gravitationally driven gas motions. Clumpy structures are, however, also detected in the rest frame optical HST NIR images \citep[e.g.][]{toft07, dasyra08, kriek09, elmegreen09, bussmann09, swinbank10a, overzier10, forster-schreiber11a, forster-schreiber11b, cameron11a, law12a, law12b}, albeit often less pronounced than in the rest-UV \citep{wuyts12}. They are clearly seen in our data as well. Combining the rest frame light distributions with the galaxies' gas kinematics enables us to take a step forward toward the attempt to establish whether galaxies on the $z\sim2$ main-sequence are mergers caught in the act or are sustained by internal dynamical processes within gas-rich disks. A merging system shows irregular morphology and disturbed kinematics: the velocity and the velocity dispersion maps will be asymmetric. A galaxy that undergoes some internal dynamic process can also show irregular morphology (such as clumps), while, in contrast to the merger, the velocity field is regular and also implies rotational support ($v_{rot}/\sigma_0\ga1.5$, where $v_{rot}$ is the rotational velocity corrected for inclination and $\sigma_0$ is the intrinsic measure of velocity dispersion; see also \citet{shapiro08} for an extended discussion). Specifically, we classify our galaxies into four categories:

\begin{itemize}
\item \textbf{Regular disks (RD)}: At rest frame optical wavelengths, the galaxies show a relatively symmetric morphology featuring a single, isolated peak light distribution and no evidence for multiple luminous components. The velocity maps show clear rotation, and the dispersion maps are centrally peaked; the kinematic maps are fitted well by a disk model. To this category belong the following galaxies: 

D3A-15504, D3A-6004, GK-2303, GK-2363, Q2343-BX610, ZC405226, ZC405501, ZC410041, ZC411737, and ZC413507.

\item \textbf{Irregular disks (ID)}: In the optical light, the galaxies have two or more distinct peaked sources of comparable magnitude. Their velocity maps show clear sign of rotation, but are less regular than in RD's (i.e., $v_{rot}/\sigma_0\ga1.5$). The dispersion maps show a peak, which is however shifted in location relative to the centers of the velocity maps. The following galaxies fulfil these criteria: 

GK-2540, K20-ID6, K20-ID7, Q2343-BX389, Q2346-BX482, ZC400528, ZC400569, ZC406690, ZC407302, ZC410123, and ZC415876.

\item \textbf{Mergers (M)}: In the rest frame optical light, two or more distinct peaked sources of comparable magnitude are detected at a projected distance $\la5~\mathrm{kpc}$ from each other. The velocity maps are highly irregular with no evidence for ordered rotation (i.e., $v_{rot}/\sigma_0<1.5$); the velocity dispersion maps show multiple peaks. The following galaxies belong to this category:

Q1623-BX599, ZC407376, and ZC412369.

\item \textbf{Unresolved systems (UNR)}: these galaxies are not well resolved in the SINFONI data. They show complex (irregular) velocity maps and dispersion maps, but it is unclear whether these high velocity dispersions reflect the physical state of the sources, or rather are contaminated by PSF-smearing of the rotation signal. In this category are the following galaxies: 

Q1623-BX502, ZC401925, ZC404221, ZC409985, and ZC413597.

\end{itemize}

This classification can also be found in Table~\ref{tbl:measurements_overview} and in Appendix~\ref{app:galaxy_by_galaxy}, where we describe each galaxy individually. From the discussion above and in the appendix, it is clear that the presented classification by eye is meant as a qualitative benchmark. Many of the UNRs will most probably belong to one of the other categories when observed at a higher resolution. Also the IDs are tricky, as some of these galaxies may be mergers that are not well resolved in the current data, or have particularly large M/L variations (e.g., young, bright clumps contributing little mass could induce us into thinking that the object is irregular while in reality its mass may be regular). 

Keeping the above caveats in mind, we classify our 29 galaxies as having different physical states. In particular, in our sample there are 10 ($25\%$) RDs, 11 ($38\%$) IDs, 3 ($10\%$) merging systems, and 5 ($17\%$) possibly dispersion-dominated systems\footnotemark[2]\footnotetext[2]{Dispersion-dominated galaxies are galaxies with random-motion-dominated kinematics. However, the determination of the physical state of a galaxy strongly correlates with its size in the sense that smaller galaxies are more likely observed to be dispersion dominated, because instrumental broadening and beam smearing are more important for such systems \citep{newman13a}.}, although we have conservatively classified them here as UNRs. Comparing the stellar population of mergers with the ones of irregular disks, we find no difference in the median age of the stellar population (404 Myr and 400 Myr, respectively). In contrast, the sSFR is lower for IDs (1.3 Gyr$^{-1}$) than for mergers (2.6 Gyr$^{-1}$), implying that the mass doubling time for mergers is half of the one of IDs. 

Our data also enable us to study similarities and differences between the morphological and kinematical classifications of galaxies. We mentioned above the importance of both AO-resolution IFS and HST rest frame optical data to be able to classify $z\sim2$ galaxies. The biggest challenge arises when looking only at HST images, where galaxies can look clumpy and disturbed. Examples for clumpy and disturbed galaxies in our sample are D3A-15504, K20-ID7, Q2346-BX482, ZC400569, ZC405226, ZC406690, and ZC412369. Based on the HST images alone, we would classify these galaxies as mergers. However, the H$\mathrm{\alpha}$ velocity maps and velocity dispersion maps of most these systems show ordered rotation. We have therefore classified all except one galaxy as a regular/irregular disk; only ZC412369 has been classified as a merger owing to no sign of rotation and a high velocity dispersion (see also Appendix~\ref{app:galaxy_by_galaxy}). 

There are also similarities between the morphological and kinematical classifications of galaxies: galaxies with a regular and smooth rest frame optical appearance are all classified as disks (exceptions are the unresolved systems). Overall, we conclude that one needs both IFS and HST data for an accurate classification only in the case where the galaxies in the light appear clumpy and disturbed.

\section{Modeling the Rest Frame Optical Light Distributions} \label{sec:GalacticShapes}

In this section we outline the procedure to obtain PSF-corrected\footnotemark[3]\footnotetext[3]{It is fundamentally important to work with PSF-corrected quantities since the PSF influences the light profile out to a large radius (at least to $r_e$) and not only changes the flux within the FWHM of the PSF \citep{schweizer79,franx89, saglia93,trujillo01,graham01}.} models of our $z\sim2$ galaxies' rest frame optical light distribution as given by the $J$- and $H$-band images. First, we compare centers of different definition and quantify how precisely the center of each galaxy can be defined. We then do a 2D profile fitting for estimating the light profile's shape (i.e. S\'{e}rsic index) and the effective (half-light) radius $r_{e}$, which we adopt as a measure for the size of the galaxies in our sample. Beside these single-component fits, we carry out double-component fits (i.e. bulge-disk decompositions) to find and constrain bulge components in our galaxies. 

Obviously, these fits are challenging: simple one- or two-component axisymmetric models can reproduce reasonably well the overall surface brightness distribution and estimate global parameters such as the effective radius or bulge-to-total ratio ($B/T$). However, one has to keep in mind that they do not capture the prominent small-scale and irregular structure of clumpy disks and merging systems. This means that automated fitting routines can lead to nonphysical results. The quality control by eye is therefore fundamentally important to ensure the goodness of the fit.

\subsection{Choosing the Origin: Kinematic vs. Light vs. Mass Centers} \label{subsec:Centers}

The choice of the center is fundamentally important for modeling the light and mass distribution of a galaxy. It sets the foundation for the physical interpretation. We focus our analysis on the following three definitions of centers: kinematic center (based on the velocity and dispersion maps), light weighted center (we refer to the $H$-band weighted center as the general light weighted center since the $H$-band and the $J$-band have a very similar light distribution), and stellar-mass-weighted center (based on the mass maps that are derived from the M/L ratios based on the observed $J-H$ colors, as presented in \citealt{tacchella15b}). These different centers all have their advantages and disadvantages. For example, the light-weighted center can be affected by dust and/or spatial age (and hence M/L ratio) variations, whereas the stellar-mass-weighted center is determined on mass maps, which depends on the alignment of the images. 

For each galaxy, the three centers are determined and compared, shown in Figures~\ref{fig:data_overview}-\ref{fig:data_overview_end} with different colors and symbols. The center of light, stellar mass and the dynamical center are plotted with a cross `x' in blue, green, and orange, respectively. In $> 60\%$ cases the three centers agree very well ($< 1$ kpc difference), and it is straightforward to determine the fiducial center (indicated by the largest red plus sign `+'). For the other $< 40\%$ cases where the centers disagree more, the kinematic center is usually separated by $\ga1$ pixel ($\ga0.4$ kpc) from the center of light and mass. This can partially be explained by the higher uncertainty in the alignment of the SINFONI data with respect to the HST imaging ($\sim1$ pixel) than between the individual $J$- and $H$-band images ($<0.2$ pixel). This higher uncertainty in the alignment of the SINFONI data is due to the small SINFONI FOV, which does not cover stars or other compact sources that could be used for more accurate cross-registration. This means that, naturally, the offsets between dynamical center and light-/mass-weighted centers are potentially larger and more uncertain. 

Since we are primarily interested in modeling the light distribution, we choose the light-weighted center as the fiducial center in most cases. Exceptions are Q2343-BX389, Q2346-BX482, and ZC406690, where we use the dynamical center, and K20-ID7, where we use the center of stellar mass. See Appendix~\ref{app:galaxy_by_galaxy} for a detailed description.

We let the center be fixed during the fitting process. In the fiducial fit, we fix the center to our fiducial center. However, to estimate the uncertainty coming from the choice of the center, we also run fits in which we fix the center to another center that is chosen from a box that scales with the spread of the different centers. With this procedure we are able to estimate how different choices of centers propagate in the measured and modeled quantities.

\subsection{Modeling Assumptions and Input Values} 

We use the program \texttt{GALFIT} \citep{peng10} to fit the two-dimensional surface brightness distribution with a \citet{sersic68} profile:

\begin{equation}
I(r)=I_0 \exp \left[ -b_n \left( \frac{r}{r_e} \right)^{(1/n)} \right], 
\end{equation}

where $I(r)$ is the intensity as a function of radius $r$ and $r_e$ is the effective radius enclosing half of the total light of the model. The constant $b_n$ is defined in terms of S\'{e}rsic index $n$ which describes the shape of the profile. Ellipticals with de Vaucouleurs profiles have a S\'{e}rsic index $n=4$, exponential disks have $n=1$, and Gaussians have $n=0.5$. In the case of single-component fits, we use one single S\'{e}rsic profile with all fitting parameters left free (S\'{e}rsic index $n$, the effective radius $r_e$, the axis ratio $b/a$, the position angle P.A. of the major axis, and the total magnitude). In the case of bulge-to-disk decompositions, we assume two S\'{e}rsic models, one for the bulge and one for the disk. All but one parameter is free: the S\'{e}rsic index $n_d$ of the disk is fixed to 1. In the next two sections we explain the fitting procedure of the single- and double-component fits in more detail. 

In the single- and double-component fits, \texttt{GALFIT} convolves the model with the effective resolution of the data (PSF) and finds the best-fit parameters with a $\chi^2$-minimization. As \texttt{GALFIT} input, we give our empirically determined PSF (Section~\ref{subsec:PSF}), as well as the noise images (sigma image). The noise image is obtained by adding to the weight image\footnotemark[4]\footnotetext[4]{The \texttt{MultiDrizzle} pipeline gives as output a weight image, which contains all the error terms associated with each pixel (i.e., including noise from accumulated dark current, detector readout, and photon noise from the background as modulated multiplicatively by the flat field and the detector gain).} the Poisson noise from the astronomical sources in the image. To reduce the impact of large-scale residuals from the flat fielding and background subtraction (Section~\ref{subsec:DataRed}), we perform the fits within a region of $10\arcsec\times10\arcsec$ centered on the sources. We remove neighboring galaxies using an object mask. In the case of very close galaxies with overlapping isophotes (ZC400569, ZC404221, ZC405501, ZC407376, ZC412369, K20-ID6, and K20-ID7), objects are fitted simultaneously.

\subsection{Single-component Fits to the J- and H-band Light Distributions} \label{subsec:GALFIT}

The $H$-band is the reddest filter available for the galaxies in our sample, i.e., it traces the light of the older stellar populations the best and is the least affected by dust. We therefore use the $H$-band as the reference/fiducial filter and start by fitting the light distribution in the $H$-band first, allowing all parameters but the center to vary freely. We fix the centers to the fiducial centers defined in the previous section. For the $J$-band fits, we fixed the axis ratio $b/a$ and position angle P.A. to the values of the $H$-band to ensure that we fit the same spatial parts. 

We fitted each galaxy $\sim100$ times, varying the input guess $n$ between 1 and 4, $r_e$ $\pm25\%$ of the half-light radius obtained by \texttt{SExtractor}, $b/a$ $\pm0.1$ of the axis ratio obtained by \texttt{SExtractor} and P.A. between $0^{\circ}$ and $90^{\circ}$. The S\'{e}rsic index $n$ is limited to the interval $[0.1,8.0]$ during the fitting. All non-physical solutions (i.e., $r_e>20~\mathrm{kpc}$ and $b/a<0.1$) have been excluded, and also all residuals have been inspected. \texttt{GALFIT} converges in $\ga90\%$ of the cases to the same solution, which indicates that changing the input guesses has only a minor influence on the final fitted parameters. 

The best-fit parameters are then taken as the $\chi^2$-weighted median of the results. For estimating the uncertainty of the best-fit parameters, we combine the errors from the choice of a specific center and a specific set of initial guesses (`fitting error') and from the reliability of the fitting procedure with \texttt{GALFIT} (`observational error'). For the fitting error, we have carried out a second set of runs where we varied the input guesses and the centers. The centers are shifted within a box of a size given by the distance between the different kind of centers (center of stellar mass, center of light, and kinematic center). The box size for each galaxy is indicated in Figure~\ref{fig:data_overview}-\ref{fig:data_overview_end}, first column in the upper right corners. For most galaxies ($\sim75\%$), the box size is $4\times4$ pixels or smaller. For galaxies Q2346-BX482, ZC405226, and ZC406690, a substantially larger box ($10\times10$ pixels) is chosen. We vary the centers in the mentioned box, letting them remain fixed during the fitting process. This run consists of $\sim1000$ \texttt{GALFIT} realizations. The errors from the fitting are then the $68\%$ confidence intervals about the median of this run. 

To quantify the observational error (such as sky background, noise, and PSF), we follow the approach described in \citet{carollo13a} and \citet{cibinel13a}: we created $\sim100'000$ mock galaxies, convolved them with the PSF, and added noise (see also Appendix~\ref{app:BD_Deco}). Then, all the model galaxies are analyzed with \texttt{GALFIT} in the same way as described above. By comparing input and output, we obtained a correction matrix, so that for each galaxy with a given magnitude, size, S\'{e}rsic index, and axis ratio (ellipticity), we can determine its `true' unbiased values. In addition to this bias correction, we also obtain the scatter around the median of this correction vector. Since most of our galaxies are bright and also substantially larger than the PSF size, the observational bias vector is small and of the order $10\%$, i.e. smaller than the scatter. We therefore do not correct for the observation bias and only add the scatter to the uncertainty.

\subsection{Bulge-disk Decompositions of the J- and H-band Light Distributions}\label{subsec:bddecomp}

Traditionally, the bulge-disk decomposition is performed on the light distribution \citep[e.g., ][]{baggett98, lackner12, bruce14}. The `bulge' component is defined as the light excess above an exponential disk profile. The modeling is generally performed with two components: an $n=1$ S\'{e}rsic profile plus an additional (free $n$, $n=2$, or $n=4$) S\'{e}rsic profile. Fitting directly on the light distribution has the advantage that it is robust and the errors are well known and describable. On the other hand, an actively star-forming thick disk can outshine the bulge component \citep{carollo15}, leading to an underestimation of the $B/T$. Furthermore, dust could play a role (e.g., higher extinction toward the center), i.e. hiding possible bulge components present. To circumvent these problems, we can do the analysis directly on the mass distribution. We will present in \citet{tacchella15b} how to convert the $J$- and $H$-band light to the mass distribution by using the observed $J-H$ color ($\propto(u-g)_{rest}$) as a mass-to-light indicator. Briefly, we use the fact that observed colors and mass-to-light ratios of stellar populations are correlated \citep[see, e.g.,][]{rudnick06, zibetti09, forster-schreiber11a}: The different stellar population model curves occupy a well-defined locus in the observed $J-H$ color versus mass-to-light ratio parameter space, reflecting the strong degeneracy between stellar age, extinction, and SFH in these properties. This degeneracy can be used to derive the mean mass-to-light ratio for a given observed color that, multiplied by the luminosity, yields a stellar mass estimate. There are then two ways to do a bulge-disk decomposition: the first one is to fit 2D mass maps (obtained from the 2D color maps); the second one is to use the fits to the individual bands. We prefer the latter approach since we find it to be more stable and robust, but both approaches converge within the errors to the same results, as we show in \citet{tacchella15b}. 

An important assumption is that we let the S\'{e}rsic index of the bulge remain free within the range $[1.0,8.0]$ during the fitting procedure. The physical motivation comes from the fact that in the local universe, we can see two kinds of bulges: `classical' and `pseudo' bulges \citep{kormendy04}. Traditionally, two different processes have been considered for the formation of galactic bulges: one is through merging, and the other is secular instability of the stellar disk. The merging channel is most likely to produce a bulge with a S\'{e}rsic index of $n_b\sim4$; the secular evolution channel may favor $n_b\sim1-3$, although even steeper slopes can also be obtained depending on the precise internal mechanisms. Since we do not know what kind of bulge our galaxies have, we let the S\'{e}rsic index remain free during fitting. In Appendix~\ref{app:BD_Deco} we show with extensive simulation tests that if a structure in the sky has a low bulge S\'{e}rsic index, imposing an $n=4$ fit leads to much larger errors on the $B/T$ than when fitting it with a free $n$. This is somewhat obvious even without simulations, but the latter demonstrate it clearly. We show that on average, the $n_b$--free fits have a $2.4$ times smaller relative error on $B/T$ than the $n_b=4$ fixed fits. 

We conduct the bulge-to-disk decomposition in both $J$- and $H$-band. Excluded from this analysis are the following five objects: Q2346-BX482, K20-ID7, ZC406690 (all not centrally peaked, i.e. no apparent bulge present), Q1623-BX502, and ZC404221 (both barely resolved). Therefore, our initial sample for the bulge-disk decomposition consists of 24 galaxies. As mentioned above, we assume two S\'{e}rsic models, one for the bulge and one for the disk ($n_d=1$ fixed). For all the fits, we use additional constraints for the following parameters of the bulge: axis ratio $(b/a)_b\in[0.6,1.0]$, S\'{e}rsic index $n_b\in[1.0,8.0]$, and $r_{e,b}\in[0.04,4.2]~\mathrm{kpc}$. As an initial guess we set $n_b=4.0$, $r_{e,b}=1~\mathrm{kpc}$, $(b/a)_b=1.0$, and $\mathrm{P.A.}_b=0^{\circ}$. The centers are again fixed to the fiducial centers. 

As in the 1 component fits, we use the $H$-band as the reference/fiducial filter and fit it first. There are clear advantages and disadvantages in performing either independent or constrained fits to the two bands. By fixing the $J$-band structural parameters to the $H$-band, one ensures that bulge and disk colors are measured consistently over the same regions: this, however, prevents the detection of structural differences and color gradients of the individual bulge and disk components. Here we follow a mixed approach, similar to \citet{cibinel13a}. For the $J$-band fits, we apply three different runs, progressively increasing the number of constraints:

\begin{itemize}
\item[$\bullet$ R1:] same assumptions as for the $H$-band, i.e. centers fixed and $n_d=1$;

\item[$\bullet$ R2:] in addition to the assumptions in R1, setting the axis ratio and the position angle of the bulge and disk of the $J$-band to the ones of the $H$-band (i.e. $(b/a)_J = (b/a)_H$ and $\mathrm{P.A.}_J = \mathrm{P.A.}_H$ of disk and bulge);

\item[$\bullet$ R3:] in addition to the assumptions in R1 and R2, assuming that the S\'{e}rsic index and the effective radius of the bulge are the same in both bands (i.e. $n_{b,J} = n_{b,H}$, and $r_{e,b,J} = r_{e,b,H}$).

\end{itemize}

We flag all unphysical models (i.e. where bulge is larger in size than the disk: $r_{e,b}>r_{e,d}$; in total 5 objects), and all residual images are visually inspected to look for possible failure of the fitting algorithm. In general, the reliable fits of R1, R2, and R3 give all similar $B/T$ (variations less than $10\%$ between R1, R2, and R3), i.e. the $B/T$ is a robust quantity. 

\begin{figure*} \begin{center} \leavevmode
\includegraphics[width=\textwidth]{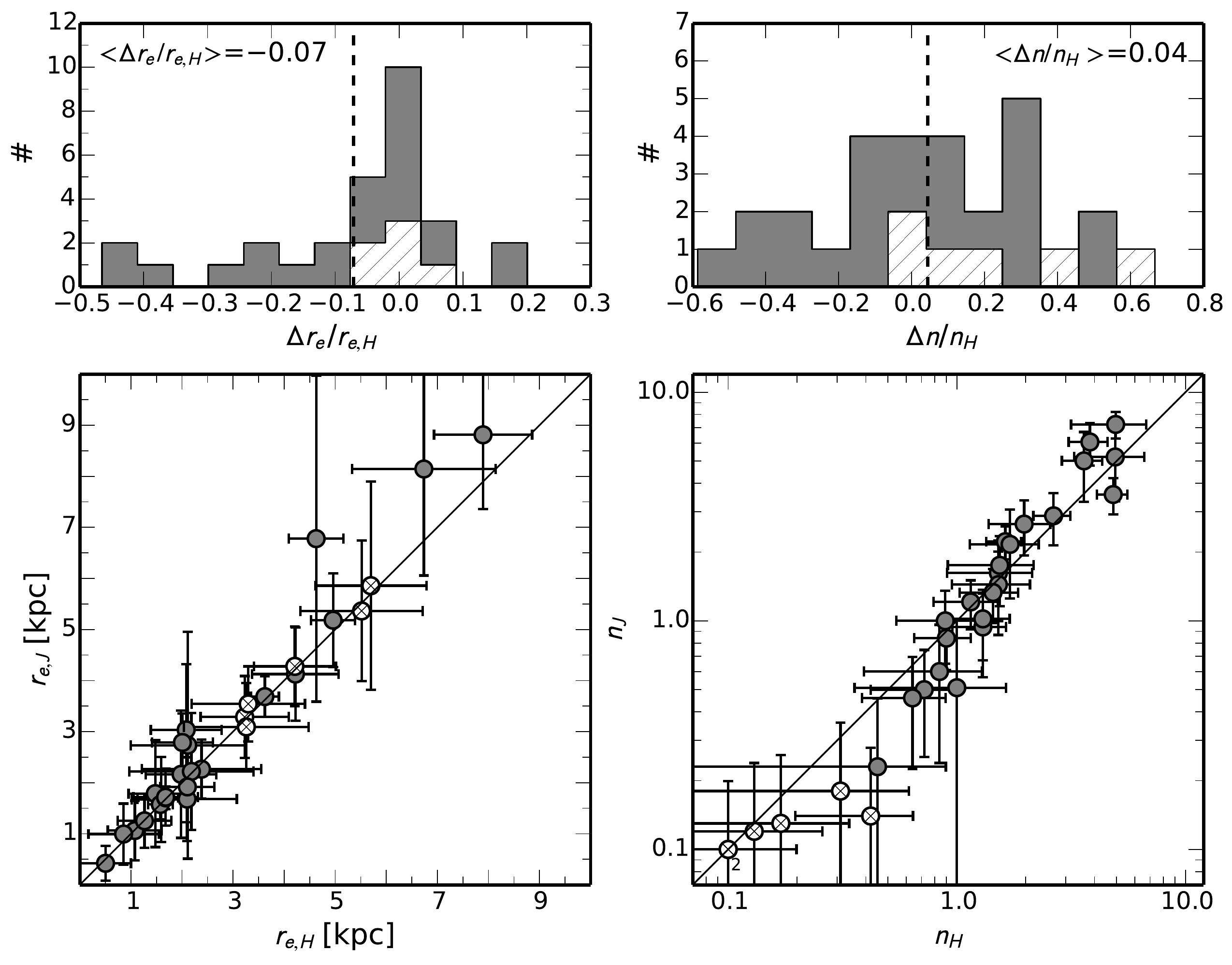} 
\caption{Comparison of \texttt{GALFIT} results for the $J$- and $H$-band. The filled gray symbols indicate our good fits, while the cross symbols (and hatched histogram) indicate the less reliable fits. Left: best-fitted effective radius $r_e$ for $J$-band (y-axis) and $H$-band (x-axis). There are only small differences on average, with the $J$-band sizes being $7\%$ larger. Right: same as the left plot, but for S\'{e}rsic index $n$. The small differences (in both quantities) between the $J$- and $H$-band imply color gradients within these galaxies.} 
\label{fig:GALFIT_JH}
\end{center}\end{figure*}

We use a quantitative procedure to select among the three different runs (R1, R2, and R3). We require that all $J$-band disk and bulge fits always have bulge and disk position angles and axis ratios within a sensible range from those of the $H$-band. The allowed range of variation for $J$ and $H$ bulge disk position angles and axis ratios is $\Delta P.A. \leq 15^{\circ}$ and $\Delta (b/a) \leq 0.15$ \citep{cibinel13a}. In addition, the fit has to reliable. When this is achieved with unconstrained fits (R1) for the $J$-band images, these unconstrained fits are retained as a fair description of the $J$-band bulge-disk decomposition. This choice maximizes the detection of possible wavelength-dependent structural difference and color gradients. For galaxies in which such a consistency requirement is not achieved with the unconstrained $J$ fits, we adopt those that satisfy such a requirement with a minimum number of parameters tied to the $H$-band fit parameters (i.e. first R2 and then R3 fits). After identifying the best run, we have added a central point source to the model to test for possible AGN contribution. The model fits do not improve for any of our galaxies by including an additional point source, nor did the final result change substantially (i.e. the results stayed within the errors for all fitted parameters). We therefore conclude that possible AGN activity does not influence the conclusions of this work.

We combine again two different sources of uncertainties to estimate the error on the $B/T$. First, for the fitting error, we have applied a variation of the fixed center in a box (same procedure as described in Section~\ref{subsec:GALFIT}) to account for possible uncertainties in the choice of different centers. On average over all galaxies, this gives an absolute error on the $B/T$ of $0.09$ for the $H$-band and $0.11$ for the $J$-band. 

The second, observational error contribution (accounting for the general fitting procedure) was estimated from the reliability of the \texttt{GALFIT} bulge-disk decomposition. See Appendix~\ref{app:BD_Deco} for details. Briefly, we have simulated galaxies and decomposed bulge and disk with the same method as described above. We found that the $B/T$ can be recovered very well, i.e. there is no systematic bias. On the other hand, other parameters of the bulge such as S\'{e}rsic index $n_{bulge}$ and size $r_{e,bulge}$ are very degenerate and cannot be reliably estimated. The main reason for this is that the bulges' sizes are at or below the resolution limit. 

\section{Results: Disks and Spheroids in Main-Sequence Galaxies at $z\sim2$}\label{sec:results}

Having described the fitting procedure and the quantification of the uncertainty in the previous section, we now describe the results from the modeling of the rest frame optical light distribution traced by the $J$- and $H$-band. We present the results on the single-component fits in Section~\ref{subsec:1comp} and on the bulge-to-disk decomposition in Section~\ref{subsec:2comp}. 

\subsection{Results from the Single-component Fits}\label{subsec:1comp}

Although galaxies have multiple components, it is nevertheless informative to treat galaxies as single entities when looking at the global parameters that describe the light distribution. The columns for `Single Component ' in Tables~\ref{tbl:measurements_overview} and \ref{tbl:measurements_overview2} show the best-fit parameters for the $J$- and $H$-band. Figures~\ref{fig:data_overview}-\ref{fig:data_overview_end} show in the first column the $H$-band image with a red ellipse that represents the best-fit parameters $r_e$, $b/a$, and P.A.. The third column in the figures shows the $J$- and $H$-band surface brightness profiles. 

About two-thirds of the galaxies in our sample are well fitted by a single S\'{e}rsic, i.e. $1\la\chi_{\mathrm{red}}^2\la2$, and our PSF-convolved model agrees well with the data. Galaxies Q2343-BX389, Q2343-BX610, Q2346-BX482, ZC405501, and ZC406690 all have bright clumps $\sim4-5$ kpc away from their centers, making the flux distributions asymmetric and not centrally peaked. This drives the S\'{e}rsic $n$ toward the lower limit of 0.1. Therefore, the fits to these galaxies have large residuals, basically because S\'{e}rsic profiles are not a good representation of the data, and the fits have to be treated with caution. Also, galaxy K20-ID7 belongs to this category with a very bright clump in its outskirts. However, in this case the clump is so bright that the center gets severely affected, and we had to model it in addition with a separate S\'{e}rsic profile. These six galaxies are marked in the following plots with a circle and cross, in contrast to the reliably fitted galaxies, which are shown with a gray filled circle. In addition, galaxies Q1623-BX502 and ZC404221 are very compact and small, i.e. are just at our resolution limit. The fits to these galaxies therefore have to be taken with care (sizes have large relative uncertainties). 

Figure~\ref{fig:GALFIT_JH} compares the best-fit results for $n$ and $r_e$ for the $J$- and $H$-band filters. The error bars are obtained from varying the centers, changing the initial guess for the fitting parameters, and simulations of the observational biases (Section~\ref{subsec:GALFIT}). Varying the center dominates the error ($\sim70~\%$). For nearly all galaxies, there is only a minor difference between the two bands. The exceptions are D3A-6004, GK-2520 and ZC400569 (all have a larger $r_e$ in $J$ than in $H$). The average normalized difference in size $(r_{e,H}-r_{e,J})/r_{e,H}=\Delta r_e/r_{e,H}=-0.07$, i.e. the $J$-band sizes are $7\%$ larger on average (left panels of Figure~\ref{fig:GALFIT_JH}). Comparing the S\'{e}rsic index $n$ of the $H$ and $J$-band shows that toward small $n$, the $H$-band predicts a larger $n$ than the $J$-band. On the other hand, toward larger values of $n$, the $J$-band has larger values. Overall, the average normalized difference is $\Delta n/n_H=0.04$, i.e. the $H$-band is slightly more concentrated, and therefore the average galaxy has blue outskirts and a red center. Even though $r_e$ and $n$ are comparable in both bands, the small differences seen in some galaxies are enough to introduce a color gradient (see Section~\ref{subsec:color_grad}).

\subsection{Results from the Bulge-disk Decompositions}\label{subsec:2comp}

\begin{figure}
\includegraphics[width=\linewidth]{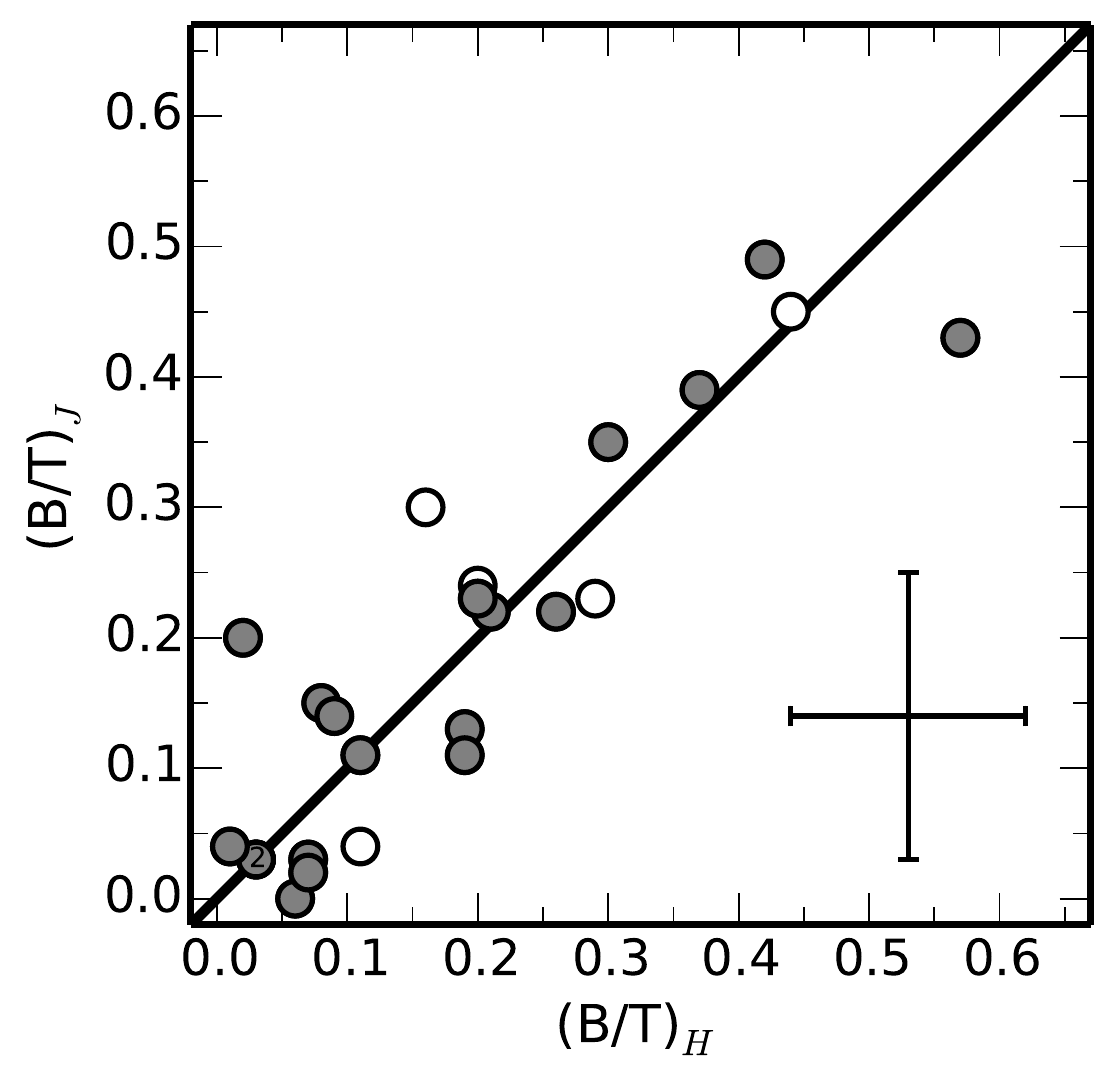} 
\caption{Comparison of the $B/T$ of the $J$- and $H$-band images for a subsample of 24 out of 29 galaxies for which the bulge-disk decomposition is performed successfully; about $40\%$ of the whole sample sample show a substantial bulge components with $B/T\ga0.2$ (see also Table~\ref{tbl:measurements_overview} and \ref{tbl:measurements_overview2}). The white symbols are galaxies that are flagged because $r_b>r_d$.} 
\label{fig:BT_compare}
\end{figure}

Table~\ref{tbl:measurements_overview} presents the results for the bulge-disk decompositions on the $H$-band, which we have fitted first. We find bulges of all sizes $r_{e,b}$ ($0.2-4.2$ kpc) and S\'{e}rsic indices $n_b$ ($1.0-8.0$). We find that the majority of galaxies are fitted the best with an $n_b\sim1$ (16 galaxies), while five and three galaxies are fitted well by $n_b=8$ and $n_b\sim4$, respectively. As discussed in Section~\ref{subsec:bddecomp} and Appendix~\ref{app:BD_Deco}, the bulge is of the size of the PSF and therefore only barely resolved in most cases, making the degeneracy between $r_{e,b}$ and $n_b$ strong. We therefore cannot make any strong statements about the shape of the bulges in our sample. 

We can now go one step further and look at the $J$-band bulge-disk decomposition. After applying the selection described in Section~\ref{subsec:bddecomp}, we end up with a total of 24 galaxies with a bulge-disk decomposition in $J$ and $H$ (16 galaxies with R1, 7 with R2, and 1 R3), out of which 5 galaxies are flagged owing to $r_b>r_d$. All results are listed in the columns `Double Component' in Tables~\ref{tbl:measurements_overview} and \ref{tbl:measurements_overview2}. In Figure~\ref{fig:BT_compare} we compare the $B/T$ for the two filter bands. There is a good agreement between the $J$- and $H$-band $B/T$ values. This is not surprising since the galaxies have very similar morphologies in the two bands. We find substantial bulge components in about half of our sample: $40\%$ of the galaxies have a $B/T\approx20-60\%$, and about 15$\%$ of galaxies show well-developed bulge components with $B/T>0.3$. About two to three galaxies of our sample (i.e., $7-10\%$) are bulge dominated with $B/T\ga0.5$, which is consistent with investigations of much larger samples: \citet{bruce14} presented H-band bulge-disk decompositions of $\sim400$ galaxies with $>10^{11}~M_{\odot}$ at $1<z<3$ and found that $11\pm3\%$ of the massive SFGs are bulge dominated. 

As discussed in the next section, the high SFR in the outskirts of our galaxies is causing the disk to outshine the inner bulge components. Therefore, the quoted numbers above are lower limits, i.e. the typical $B/T$ in stellar mass will be higher than what we measure in light.


\section{Color Distributions within Galaxies}\label{sec:color}

In this section we analyze the observed $J-H$ color distribution within the galaxies, presenting 1D color profiles and 2D color maps. We also convert the observed $J-H$ color to the rest frame $(u-g)_{rest}$ color to be able to compare our $z\sim2$ galaxies with local, $z\sim0$ ones. For constructing the color profiles and color maps, we have mainly followed the careful analysis of \citet{cibinel13b}. For our galaxies at redshift $2<z<2.5$, the $H$-band probes the emission redward of the age-sensitive Balmer/4000$\mathrm{\AA}$-break and the bulk of stellar mass, and the $J$-band probes blueward of the break. Exploiting the degeneracy between stellar population's age, dust extinction, metallicity, and star formation history, we can derive from the $J-H$ color the mass-to-light ratio variations without requiring any knowledge about these parameters. The $J-H$ color is therefore extremely useful for galaxies at $z\sim2$.

\subsection{Color Profiles}\label{subsec:color_prof}

The $J-H$ color profiles are based on the difference of the individual surface brightness profiles $I(r)$ of the $J$- and $H$-band. We use the single-component S\'{e}rsic models, which are described in Section~\ref{subsec:GALFIT}. The advantage of this approach is the substantial removal of the observational biases, including PSF smearing. The PSF-convolved color profiles are shown in Figure~\ref{fig:data_overview}-\ref{fig:data_overview_end}, including their uncertainties, which are substantial at large radii. 

In a large fraction of galaxies (19 out of 29), the color profiles show no significant color variation from the centers to their outskirts ($\la0.3$ mag). For 10 galaxies ($\sim 1/3$ of the sample) we find negative color gradients such that their cores are redder. These color gradients can be explained by both, radial variation in dust content, age of the stellar populations, or a combination of both. To be able to disentangle the contribution from dust and age to the reddening of the stellar population, we would need additional passbands in the blue spectral range\footnotemark[5]\footnotetext[5]{In Cycle 22 (GO13669, PI Carollo), we will obtain WFC3 F438W images, sampling the rest frame far-UV ($\sim1400~\mathrm{\AA}$), to measure the slope of the UV continuum throughout the galaxies and to break the degeneracy between the stellar population's age, star formation history, and dust extinction.}. In addition, there are six galaxies (GK-2303, ZC404221, ZC407302, ZC409985, ZC413597, and ZC415876) that show a peculiar color distribution: the color profile is slightly increasing out to $\sim~r_e$ and then declining. Such trends indicate the presence of red clumps in the outskirts, as seen in some color maps that we present next.

\subsection{Derivation and Analysis of the Color Maps}\label{subsec:color_map}

\begin{figure}
\includegraphics[width=\linewidth]{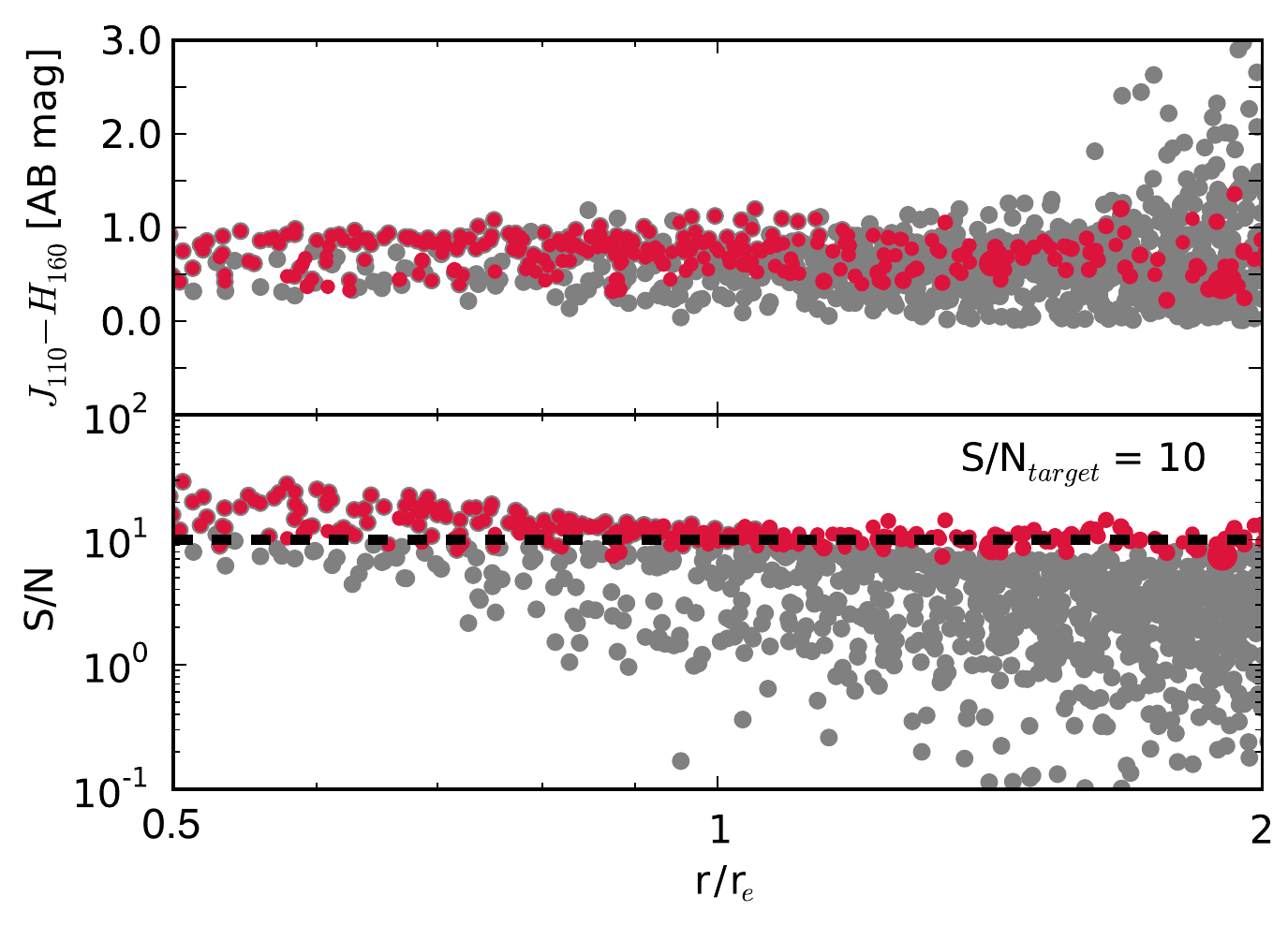} 
\caption{Effect of the Voronoi Tessellation (VT) on a pixel by pixel analysis for the galaxy ZC400569. The top panel shows the $J-H$ as a function of normalized radius. The gray points indicate the data before VT, the red ones after applying VT with a target S/N of 10 (the size of the points corresponds to the size of the bin). In the bottom panel, the S/N is plotted as a function of normalized radius. } 
\label{fig:VT_effect}
\end{figure}
\begin{figure*} \begin{center} \leavevmode
\includegraphics[width=\textwidth]{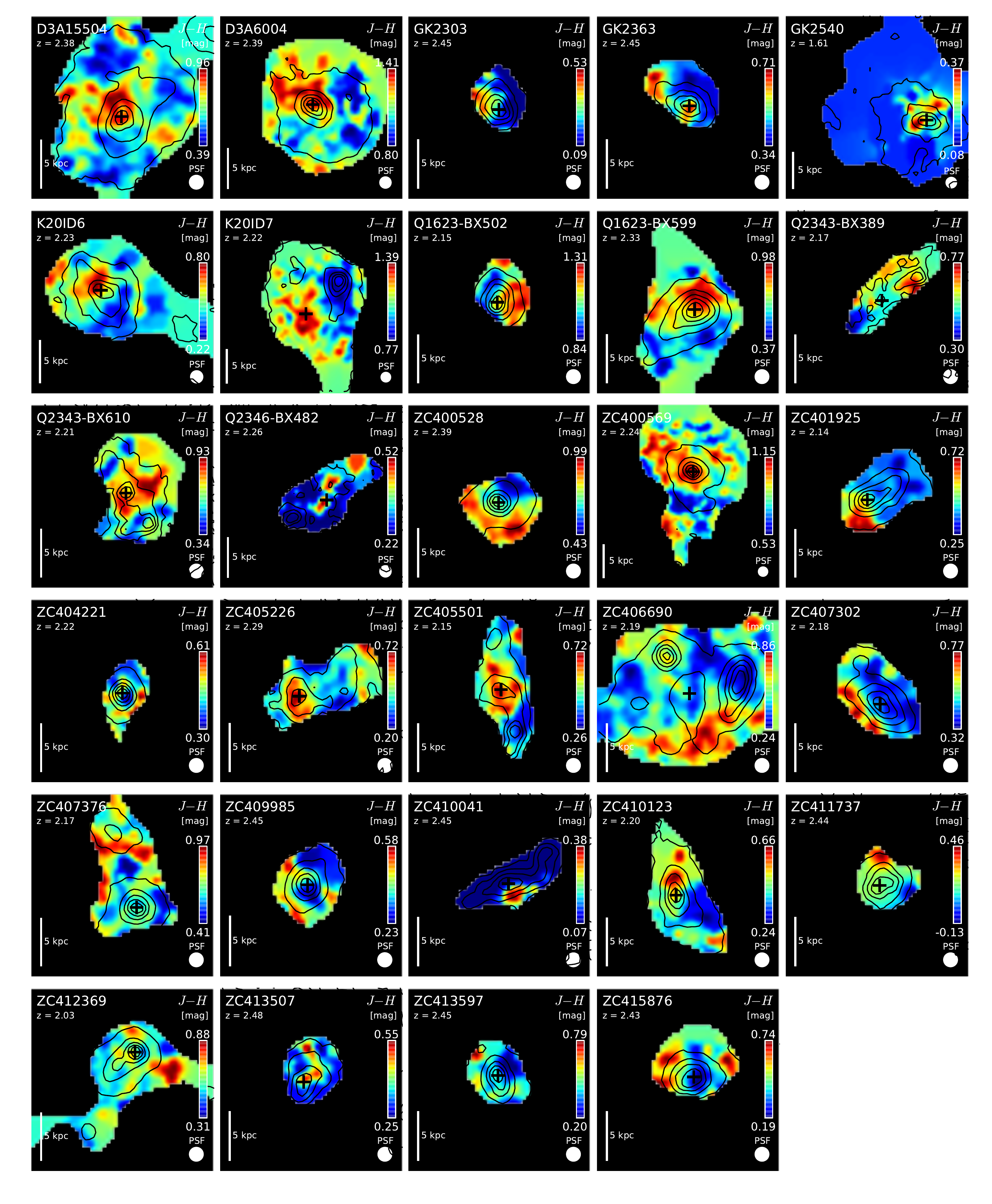} 
\caption{Observed $J-H$ color maps. The color scale is linear and is different for each galaxy (indicated by the color bar to the right). In each stamp, name, redshift, scale of 5 kpc, and PSF size are indicated. The black plus sign `+' marks the fiducial center of the galaxy, and the contour shows the surface brightness distribution of the $H$-band image. Several galaxies show red centers in comparison with their outskirts.} 
\label{fig:col_MS}
\end{center}\end{figure*}
\begin{figure} 
\includegraphics[width=\linewidth]{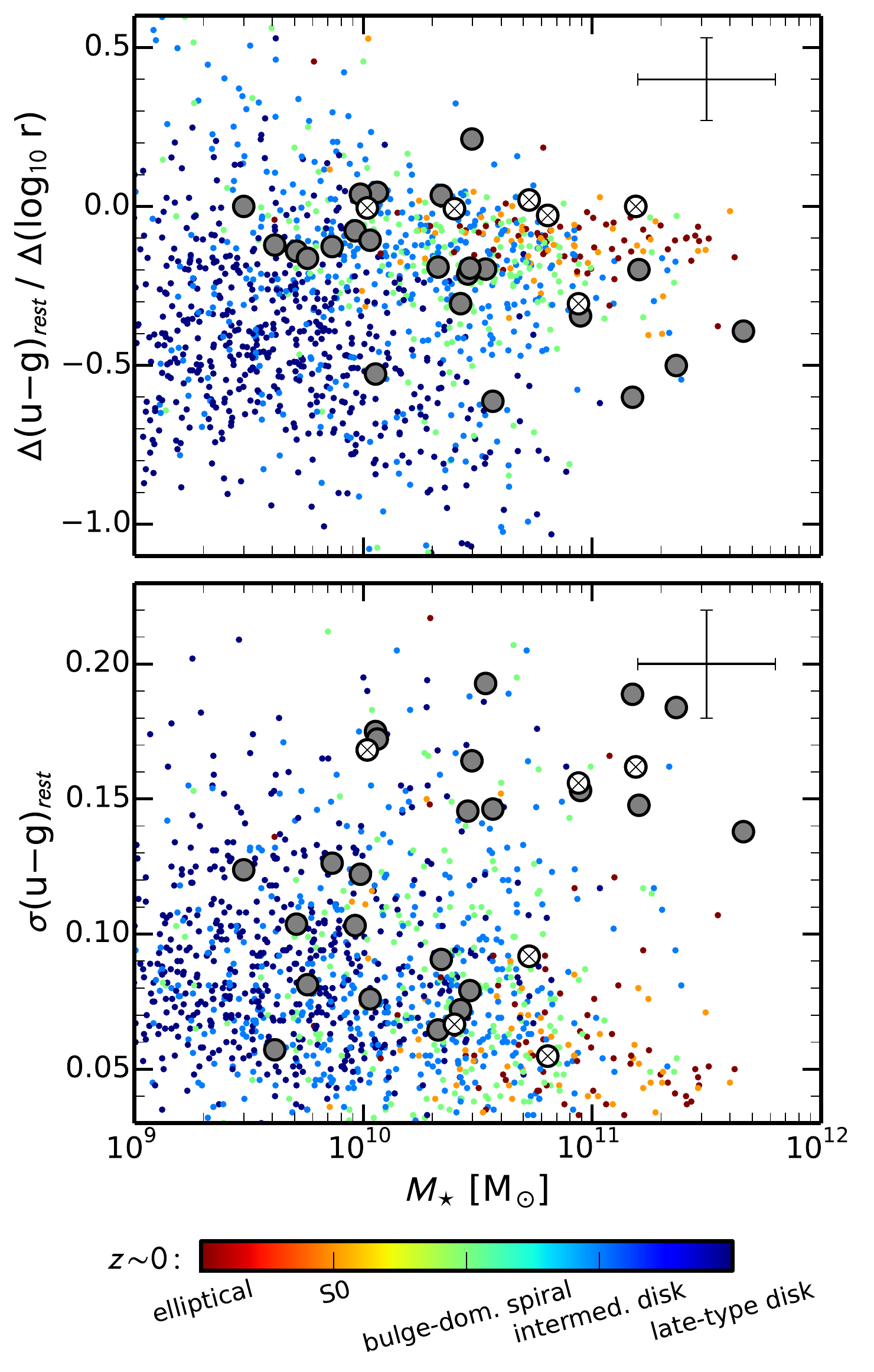} 
\caption{Color gradient (top) and color dispersion (bottom) as a function of stellar mass $M_{\star}$. We measure a wide range in PSF-corrected color gradients $\Delta(u-g)_{rest}/\Delta(\log_{10}r)=[-0.66,0.21]$, where the most massive systems have the steepest negative gradient, i.e. have redder centers in comparison to their blue outskirts. The color dispersion around the average color profile, $\sigma(u-g)_{rest}$, increases with stellar mass, i.e. hinting at clumpier features in more massive galaxies. We compare our $z\sim2$ sample with the local $z\approx0$ ZENS sample \citep{carollo13, cibinel13b, cibinel13a}, color-coded based on morphology. } 
\label{fig:col_grad}
\end{figure}

First, the $J-H$ color maps are obtained by taking the difference between the $J$- and $H$-band image (in magnitudes). A good alignment of the two images is crucial, as offsets down to a fraction of a pixel can generate a fake color gradient. As mentioned in Section~\ref{subsub:astrometry}, we pay careful attention to the alignment of the images in the data reduction: we find a mean offset below 0.20 pixels, with an rms below 0.10. If we use a S\'{e}rsic profile and take the worst-case scenario, i.e., a 0.30 pixel offset with $r_{e} = 0\farcs5$ (10 pixels), we get a maximum color offset below 0.07 AB mag. Hence, we can safely discard misalignments as a possible source of systematic uncertainty in the color gradients. 

Another key step for obtaining the color map image of an extended source is to quantitatively select the pixels in the color map with reliable color determination. For this, we compute the S/N of the color map image pixel by pixel. Adopting the approximation that the Poisson noise distribution function in an HST image is close to a lognormal law, one can obtain that for two images with signals $\mu_{F_J}$, $\mu_{F_H}$ and noises $\sigma^2_{F_J}$, $\sigma^2_{F_H}$, the noise of the $J-H$ color image satisfies (see \citealt{zheng04} for details)

\begin{equation}
\sigma^2 = \log_{10} \left(\frac{\sigma^2_{F_J}}{\mu^2_{F_J}}+1\right) + \log_{10} \left(\frac{\sigma^2_{F_H}}{\mu^2_{F_H}}+1\right) .
\end{equation}

To increase the S/N of the color maps in the outer regions of a galaxy, where the flux from the sky background is dominant, we perform an adaptive local binning of pixels using a Voronoi tessellation (VT) approach. The main idea behind the method is to group adjacent pixels into bigger units that have a minimum scatter around a desired S/N. We perform the VT on the $J-H$ color maps by using the publicly available IDL code of \citet{cappellari03} and adopting the generalization (weighted VT) proposed by \citet{diehl06} (see also \citealt{cibinel13a}). 

Pixels with very low S/N, which would affect the robustness of the algorithm, are excluded from the binning by imposing a minimum threshold of $\mathrm{S/N}=1.0$. A target S/N of 5-10 was chosen to construct the binned color maps, depending on the average S/N of the input color map. The final color maps are obtained by extrapolating the nontessellated pixels from their neighbors, as well as at very large distance from the center, by extrapolating toward the median of the tessellated color maps. The effect of the VT is shown in Figure~\ref{fig:VT_effect}, where we show the S/N as a function of radius before and after VT. The VT clearly enables a more robust measurement of the color distribution of galaxies at large radii and removes the high-frequency fluctuation associated with the noise in the original color maps, while retaining substantial information on lower-frequency, physical color variation within galaxies. 

An overview of the $J-H$ color maps is given in Figure~\ref{fig:col_MS}. A large fraction of galaxies (e.g., D3A-15504, D3A-6004, Q2343-BX389, ZC405501) show a red center. Several galaxies also show blue and red substructures in the outskirts that are $\la0.3$ mag bluer and redder, respectively. In \citet{forster-schreiber11b} we found that -- consistently with our findings in Figure~\ref{fig:col_MS} -- clumps identified at different wavelengths do not fully overlap: NIR-identified clumps tend to be redder/older than $I$-band or H$\alpha$ identified clumps without rest frame optical counterparts. We further discuss this clumpy structure in the $J-H$ color maps in Section~\ref{subsec:color_grad} (and Figure~\ref{fig:col_grad}).

\subsection{Color Gradients and Color Dispersions}\label{subsec:color_grad}

Since we are interested in comparing our $z\sim2$ galaxies with local $z\sim0$ galaxies, we convert the observed $J-H$ color to the rest frame $u-g$ color (SDSS filter bands), using the best-fit SED of each galaxy individually. The color conversion amounts, on average, to $+0.14$ mag. From the azimuthally averaged $u-g$ color profile, we determine the color gradient, $\Delta(u-g)_{rest}/\Delta(\log_{10}r)$, i.e. change in color per decade in radius. Logarithmic color gradients are calculated by fitting the linear relation $(u-g)_{rest} = (u-g)_{rest,r_{e}}+\alpha\cdot\log_{10}(r/r_{e})$ to the color radial profiles. The slope $\alpha=\Delta(u-g)_{rest}/\Delta(\log_{10}r)$ defines what we will refer to as the `radial color gradient'; $(u-g)_{rest,r_{e}}$ defines the color at the galaxy's effective radius $r_{e}$. Fits are performed within the radial range $0.5r_{e}-1.5r_{e}$. The error bars of the color gradients are estimated by varying the fitting range: the upper limit of the fitting range is varied from $0.7r_{e}$ to $3.0r_{e}$ while letting the lower limit remain constant.

The color maps contain, of course, more information than 1D profiles and gradients; in particular, they enable us to study also the color rms dispersion (scatter) around the smooth average color profiles within galaxies. This is defined and computed as $\sigma(J-H)=\sqrt{\sum_i \xi_i^2/N}$, with $\xi$ being the residuals with respect to the azimuthally smoothed radial color profile \citep{cibinel13b}. The observed internal color structure and dispersion in galaxies at high redshift were considered previously. For example, \citet{papovich03, papovich05} investigated this issue in a quantitative fashion by using a dedicated statistical tool.

The results of the color gradients and dispersions as a function of stellar mass are shown in Figure~\ref{fig:col_grad}. The white crossed points indicate again the objects with less reliable fits. We measure a mean color gradient of $\Delta(u-g)_{rest}/\Delta(\log_{10}r)=-0.17$ mag per dex with a wide range of $[-0.61, 0.21]$. In our sample, there is a weak trend that for more massive systems the color gradient is steeper and more negative, i.e. they have redder centers and bluer outskirts. A simple linear regression (mass in log) gives a marginally negative slope of $-0.16\pm0.06$. To check whether the low-mass ($M_{\star}<2\cdot10^{10}~\mathrm{M_{\odot}}$) and high-mass ($M_{\star}>7\cdot10^{10}~\mathrm{M_{\odot}}$) galaxies are drawn from the same distribution, we apply the \citet{anderson52} test to falsify the null hypothesis that the high- and low-mass samples are drawn from the same population. We find a $p$-value of 0.02, i.e. we can reject at about $2.3\sigma$ confidence level the null hypothesis. 

Simultaneously, these massive galaxies have also a higher color dispersion, i.e. these galaxies have more variation in their color maps, showing that these galaxies have substructure in the color distribution. The mean color dispersion amounts to 0.12 mag with a range of $[0.05, 0.19]$. We find a positive slope of $0.03\pm0.01$. Using again the \citet{anderson52}, we find a $p$-value of 0.02, i.e. we can reject at about $2.3\sigma$ confidence level the null hypothesis that low- and high-mass galaxies at $z\sim2$ are drawn from the same population. Clearly, these results presented here have to be verified on larger and more homogeneously selected samples in the future. A comparison with the underlying $z\sim0$ sample is done in Section~\ref{subsec:discusscolors}.

\begin{figure*} \begin{center} \leavevmode
\includegraphics[width=\textwidth]{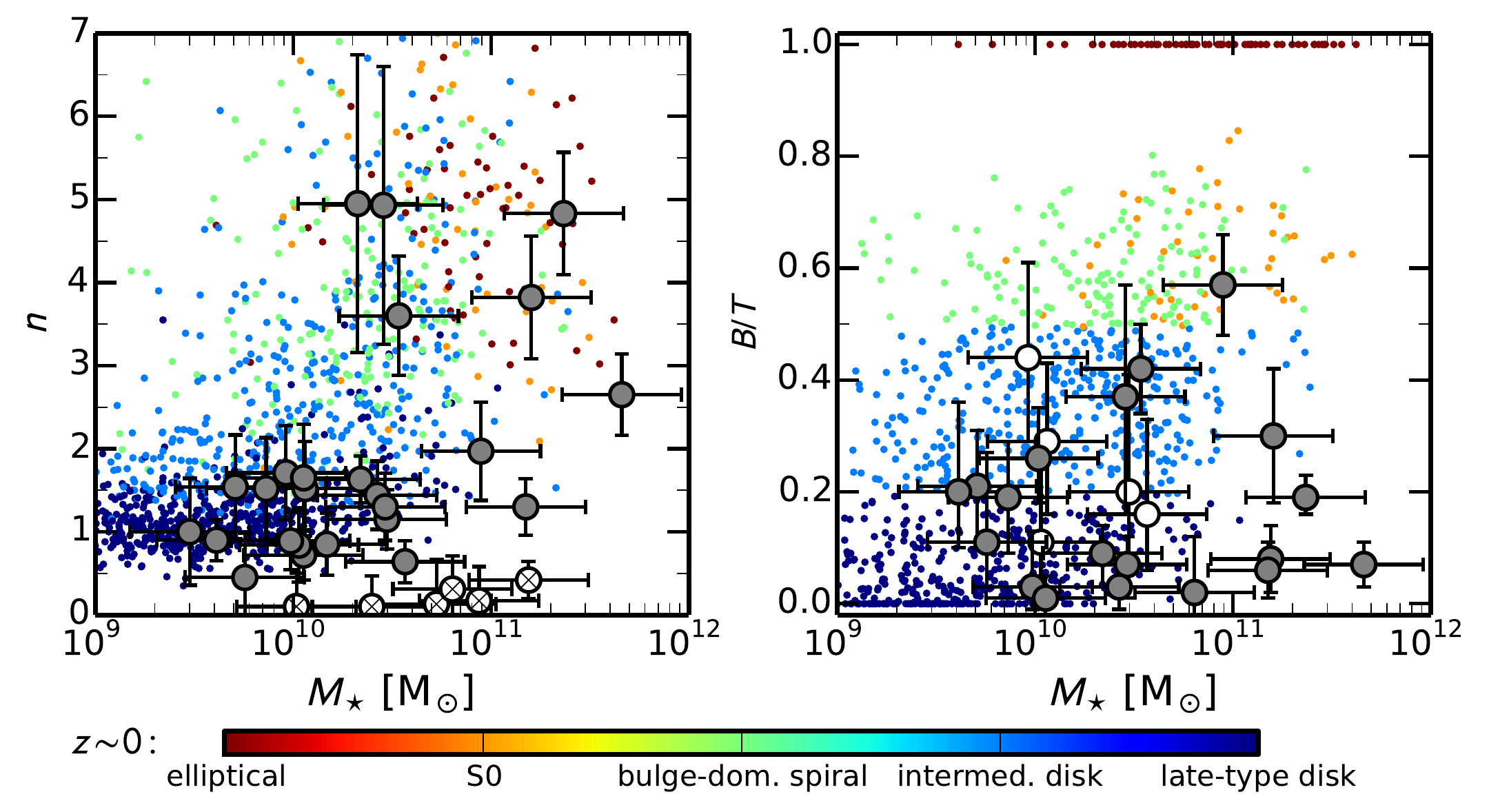} 
\caption{Left: $H$-band S\'{e}rsic indices $n$ as a function of stellar mass $M_{\star}$. The cross symbols indicate the galaxies with less reliable fits. Right: $H$-band $B/T$ as a function of $M_{\star}$. The white symbols indicate the galaxies where $r_b>r_d$. We compare our $z\sim2$ sample with the local $z\approx0$ \textit{ZENS} sample of \citet{carollo13}, which is color-coded based on galaxy morphology. Most of the more massive $z\sim2$ systems ($\ga10^{10}~M_{\odot}$) have $n$ similar to local early-type galaxies, while the less massive galaxies have $n$ values comparable to local late-type disks. The absence of the upturn of $B/T$ in the high-mass regime can be explained by the outshining of the quiescent central component by the bright, star-forming disk. } 
\label{fig:GALFIT_star}
\end{center}\end{figure*}

\section{Discussion: Comparisons with other $z=2$ samples and with the $z=0$ galaxy population}\label{sec:discussion}

In this section we briefly investigate the global relations followed by all galaxies in our sample, in comparison with other $z\sim2$ samples and also with the bulk galaxy population in the local Universe. Specifically, we compare the light-profile parameter S\'{e}rsic index $n$, $B/T$, size $r_e$, color gradient, and color dispersion for our galaxies with independent samples at similar redshifts and $z\sim0$ galaxies of similar mass. As mentioned in Section~\ref{sec:GalaxySample}, our galaxy sample is a representative sample of the $z\sim2$ star-forming main-sequence, but missing the population of quiescent galaxies that are already in place at those epochs \citep[e.g.,][]{ilbert13, muzzin13}. 

The motivation for comparing our $z\sim2$ with $z\sim0$ galaxies lies in the investigation of the parameter space of a representative sample of $z\sim2$ main-sequence galaxies with respect to a sample of their $z\sim0$ descendants. In particular, we want to know where in parameter space the two populations overlap and where they are distinct, i.e. what kind of $z\sim0$ galaxies occupy the same parameter space as our $z\sim2$ systems. From this, we can learn how they have to evolve over 10 billion years from $z\sim2$ to $z\sim0$.

We use our \textit{Zurich ENvironmental Survey} \citep[\textit{ZENS}; ][]{carollo13, cibinel13b, cibinel13a} as our main $z=0$ comparison sample. The ZENS sample consists of about 1600 galaxies within the narrow redshift range $0.0500<z<0.0585$. We use ZENS as our $z=0$ benchmark, rather than cataloged data for a much larger sample, because of (\textit{i}) our start-to-end understanding of its data quality, and all assumptions behind the measured quantities; and (\textit{ii}) the fact that its manageable sample size has allowed us to inspect and cross-check, for each galaxy individually, all parameter measurements, thereby minimizing the impact of systematic errors in the data processing and resolving any inconsistency in such parameters. The overall trends of basic measurements in ZENS (such as size-mass comparisons) agree well with measurements for larger samples (see Figure~\ref{fig:mass_size} and \citealt{cibinel13b}).

\subsection{The S\'{e}rsic index and $B/T$ versus Stellar Mass} \label{subsec:serBT_M}

In Figure~\ref{fig:GALFIT_star} we show the best-fit S\'{e}rsic index $n$ of the single-component fits (left panel) and $B/T$ (right panel) of the light distribution in $H$-band, as a function of stellar mass $M_{\star}$. The color-coded points show the local $z\approx0$ galaxies from the \textit{Zurich ENvironmental Survey} \citep[\textit{ZENS}; ][]{carollo13, cibinel13b, cibinel13a}. The color-coding of the \textit{ZENS} sample is based on the galaxies' morphological classification: ellipticals, S0, bulge-dominated spirals, intermediate disks, and late-type disks. 

At the lower masses, $z\sim2$ galaxies have S\'{e}rsic indices $n\approx0.5-2.0$, i.e., typical of $z=0$ late-type disks; the galaxies with $n\ga3$, i.e., with structural properties that are similar to those of local early-type galaxies, are all more massive than $>10^{10}~M_{\odot}$. Overall, the $z\sim2$ galaxies cover the same $n-M_{\star}$ space as the local $z\sim0$ galaxies. When we analyze, however, the $B/T$ distribution of the $z\sim2$ galaxies (right panel of Figure~\ref{fig:GALFIT_star}), we find several galaxies with high $M_{\star}$, but only small $B/T<0.10$, which are very rare at $z\sim0$. A first reason for this, as mentioned in Section~\ref{subsec:2comp}, is that these massive galaxies at $z\sim2$ are still heavily star-forming in comparison with their local counterparts. Since most of the star formation takes place in the outskirts (see next section), the outer parts are substantially brighter than the central component, leading to outshining of the central bulge and an underestimation of $B/T$ for the most massive galaxies \citep{carollo13a, carollo15}. The difference between the $z\sim2$ and $z\sim0$ population can also be explained by our selection of only SFGs: adding passive or nearly passive galaxies to our $z\sim2$ sample will add several galaxies in the high-$M_{\star}$ and high-$B/T$ regime \citep[e.g.,][]{bruce14}. For our intermediate-mass galaxies -- $M_{\star}\sim10^{10}~M_{\odot}$  -- the central regions are still star forming and therefore very bright. This explains why the bulge-to-disk decomposition performed on the light distributions converges to relatively high $B/T$ values of $\sim0.3-0.6$.

\subsection{Size versus Stellar Mass Relations} \label{subsec:mass-size}

\begin{figure}
\includegraphics[width=\linewidth]{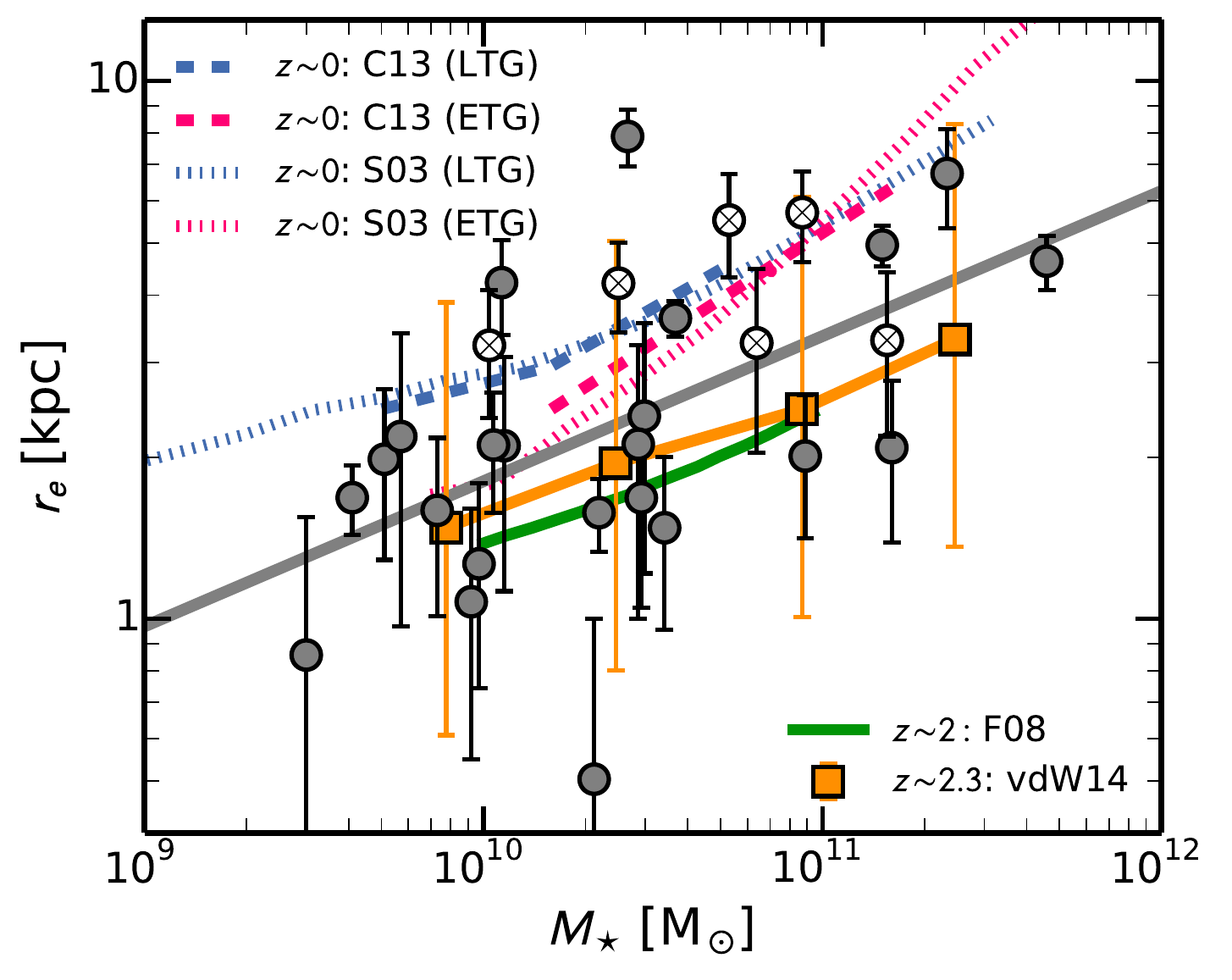} 
\caption{Stellar mass vs. size relation ($M_{\star}-r_e$) for our galaxy sample in comparison with other samples in the literature. The gray points show our data points, with cross symbols indicating galaxies with less reliable fits. The gray line shows the fit to our data points. At redshift 0, we compare with the relations of \citet[S03]{shen03} and \citet[C13]{cibinel13a} that are shown as dotted and dashed lines, respectively. The late-type galaxies (LTGs) and early type galaxies (ETGs) are highlighted in both samples with magenta and blue, respectively. The other observational data correspond to the published $z\sim2$ samples of \citet[F08]{franx08} and \citet[vdW]{van-der-wel14}. The selection criteria of each of these studies are discussed in the text. The small systematic shift between the different high-$z$ samples highlights the impact of different definitions of quantities and sample selection criteria on this relation.} 
\label{fig:mass_size}
\end{figure}

The rest frame optical sizes $r_e$ obtained by fitting with \texttt{GALFIT} the HST $H$-band images are shown as a function of stellar mass in Figure~\ref{fig:mass_size}. Specifically, the $M_{\star}- r_e$ relation for our galaxies is compared with the $z\sim2$ estimates of \citet{franx08} and \citet{van-der-wel14} and also with the $z=0$ relations for early- and late-type galaxies of \citet{shen03} and \citet{cibinel13a}. When appropriate, results have been put on a common ground by converting all values to a Chabrier IMF and to circularized effective half-light radii; stellar masses defined as `actual' masses (excluding mass returned to the ISM) have been shifted upward by 0.14 dex (for SFGs; e.g., \citealt{bruzual03}), so as to be comparable with our masses (which are defined as the integral of the SFR). The \citet{franx08} sample is a $K_s$-selected sample with photometric redshifts. The sizes have been determined in the band redward of the redshifted 4000$\mathrm{\AA}$-break and closest to the rest frame $g$-band. The \citet{van-der-wel14} sample is extracted from the HST CANDELS survey; these authors used a similar approach to size measurements to the one used in this work; therefore, their sizes should be directly comparable with ours. We limit the comparison to their SFGs (based on UVJ-selection) in the redshift bin $2.0<z<2.5$. The error bars indicate the $16\%-84\%$ range and show a large spread in the measured sizes for a given stellar mass. 

A fit to our data in Figure~\ref{fig:mass_size} gives $r_e\propto M^{0.3\pm0.1}$ (including only the reliable fits), which is close to the size-mass relation of \citet{van-der-wel14}. At any given stellar mass, there is a large dispersion of measured sizes: the most massive galaxies in our sample have sizes similar to local $z\sim0$ early-type galaxies of the same mass, while others lie clearly below the local relation. It is well known that there is evolution with cosmic time of the median size-mass relation for both star-forming and quenched galaxies, although high-$z$ galaxies as large as correspondingly massive $z=0$ counterparts have also been identified \citep[e.g.,][]{onodera10b,mancini10}. The origin of this evolution is debated \citep[e.g., ][]{daddi05, trujillo07, mcgrath08, van-dokkum08, szomoru11, newman12a, barro13, dullo13, poggianti13, shankar13, carollo13, cassata13, van-der-wel14}. As argued by e.g., \citet{carollo13, cassata13, poggianti13}, understanding the causes of this evolution requires taking into account the evolution of the number densities of galaxies of a given size at each epoch, which is beyond the scope of this paper. Here we simply highlight two important points related to our sample: $(i)$ our galaxies are $\sim0.13$ dex above, but still within the scatter, the size-mass relation of other $z\sim2$ samples (our galaxies are on average $1.3$ times larger at a given mass); and $(ii)$ at the same time, the small shifts between the different samples highlight the impact of difference in the measurement definitions of sizes, star formation flag (e.g., based on H$\alpha$ flux in our study and on a color selection criterion in van der Wel et al.), and masses (e.g., through different galaxy template libraries). The small offset toward larger size comes from selecting SINFONI-AO targets with slightly larger sizes so as to be able to resolve the dynamics better.

\subsection{Information from the resolved Color Properties}\label{subsec:discusscolors}

In Figure~\ref{fig:col_grad} we compare the $(u-g)_{rest}$ color gradients and color scatter measurements for our sample with the corresponding measurements for the $z\sim0$ of the \textit{ZENS} sample split into different morphological types. 

Our $z\sim2$ sample spans a similar broad range of color gradients as the $z=0$ late-type disk galaxies. The galaxies with the highest masses ($M_{\star}>7\cdot10^{10}~\mathrm{M_{\odot}}$) in our sample show slightly steeper color gradients than correspondingly massive galaxies at $z=0$: the median color gradient of our $z\sim2$ galaxies is $-0.34\pm0.10$, while the one of $z\sim0$ galaxies is $-0.16\pm0.01$. For the massive galaxies ($M_{\star}>7\cdot10^{10}~\mathrm{M_{\odot}}$), the \citet{anderson52} test indicates that the null hypothesis that the high- and low-$z$ samples belong to the same parent population concerning their color gradient can be rejected with a low confidence level  ($\sim 2.1\sigma$). At such high masses, the local counterparts are, however, in contrast with our $z\sim2$ systems, quenched galaxies with an early-type morphology. 

The central $(u-g)_{rest}$ colors within 1 kpc of our massive $z\sim2$ galaxies are $\sim0.47$ mag bluer on average than those of the $z=0$ massive early-type galaxies (median values of $1.22\pm0.16$ and $1.69\pm0.14$ at $z\sim2$ and $z=0$, respectively, with the error quantifying the scatter). This color difference is most likely a combination of active star formation and younger stellar ages in our $z\sim2$ galaxies. The negative color gradients at $z=2$ (steeper than for equivalent massive $z\sim0$ galaxies) suggest, however, that star formation at these early epochs is sustained in the galaxy outskirts. This is also indicated by \citet{genzel14a} and discussed in detail in \citet{tacchella15b}. In terms of different stellar age, we estimate the $(u-g)_{rest}$ color to redden owing to ageing from $z\sim2$ to $z\sim0$ by about $0.5\pm0.1$ mag. 

At lower masses, the color dispersion from the color maps of our $z\sim2$ galaxies is similar to what is measured at $z\sim0$. In contrast, our massive $z\sim2$ galaxies have a higher color dispersion (median value of $0.15\pm0.01$) than the correspondingly massive galaxies at $z\sim0$ ($0.056\pm0.002$). Using again the \citet{anderson52} test, we find that null hypothesis that the high- and low-$z$ samples belong to the same parent population concerning their color dispersion can be rejected at the $4.2\sigma$ level. The higher color dispersion in higher-$z$ galaxies is not unexpected, given that the massive $z\sim2$ galaxies in our sample are star-forming and show clumpy substructure, while the $z=0$ massive counterparts are quenched early-type galaxies. The comparison is nevertheless interesting, since it highlights that quenching of star formation in the massive $z\sim2$ main-sequence population should leave behind similarly massive quenched remnants with patchy, inhomogeneous stellar populations in their outer regions. In addition, when restricting the local sample to SFGs, the color dispersion of high-$z$ systems, particularly at the massive end, still lies among the high color dispersion tail of nearby SFGs. This implies that the star formation distribution in $z\sim2$ galaxies has more substructure and is not as uniformly distributed as at $z\sim0$. As mentioned previously, the presented trends have to be confirmed on larger and more homogeneously selected samples.

\section{Summary}\label{sec:Summary}

We have presented the analysis of $J$- and $H$-band HST NIR imaging for the sample of 29 of the 35 $z\sim2$ SFGs of our SINS/zC-SINF program for which we have acquired SINFONI Adaptive Optics rest-H$\alpha$ integral field spectra at $\sim1-2$ kpc spatial resolution. Together, the HST and SINFONI data provide simultaneous information, for each galaxy on such subgalactic scales, on its old stellar population and thus stellar mass distribution and on the distribution of its ongoing star formation. At a visual inspection of the HST images, the galaxies show a wide range of rest frame optical morphologies. Combining the information on the rest frame appearance of the galaxies with their H$\alpha$ kinematic properties reveals 10 ($35\%$) regular disks, 11 ($38\%$) irregular disks, 3 ($10\%$) merging systems, and 5 ($17\%$) unresolved systems (possibly, but not necessarily, genuine dispersion-dominated systems; see Section~\ref{sec:LinkingSINFHST}). This classification captures well the galaxy properties as observed with the currently available spatial resolution of $1-2$ kpc; it will be interesting to see what additional insights will be gained at yet higher resolution with future 20-40m-class telescopes and James Webb Space Telescope. 

Single S\'{e}rsic analytical fits to the rest frame optical light distributions return for most galaxies profiles with an index $n\sim1$, which is typical of late-type, disk galaxies. About $15\%$ of the systems require, however, substantially higher values, $n>3$, which are shown by early-type galaxies in the local Universe. Such high-$n$ galaxies are all massive systems with $\ga10^{10}~M_{\odot}$ (see Figures~\ref{fig:GALFIT_JH} and \ref{fig:GALFIT_star}). Similar conclusions were drawn from much larger samples, i.e. with the aid of multiwavelength CANDELS data, it has been shown that SFGs have $n\sim1$ and typically low $B/T$, whereas massive, quenched galaxies have $n\sim4$ and higher B/T \citep[e.g.,][]{wuyts11,bruce14,lang14}. 

Two-component bulge+disk fits return measurable bulge components in the {\it light} distributions, i.e., $H$-band bulge-to-total ratios of $B/T\approx20-60\%$, for about $40\%$ of the galaxies (see Figure~\ref{fig:BT_compare}); i.e. about 15$\%$ of galaxies show well-developed bulge components with $B/T>0.3$, in agreement with works on larger samples \citep[e.g.,][]{bruce14}. As found in our analysis and summarized below, the high SFRs in such massive galaxies are most likely distributed (in disks) at large radii, causing the disk to outshine the inner bulge components. Therefore, the frequency of massive $z\sim2$ SFGs with a substantial bulge component in {\it stellar mass} will be higher than the frequency of such galaxies for which we measure a bulge in the light distributions \citep[e.g.,][]{carollo15,lang14}. 

The rest frame half-light radii of the galaxies span the range 1-7 kpc, with a median of 2.1 kpc and mean of 2.6 kpc. On the size-mass plane, these measured sizes imply a relation that overlaps largely with a small bias toward larger sizes, which has previously been measured for other independent $z\sim2$ star-forming samples (see Figure~\ref{fig:mass_size}). This global shift of the relation, by about 0.13 dex, may arise from different systematics in either the size or the mass measurements (or both), and/or on different sample selections, highlighting that care must be paid in comparing samples at different epochs to infer the evolution of the relation with redshift. In the case of our galaxies, larger galaxies were favored in the selection of SINFONI-AO follow-up because for these we are able to resolve the kinematics better. We also emphasize that the size-mass relation that we measure shows a large scatter at all masses, a fact that has already been emphasized by other authors \citep[e.g.,][]{onodera10b, mancini10, van-der-wel14} and that is likely important for understanding how main-sequence SFGs grow in stellar mass and size with cosmic time. 

Averaged across the entire sample, the typical $J-H$ (i.e., $(u-g)_{rest}$) color gradient is negative and about $-0.14$ mag per dex. The most massive galaxies have rather strong (negative) color gradients and central colors within 1 kpc that are consistent, assuming simple ageing of the stellar population, with the central colors of local quenched galaxies of similar mass. This suggests that the high SFRs of such massive galaxies at $z\sim2$ are mostly distributed at large galactocentric distances, and supports independent evidence for an inside-out growth scenario of galaxies (see Figures~\ref{fig:col_MS} and \ref{fig:col_grad}). The two-dimensional $J-H$ color maps indeed show generally red cores and blue outskirts, together with a highly structured, clumpy morphology; many clumps are redder than their surroundings. The color rms dispersion in the lower-mass galaxies spans the whole range that is observed in local galaxies of similar stellar masses. The more massive galaxies show, at large radii, a prominent clumpy structure and, correspondingly, a large color r.m.s. dispersion around their azimuthally averaged color profiles. Quenching of these massive SFGs at $z\sim2$ out of the main-sequence should thus lead to inhomogeneities in the outer stellar populations of today's quenched remnants of similar mass. This is at odds with observations in the local Universe, which show low rms dispersions in the rest-optical color maps of quenched, high-mass early-type galaxies. We speculate that saturation of rest-optical colors at stellar ages $>10$ Gyr and dynamical mixing may be hiding such inhomogeneities in the $z=0$ relics of the quenching process. 

The measurements that we have derived and presented in this paper are collected in a comprehensive catalog, which we publish here, and enable us to address some important questions concerning the growth of bulges and disks around the $z\sim2$ cosmic epoch (including \citealt{genzel14a} and \citealt{tacchella15b}).

\acknowledgments
We thank the referee for a detailed report. It is a pleasure to acknowledge insightful conversations with A. Faisst, M. Onodera, and B. Trakhtenbrot. We acknowledge support by the Swiss National Science Foundation. This research made use of NASA's Astrophysics Data System (ADS), the arXiv.org preprint server, the Python plotting library \texttt{matplotlib} \citep{hunter07}, and \texttt{astropy}, a community-developed core Python package for Astronomy \citep{astropy-collaboration13}.

{\it Facility:} \facility{HST (WFC3, ACS)}, \facility{VLT (SINFONI)}


\begin{appendix}
\section{Global properties of individual galaxies}\label{app:galaxy_by_galaxy}

For each galaxy individually, we briefly comment below on its global properties (stellar mass, SFR, AGN contribution), the fiducial isophotal center that we adopted for modeling its light distribution, and its overall classification (the latter obtained by combining the information on its rest frame optical morphology and on its ionized gas kinematic classification). For more details about the kinematic properties, we refer the reader to to \citep{genzel06, genzel11, newman13a}; about evidence for AGNs from the rest-optical line ratios and profiles, to \citep{daddi03,shapley03,newman14, forster-schreiber14}. As mentioned in Section~\ref{subsec:Centers}, the different centers (light, stellar mass, and kinematic) agree in $> 60\%$ of cases very well ($< 1$ kpc difference). In the cases where the centers disagree, the kinematic center is usually separated by $\ga1$ pixel ($\ga0.4$ kpc) from the center of light and mass. This can partially be explained by the higher uncertainty in the alignment of the SINFONI data with HST ($\sim1$ pixel) than between the individual $J$- and $H$-band images ($<0.2$ pixel).

\paragraph{D3A-15504}
This galaxy has a stellar mass of $15.0\cdot10^{10}~M_{\odot}$ and is therefore one of the most massive ones in our sample. Rest-UV and rest-optical spectral features point to AGN activity. As one sees in Figure~\ref{fig:data_overview}, the galaxy is very extended ($\sim15$ kpc). The center of light and the center of mass agree very well ($\la0.1$ kpc), which we therefore take as the fiducial center for analysis. The dynamical center is off by 1.5 pixels. Looking at the $J$- and $H$-band images (i.e. rest frame optical), the overall structure is smooth, regular, and centrally concentrated. The kinematic classification reveals a rotation-dominated system, with a velocity map that shows nice rotation and a velocity dispersion map that is centrally peaked. We therefore classify this galaxy as a rotating disk. 

\paragraph{D3A-6004}
This is the most massive galaxy ($45.8\cdot10^{10}~M_{\odot}$) in our sample. It is possible that this galaxy has an AGN, since it is lying in the composite region of the BPT diagram \citep{baldwin81}, encompassing the range between the boundaries of local normal SFGs and AGN-dominated systems. The light and mass centers agree perfectly; the dynamical center is shifted by 0.4 kpc. We again assume as fiducial center the center of light. This galaxy has a rather smooth light distribution, with some weak clumpy feathers to the west (right). Note that in the H$\alpha$ light, the narrow H$\alpha$ emission is very faint in the center, while the western clump complex is very bright. We classify this galaxy also as disk-like (rotating disk), in agreement with the kinematic data. With a value of $v_{rot}/\sigma_{0}=9.0\pm2.4$, this galaxy is clearly rotation dominated.

\paragraph{GK-2303}
With a stellar mass of $0.97\cdot10^{10}~M_{\odot}$, this is one of the least massive galaxies in this sample. The center of light and the dynamical center are only 0.6 kpc apart from each other, while the center of mass is about 1.3 kpc away. We chose the center of light as our fiducial center. This galaxy show a compact morphology with a size (diameter) of $\sim5$ kpc. The kinematic data shows a nice rotation pattern. We therefore classify this galaxy as a rotating disk. 

\paragraph{GK-2363}
This galaxy has a stellar mass of $2.92\cdot10^{10}~M_{\odot}$. The center of light and center of mass coincide ($\la0.1$ kpc); the dynamical center is at a distance by 0.6 kpc. As the fiducial center we therefore assume the light center for all the further analysis. This galaxy shows a similar morphology to GK-2303 but is somewhat more extended. This galaxy shows a rotation pattern and is therefore a rotating disk ($v_{rot}/\sigma_{0}=4.2\pm3.2$). 

\paragraph{GK-2540}
GK-2540 has the lowest redshift ($z=1.61$) of all the galaxies in our sample ($z\sim2.2$) and a stellar mass of $2.66\cdot10^{10}~M_{\odot}$. All three centers differ from each other by $\sim0.5$ kpc. We adopt as fiducial center the light-weighted center. This galaxy is very extended, with much low surface brightness structure around the main components. The H$\alpha$ map also shows very extended star formation in the outskirts. With a value of $v_{rot}/\sigma_{0}=8.2\pm5.4$, this galaxy is rotation dominated. The `substructures' show rotation, but possibly disturbed. We classify this galaxy as an irregular disk. 

\paragraph{K20-ID6}
This galaxy has a stellar mass of $3.68\cdot10^{10}~M_{\odot}$. In the rest frame optical light, the galaxy consists of two distinct components: a larger component to the east (left) and a small component to the west (right). On the other hand, the H$\alpha$ line emission is connected to these two components, and consequently, the dynamical center lies farther to the west than the center of light and mass, which themselves are distanced by 0.7 kpc. We set the light-weighted center as the fiducial center. The kinematic of the main northeastern part shows ordered rotation ($v_{rot}/\sigma_{0}=5.8\pm1.8$, i.e. rotation dominated), with some irregularity in the H$\alpha$ velocity field toward the western side. We therefore classify this galaxy as an irregular disk. The southwestern, fainter component, which is well seen in the HST images, lies at the edge of the SINFONI FOV, i.e. the velocity maps are too nosy to make any reliable statement about this part. 

\paragraph{K20-ID7}
This galaxy has a stellar mass of $8.71\cdot10^{10}~M_{\odot}$ and an SFR of $110~M_{\odot}~yr^{-1}$. This galaxy shows a bright clump in the northwest (upper-right), which makes it difficult to model. The mass map reveals that this clump (with the underlying disk) has a stellar mass $\sim5\cdot10^{9}~M_{\odot}$ (6 \%). Therefore, the center of mass and the center of light (as well as the kinematic center) are far away from each other. We use the center of stellar mass, since we model the clump in the northwest as an extra component. The galaxy shows rotation ($v_{rot}/\sigma_{0}=8.5\pm2.5$), with some disturbance. Hence, we classify this galaxy as an irregular disk. 

\paragraph{Q1623-BX502}
This galaxy has the lowest stellar mass of all the galaxies in our sample ($0.3\cdot10^{10}~M_{\odot}$). The center of light and center of mass are in very good agreement ($\la0.1$ kpc). The dynamical center is off by 1 kpc. We assume the light-weighted center as the fiducial center. It has a very compact morphology: this galaxy is a single-peaked source and compact. The velocity maps shows an irregular velocity field, and we therefore classify this galaxy as a dispersion-dominated system. 

\paragraph{Q1623-BX599}
This galaxy has a stellar mass of $8.89\cdot10^{10}~M_{\odot}$ and an SFR of $34~M_{\odot}~yr^{-1}$; therefore, it lies a factor of 2 below main-sequence. In the $H$-band image, the galaxy shows an elongated shape, with low surface brightness emission toward north (top). The light and mass centers agree (difference less than 0.4 kpc), while the dynamical center is 1.1 kpc away from the center of light. We use the center of light as our fiducial center. We classify this galaxy as a merger based on the visual inspection of the rest frame optical images and kinematic data. 

\paragraph{Q1623-BX389}
This galaxy, with a stellar mass of $6.39\cdot10^{10}~M_{\odot}$, has a very elongated shape, indicating that the system is viewed in an edge-on perspective. The small source to the south is a small companion \citep{forster-schreiber11a, forster-schreiber11b}. The dynamical center lies about 5 kpc away from the bright northwestern clump, which is the center of mass and center of light. The dynamical center is the fiducial center because the galaxy is rotation supported ($v_{rot}/\sigma_{0}=4.9\pm0.9$) and classified as an irregular disk.

\paragraph{Q1623-BX610}
This galaxy has a stellar mass of $15.5\cdot10^{10}~M_{\odot}$ and is therefore on the massive side in our sample. For its mass, it has a rather low SFR, $60~M_{\odot}~yr^{-1}$, and is therefore a factor of $\sim3$ below the main-sequence. It shows signatures of AGN activity. It has a clumpy morphology in the $H$-band, with the brightest clump in the southwest (bottom right). All the three centers agree well, i.e. are distanced less than 0.5 kpc from one another. We use the center of light as the fiducial center, which lies between the mass and the dynamical center. Looking at the kinematics, this galaxy is a perfect example for a rotating disk (rotation dominated with $v_{rot}/\sigma_{0}=3.0\pm1.3$). We therefore classify this galaxy as a rotating disk with clumps in its outskirts. 

\paragraph{Q1623-BX482}
Q1623-BX482 ($M_{\star}=2.5\cdot10^{10}~M_{\odot}$) with an SFR of $80~M_{\odot}~yr^{-1}$, has a bright clump in the outskirts, similarly to Q1623-BX389 and K20-ID7. Since the kinematic data reveal much rotation ($v_{rot}/\sigma_{0}=4.1\pm0.9$) and the velocity field is regular, with only some perturbation toward the northwest, we classify this galaxy as an irregular disk and choose the dynamical center as our fiducial center. 

\paragraph{ZC400528}
ZC400528 hosts an AGN and has a large stellar mass of $16.0\cdot10^{10}~M_{\odot}$. All three centers lie close to each other ($\la0.2$ kpc). This galaxy is compact and shows no sign of interaction in the rest frame optical light. The velocity maps shows rotation ($v_{rot}/\sigma_{0}=4.9\pm1.7$), but with some irregularity. This galaxy is therefore a irregular disk.

\paragraph{ZC400569}
The second most massive galaxy in our sample ($23.3\cdot10^{10}~M_{\odot}$) is very extended, with $\sim20$ kpc, and possibly hosts an AGN. The analyzed centers agree perfectly ($\la0.1$ kpc). In the optical light, the inner region shows a smooth disk, which is also revealed by our AO kinematics. Toward the outskirts, there are bright clumps. This galaxy is rotation dominated with $v_{rot}/\sigma_{0}=5.1\pm2.8$ and is classified as a irregular disk. 

\paragraph{ZC401925}
This low-mass ($0.73\cdot10^{10}~M_{\odot}$) galaxy has an elongated morphology. The dynamical center of light lies in the `middle' of the galaxy, while the center of light and center of mass are offset by $\sim2.5$ kpc. Since the center of mass and light agree within 0.2 kpc, we adopt the center of light as our fiducial center. The light distribution is regular but very elongated. The kinematic data therefore reveal much dispersion ($v_{rot}/\sigma_{0}=1.0\pm0.6$), and we classify it as unresolved system. 

\paragraph{ZC404221}
This galaxy has a stellar mass of $2.12\cdot10^{10}~M_{\odot}$ and very compact morphology. The three centers all agree within 0.2 kpc. This galaxy is clearly dispersion dominated with $v_{rot}/\sigma_{0}=0.5\pm0.1$ and is classified as unresolved. 

\paragraph{ZC405226}
With a stellar mass of $1.13\cdot10^{10}~M_{\odot}$ and an SFR of $117~M_{\odot}~yr^{-1}$, this galaxy lies a factor of 7 above main-sequence. It has a large extent, with some indication and hints of spiral structure. The center of light and center of mass are distanced by only 0.1 kpc, while the dynamical center is $\sim1$ kpc off. We take the center of light as our fiducial center. The galaxy shows rotation ($v_{rot}/\sigma_{0}=2.1\pm1.0$) and is classified as disk. 

\paragraph{ZC405501}
This galaxy has a stellar mass of $1.04\cdot10^{10}~M_{\odot}$ and a clumpy morphology. All the three centers agree perfectly. It is only minimally rotation dominated with $v_{rot}/\sigma_{0}=1.6\pm0.5$. We classify this galaxy as a rotating disk. 

\paragraph{ZC406690}
ZC406690 is a highly interesting case. With a stellar mass of $5.30\cdot10^{10}~M_{\odot}$, it shows very extended H$\alpha$ emission. There are several bright clumps allocated in a ring-like structure. The dynamical center, center of light and center of mass disagree. The dynamical center is taken as the fiducial center. Looking at the H$\alpha$ emission line data, the kinematic data reveal large-scale rotation with $v_{rot}/\sigma_{0}=4.7\pm1.2$. We therefore classify this galaxy as an irregular disk.

\paragraph{ZC407302}
The stellar mass of this galaxy is $2.98\cdot10^{10}~M_{\odot}$. The three centers agree reasonably well: the light and mass centers are distanced by 0.3 kpc, the dynamical by $\sim0.6$ kpc. This galaxy has a small clump in the north. The kinematics shows rotation ($v_{rot}/\sigma_{0}=3.0\pm1.1$); however, the clump in the north perturbs the systems, i.e., it is an irregular disk.

\paragraph{ZC407376}
This object consists of two well-separated components, as visible in the $H$-band image and also resolved by the SINFONI data. We choose here to characterize the brighter southern component that is interacting with the northern one. The total stellar mass is $3.42\cdot10^{10}~M_{\odot}$. The dynamical center lies between the two components and is therefore $\la4$ kpc away from the light and mass centers, which lie on the top of each other. We take the center of light as our fiducial center. The galaxy shows no rotation (slightly dispersion dominated with $v_{rot}/\sigma_{0}=1.3\pm0.6$) and regions with different velocity components. We classify this object as a merger. 

\paragraph{ZC409985}
ZC409985 has a stellar mass of $2.19\cdot10^{10}~M_{\odot}$. This galaxy has a compact morphology, with several other systems in its neighborhood. We do not have spectroscopic redshifts for these neighboring systems since they lie outside of SINFONI's FOV. The dynamical center and the center of light agree very well (offset is less than 0.1 kpc). However, the center of mass is distanced by 0.8 kpc. We take the center of light as our fiducial center. The galaxy shows no rotation or strong velocity peaks and has $v_{rot}/\sigma_{0}=0.4\pm0.1$. We therefore classify this galaxy as unresolved.

\paragraph{ZC410041}
This elongated galaxy has a stellar mass of $0.57\cdot10^{10}~M_{\odot}$. The three centers all disagree with one another by $\sim1.5$ kpc. We take again the center of light. In the optical light, the galaxy shows no sign of interaction. ZC410041 also shows a nice rotation pattern ($v_{rot}/\sigma_{0}=2.3\pm1.0$) with a centrally peaked dispersion map. Hence, ZC410041 is classified as a rotating disk (edge-on).

\paragraph{ZC410123}
This galaxy has a stellar mass of $0.51\cdot10^{10}~M_{\odot}$. The center of light and center of mass are lying on the top of each other, while the dynamical center is off by 1.6 kpc. We take the center of light as the fiducial center. The galaxy shows a slight asymmetric shape; however, there is no clear sign of multiple nuclei (probably postmerger system). The velocity map shows rotation but also irregularities toward northeast (dispersion dominated with $v_{rot}/\sigma_{0}=1.3\pm0.7$). We classify this galaxy as an irregular disk. 

\paragraph{ZC411737}
This compact galaxy has a stellar mass of $0.41\cdot10^{10}~M_{\odot}$ and is at the lower end of the mass spectrum of our sample. This galaxy has a very similar light distribution in the $J$- and $H$-band. Therefore, the center of light and center of mass coincide. The dynamical center is offset by 0.8 kpc. The velocity shows ordered rotation ($v_{rot}/\sigma_{0}=1.6\pm0.8$), i.e., it is a disk. 

\paragraph{ZC412369}
This galaxy has a stellar mass of $2.86\cdot10^{10}~M_{\odot}$. It is extended and has several close neighboring clumps. The ``tail'' to the southeast has H$\alpha$ emission at about same redshift, confirming the physical association. The other clumps are out of SINFONI's FOV and we therefore do not have spectral information. The center of mass and center of light agree well, while the dynamical center is offset by $\sim4$ kpc. Based on the velocity and dispersion map, this galaxy is dispersion dominated with $v_{rot}/\sigma_{0}=1.3\pm0.4$, and we classify this galaxy as a merger.

\paragraph{ZC413507}
With a stellar mass of $1.07\cdot10^{10}~M_{\odot}$ and a fluffy morphology, the light and mass centers agree well, while the kinematic center is off by $\sim1$ kpc. 	We take the center of light as our fiducial center. The galaxy shows no sign of interaction in its optical images. The kinematics shows a rotation-dominated galaxy with $v_{rot}/\sigma_{0}=2.1\pm1.2$. Hence, the galaxy is classified as a rotating disk. 

\paragraph{ZC413597}
This galaxy has a stellar mass of $0.92\cdot10^{10}~M_{\odot}$. All the centers agree well. There are several other systems surrounding this galaxy. Based on its kinematic data, we mark it as unresolved ($v_{rot}/\sigma_{0}=1.2\pm0.7$).

\paragraph{ZC415876}
This galaxy has a stellar mass of $1.15\cdot10^{10}~M_{\odot}$. The dynamical and light centers agree well (offset by 0.4 kpc); however, the stellar mass is offset by $\sim3$ kpc owing to a clump. We choose the center of light as our fiducial center. This galaxy is rotation dominated ($v_{rot}/\sigma_{0}=2.4\pm0.8$), and the velocity reveals some rotation as well. We classify this galaxy as an irregular disk.

\section{On Bulge/Disk Decomposition}\label{app:BD_Deco}

\subsection{Bias and Error Estimation of Bulge/Disk Decompositions}

The resolution of the HST/WFC3 data for the $J$- and $H$-bands is $1-2$ kpc. This implies that bulge components are barely resolved. To test whether it is feasible to do a bulge-disk decomposition (double-components fit) or not, we run simulations. In summary, we are able to recover the $B/T$ very well without any bias. However, we are not able to estimate the size and/or S\'{e}rsic index of the bulge, i.e. we are not able to derive any corrections for accounting for any biases.

In the simulations, we create $\sim100'000$ mock galaxies with \texttt{GALFIT}: each galaxy consists of two S\'{e}rsic components (bulge and disk). We vary the $B/T$ (between 0 and 0.9), sizes ($r_{e,d}\in[2.0,4.8]~\mathrm{kpc}$ and $r_{e,b}\in[0.4,2.4]~\mathrm{kpc}$), axis ratios ($b/a\in[0.6,1.0]$), and position angles ($\mathrm{P.A.}\in[-90^{\circ}, 90^{\circ}]$) of the disk and bulge. In addition, we vary the S\'{e}rsic of the bulge $n_b$ between 1 and 6 and the integrated magnitude of the disk between 21.5 and 24.5. These simulated galaxies are convolved with the PSF and added on real background images. Then all the model galaxies are processed in exactly the same way as the real galaxies.

\begin{figure*}
\includegraphics[width=\textwidth]{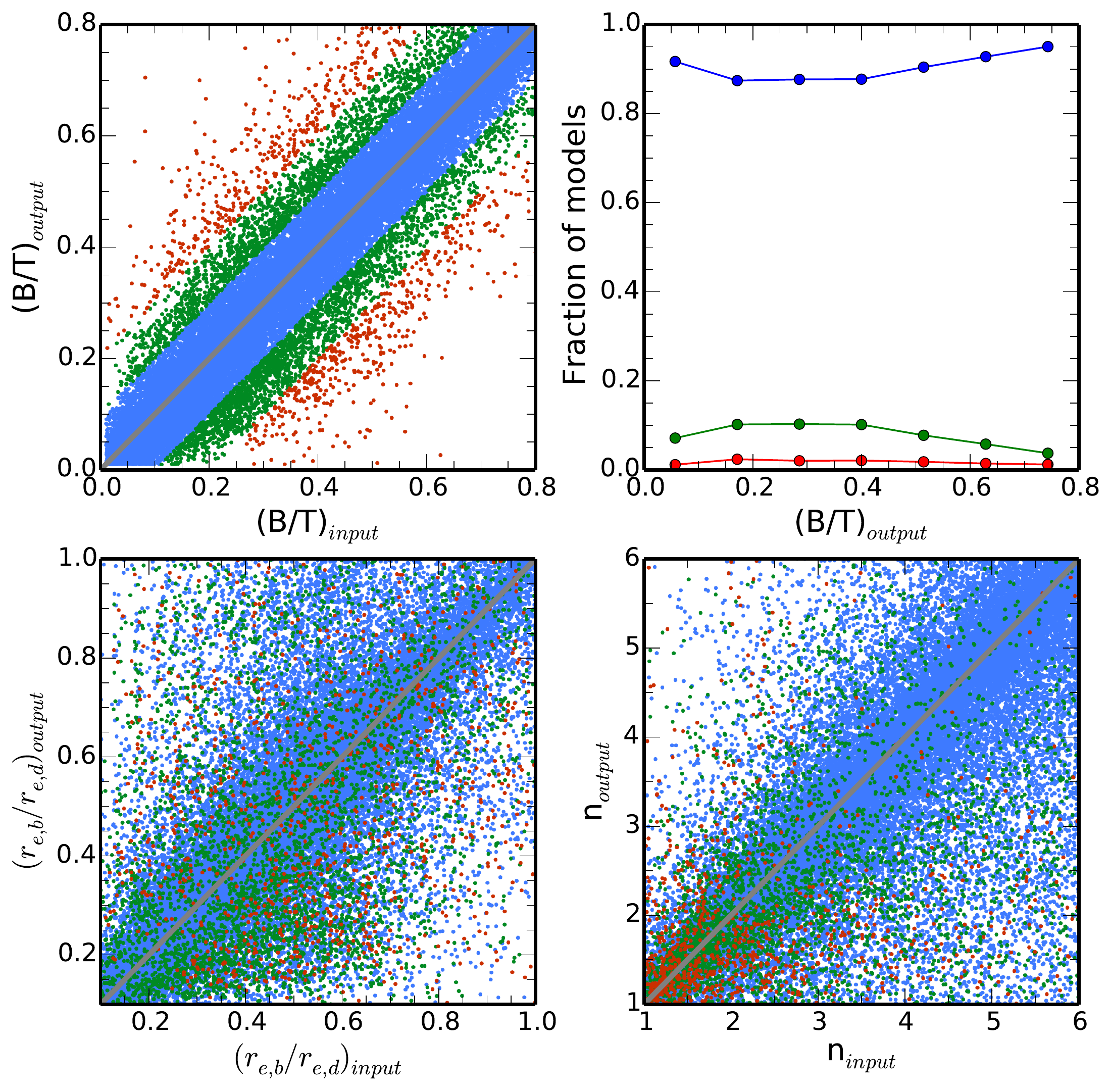} 
\caption{Comparison of input and output obtained from our simulations of the bulge-to-disk decompositions. Top left panel: Comparison of intrinsic and measured $B/T$ for the model galaxies. The gray line shows the one-to-one correspondence. Blue points (in all panels) indicate models for which $|\Delta(B/T)|<0.15$, green those with $0.15<|\Delta(B/T)|<0.3$ and red is for models for which the measured B/T deviates more than 0.3 from the intrinsic value. Our simulation shows that we are able to recover the intrinsic $B/T$ very robustly, i.e. the fraction of wrong recovered $B/T$ is very small, $<10~\%$ (see top right panel). In the two bottom panels, we display the intrinsic and measured ratio of the size of the bulge to the disk $r_{e,b}/r_{e,d}$, and S\'{e}rsic index of the bulge, respectively. There is a large scatter around the one-to-one relation, implying that the uncertainties in the bulge size and shape are very large. Furthermore, there is evidence for overestimating bulge sizes. However, owing to large scatter, we are not able to derive any correction function. Furthermore, we see that the points with the largest difference in intrinsic and measured $B/T$ values are the ones with the lowest bulge S\'{e}rsic index, i.e. in these cases, the bulges are indistinguishable from the disk.} 
\label{fig:BT_simulation}
\end{figure*}

The result of the simulations, i.e. the comparison of the measured to the intrinsic quantities, is shown in Figure~\ref{fig:BT_simulation}. We find that we are able to recover the $B/T$ values well, i.e. $B/T$ - the quantity we are interested in - is very robustly determined with our fitting procedure. The top left panel of Figure~\ref{fig:BT_simulation} shows in red and green the points for which $(B/T)_{input}$ and $(B/T)_{output}$ differ more than 0.3 and between 0.15 and 0.3, respectively. We see that only in $2-3\%$ of all the runs is $\Delta(B/T)$ larger than 0.3. In $90\%$ of the cases, we are able to recover $(B/T)_{input}$ better than 0.15. As one sees in the bottom right panel of Figure~\ref{fig:BT_simulation}, the cases where we fail to recover $B/T$ are the ones with a disk-like bulge ($n\approx1-2$, i.e. pseudo-bulge). 

However, the bulge properties are subject to relatively large uncertainties, especially $r_{e,b}$ and $n_{b}$ (see Figure~\ref{fig:BT_simulation}, bottom panels). The scatter is much larger, and there is not a clear trend. This is also due to the well-known degeneracy between size and S\'{e}rsic index. Therefore, we are not able to make any conclusive statements about these two quantities of the bulge. 

\begin{figure}
\includegraphics[width=\textwidth]{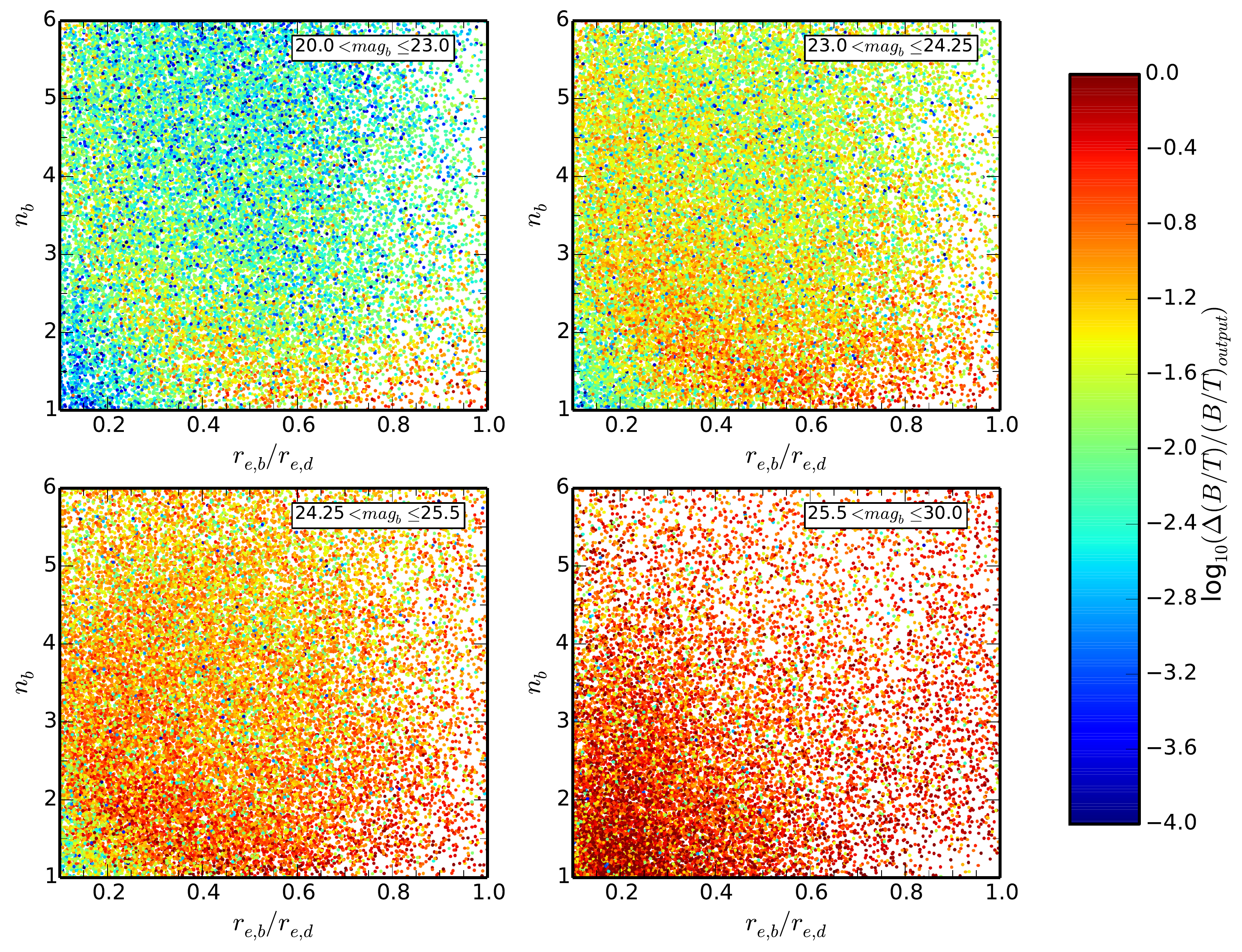} 
\caption{Estimation of the error in the $B/T$ for given measured properties of the disk and bulge. The relative error in $B/T$ is shown as a function of the magnitude of the bulge $mag_b$ (panel-wise), S\'{e}rsic index of the bulge $n_b$, and the ratio of the size of the bulge and disk $r_{e,b}/r_{e,d}$. The uncertainty in $B/T$ clearly increases for fainter bulges and disk-like bulges. } 
\label{fig:BT_err}
\end{figure}

For estimating the error on the $B/T$ itself, we use for a given magnitude of the bulge, bulge S\'{e}rsic index, and the ratio of the size of the disk to the bulge ($r_{e,b}/r_{e,d}$) the dispersion in $B/T$ values, as shown in Figure~\ref{fig:BT_err}. We get larger errors for fainter and more disk-like bulges. 

\subsection{Fitting Bulges with free vs fixed Sersic Index}

After the analysis of the error of the $B/T$, we would like to shed more light on the modeling assumptions. Specifically, we investigate how the relative error in the $B/T$ changes when fixing the S\'{e}rsic index of the bulge to 4, $n_b=4$, in comparison with letting $n_b$ remain free during the fitting process. As described in Section~\ref{subsec:2comp}, the main physical motivation for these two modeling approaches is directly linked to the two different processes for the formation of galactic bulges: from merging one expects $n_b\sim4$ (`classical' bulge), while from secular instability of the stellar disk $n_b\sim1-3$ (`pseudo' bulge). At $z\sim2$, we do not know whether the galaxies have a classical or pseudo-bulge. Hence, we simulate galaxies with a variety of physically sensible bulge S\'{e}rsic indices ($n_b\in[0.0,4.5]$), recovering each galaxy with the two different fitting methods. 

In the spirit of the section before, we simulate $\sim25'000$ mock galaxies, recovering each galaxy with the $n_b=4$ fixed and $n_b$ free (initial value is set to $n_b=4$) methods. As before, for the mock galaxies, we vary the sizes and brightnesses of the bulge and disk. The S\'{e}rsic index of the bulge is also varied ($n_b\in[0.0,4.5]$), while the S\'{e}rsic index of the disk is fixed to 1. Furthermore, in contrast to the simulations above, we fix the axis ratio to 1.0 and the position angle to $0^{\circ}$ to minimize the parameter space. We therefore expect -- for both fitting methods -- smaller relative errors than from the analysis above. 

\begin{figure}
\includegraphics[width=\textwidth]{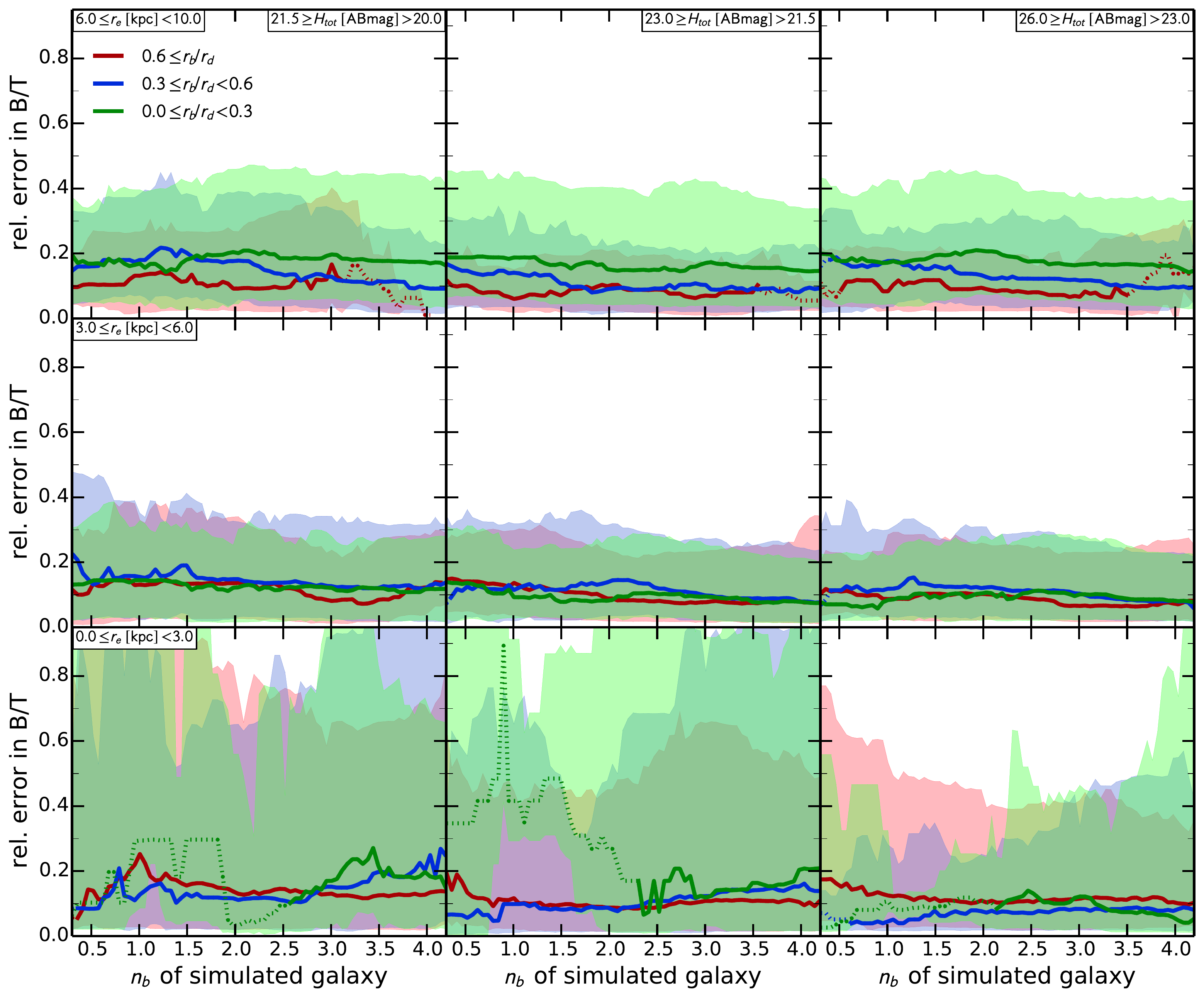} 
\caption{Relative error in $B/T$ as a function of the S\'{e}rsic index of the bulge $n_b$ of the simulated galaxy when we fix $n_b=4$ during the fitting. In each panel, the red, blue, and green lines (shaded area shows $1\sigma$ scatter) indicate size ratio of the bulge to the disk $r_b/r_d$ in the intervals $[0.6, 1.0]$, $[0.3, 0.6[$, and $[0.0, 0.3[$, respectively. The dashed lines show bins with $<20$ galaxies, while the solid lines show bins with $>20$ galaxies (on average 140 galaxies per bin). The panels show bins of higher magnitudes (fainter galaxies) to the right and smaller sizes to the bottom, i.e. the brightest and largest galaxies are in the top left panel, while the faintest and smallest galaxies are in the bottom right. The relative error is, on average, $\sim0.19$, increasing toward smaller and fainter galaxies as expected. } 
\label{fig:BT_err_fix}
\end{figure}

\begin{figure}
\includegraphics[width=\textwidth]{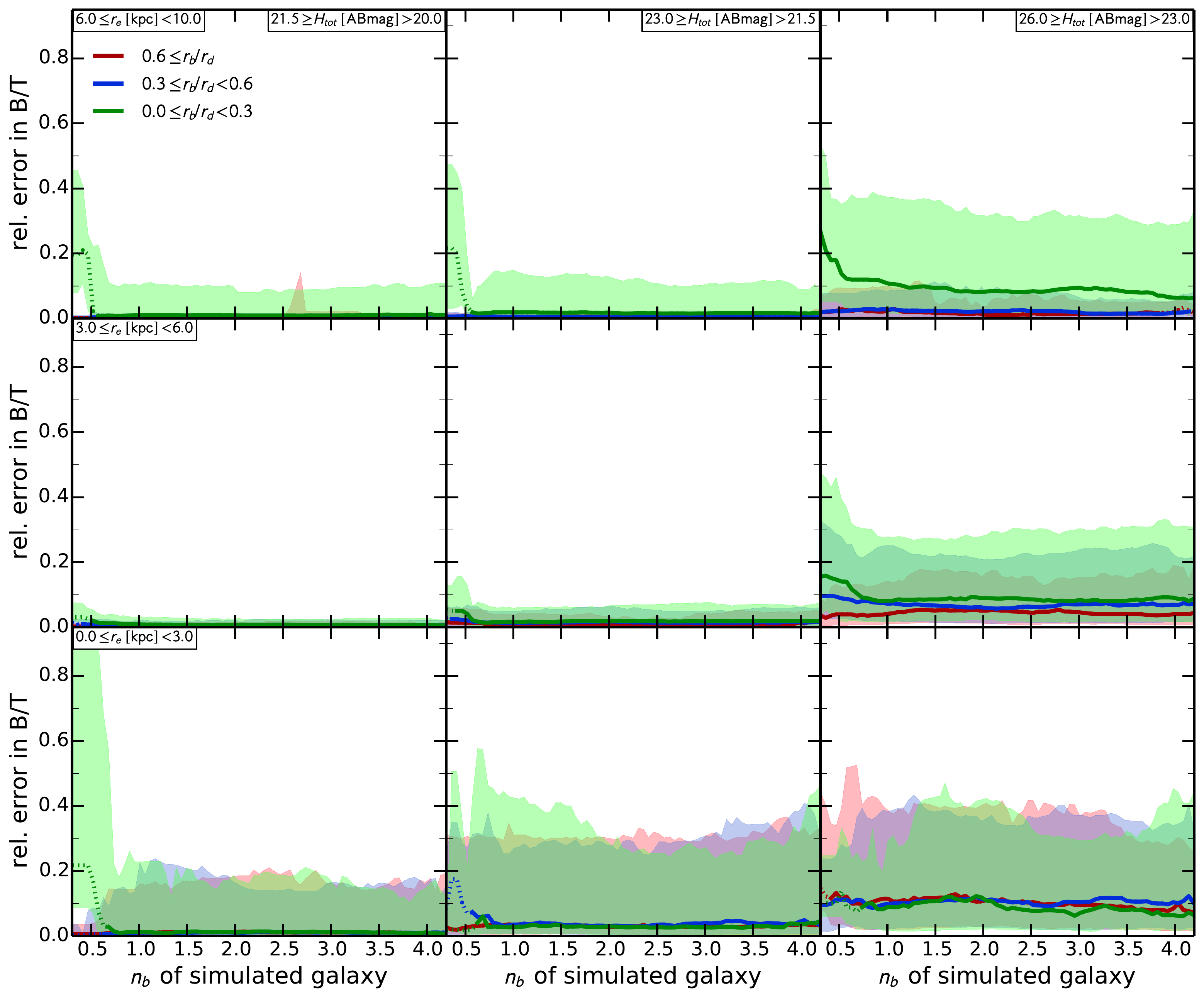} 
\caption{Same as Figure~\ref{fig:BT_err_fix}, but with a free S\'{e}rsic index of the bulge $n_b$ during fitting. The relative error is, on average, $\sim0.08$, i.e. smaller than when letting $n_b$ remain fixed to 4 during the fit. This indicates that we can recover the $B/T$ better when letting the $n_b$ remain free during the fitting. } 
\label{fig:BT_err_free}
\end{figure}

The results of the simulations for $n_b=4$ fixed and $n_b$ free methods are shown in Figures~\ref{fig:BT_err_fix} and \ref{fig:BT_err_free}, respectively. The galaxies get fainter (total magnitude $H_{tot}$) and smaller (half-light radius $r_e$ of the total light profile, i.e. bulge and disk) in panels toward to the right and bottom. Each panel displays the relative error in $B/T$ as a function of the S\'{e}rsic index $n_b$ of the simulated galaxy. The three different colored lines show galaxies with different size ratio of the bulge to the disk $r_b/r_d$: red, blue, and green indicate the  intervals $[0.6, 1.0]$, $[0.3, 0.6[$, and $[0.0, 0.3[$, respectively. The dashed lines show the bins with $<20$ galaxies, while the bins indicated by the solid lines have $>20$ galaxies (on average 140 galaxies). The shaded areas show the $1\sigma$ scatter. 

We find smaller relative errors in $B/T$ with the $n_b$ free method than with the $n_b=4$ fixed method, for all S\'{e}rsic indices $n_b$, brightnesses $H_{tot}$, sizes $r_e$, and size ratios $r_b/r_d$. On average, the relative error in $B/T$ for the $n_b$-free method is $\sim0.08$, in contrast to the $n_b=4$ fixed method, which gives $\sim0.19$. For both methods, the relative errors vary only a little for different size ratios. As expected, the relative errors increase toward fainter and smaller galaxies. In summary, we have shown that the $n_b$ free method is able to recover the $B/T$ of galaxies with a variety of bulge S\'{e}rsic indices much more reliable than the $n_b=4$ fixed method.

\end{appendix}

\clearpage
\capstartfalse 
\begin{deluxetable}{lrrcccrrrcc}
\tabletypesize{\scriptsize}
\tablecolumns{11}
\tablewidth{0pc}
\tablecaption{Sample galaxies with their $K$-band magnitude, H$\alpha$ redshifts, and main stellar properties. \label{tbl:aux_data_stel}}
\tablehead{
\colhead{Source} & \colhead{R.A.} & \colhead{Decl.} & \colhead{$K_{\mathrm{AB}}~^\mathrm{a}$} & \colhead{$z_{\mathrm{H\alpha}}~^\mathrm{b}$} & \colhead{$M_{\star}$} & \colhead{SFR} & \colhead{sSFR} &\colhead{Age} & \colhead{$A_{\mathrm{V}}$} & \colhead{$(U-V)_{\mathrm{rest}}~^\mathrm{c}$} \\
\colhead{} & \colhead{} & \colhead{} & \colhead{[mag]} & \colhead{} & \colhead{[10$^{10}$ M$_{\sun}$]} & \colhead{[$\frac{\mathrm{M}_{\sun}}{yr}$]} & \colhead{[Gyr$^{-1}$]} & \colhead{[Myr]} & \colhead{[mag]} & \colhead{[mag]} 
}

\startdata

D3A-15504 & 11:24:15.6 & -21:39:31.2 & 21.28 & 2.3826 & 15.0 & 150 & 1.0 & 454 & 1.0 & 0.71 \\
D3A-6004 & 11:25:03.8 & -21:45:32.8 & 20.96 & 2.3867 & 45.8 & 210 & 0.5 & 641 & 1.8 & 1.52 \\
GK-2303 & 03:32:38.9 & -27:43:21.5 & 22.78 & 2.4501 & 0.97 & 21 & 2.2 & 286 & 0.4 & 0.61 \\
GK-2363 & 03:32:39.4 & -27:42:35.7 & 22.67 & 2.4518 & 2.92 & 64 & 2.2 & 286 & 1.2 & 0.69 \\
GK-2540 & 03:32:30.3 & -27:42:40.5 & 21.80 & 1.6146 & 2.66 & 21 & 0.8 & 509 & 0.6 & 0.78 \\
K20-ID6 & 03:32:29.1 & -27:45:21.1 & 22.14 & 2.2345 & 3.68 & 45 & 1.2 & 404 & 1.0 & 0.72 \\
K20-ID7 & 03:32:29.1 & -27:46:28.5 & 21.47 & 2.2241 & 8.71 & 110 & 1.3 & 509 & 1.0 & 0.65 \\
Q1623-BX502 & 16:25:54.4 & +26:44:09.3 & 23.90 & 2.1556 & 0.30 & 14 & 4.7 & 227 & 0.4 & 0.14 \\
Q1623-BX599 & 16:26:02.5 & +26:45:31.9 & 21.79 & 2.3313 & 8.89 & 34 & 0.4 & 2750 & 0.4 & 0.93 \\
Q2343-BX389 & 23:46:28.9 & +12:47:33.6 & 22.04 & 2.1733 & 6.39 & 25 & 0.4 & 2750 & 1.0 & 1.29 \\
Q2343-BX610 & 23:46:09.4 & +12:49:19.2 & 21.07 & 2.2103 & 15.5 & 60 & 0.4 & 2750 & 0.8 & 0.93 \\
Q2346-BX482 & 23:48:13.0 & +00:25:46.3 & 22.34$^d$ & 2.2571 & 2.50 & 80 & 3.2 & 321 & 0.8 & 0.77 \\
ZC400528 & 09:59:47.6 & +01:44:19.1 & 21.08 & 2.3876 & 16.0 & 148 & 0.9 & 1140 & 0.9 & 0.84 \\
ZC400569 & 10:01:08.7 & +01:44:28.3 & 20.69 & 2.2405 & 23.3 & 241 & 1.0 & 1010 & 1.4 & 1.29 \\
ZC401925 & 10:01:01.7 & +01:48:38.2 & 22.74 & 2.1411 & 0.73 & 47 & 6.4 & 160 & 0.7 & 0.4 \\
ZC404221 & 10:01:41.3 & +01:56:42.8 & 22.44 & 2.2201 & 2.12 & 61 & 2.8 & 360 & 0.7 & 0.4 \\
ZC405226 & 10:02:19.5 & +02:00:18.1 & 22.33 & 2.2872 & 1.13 & 117 & 10.3 & 100 & 1.0 & 0.56 \\
ZC405501 & 09:59:53.7 & +02:01:08.9 & 22.25 & 2.1543 & 1.04 & 85 & 8.2 & 130 & 0.9 & 0.33 \\
ZC406690 & 09:58:59.1 & +02:05:04.3 & 20.81 & 2.1949 & 5.30 & 200 & 3.8 & 290 & 0.7 & 0.56 \\
ZC407302 & 09:59:55.9 & +02:06:51.3 & 21.48 & 2.1814 & 2.98 & 340 & 11.4 & 100 & 1.3 & 0.5 \\
ZC407376 & 10:00:45.1 & +02:07:05.0 & 21.79 & 2.1733 & 3.42 & 89 & 2.6 & 400 & 1.2 & 0.8 \\
ZC409985 & 09:59:14.2 & +02:15:47.0 & 22.30 & 2.4577 & 2.19 & 51 & 2.3 & 450 & 0.6 & 0.64 \\
ZC410041 & 10:00:44.3 & +02:15:58.5 & 23.16 & 2.4539 & 0.57 & 47 & 8.2 & 130 & 0.6 & 0.36 \\
ZC410123 & 10:02:06.5 & +02:16:15.5 & 22.80 & 2.1987 & 0.51 & 59 & 11.6 & 100 & 0.8 & 0.31 \\
ZC411737 & 10:00:32.4 & +02:21:20.9 & 22.81 & 2.4443 & 0.41 & 48 & 11.7 & 100 & 0.6 & 0.53 \\
ZC412369 & 10:01:46.9 & +02:23:24.6 & 21.39 & 2.0283 & 2.86 & 94 & 3.3 & 320 & 1.0 & 0.88 \\
ZC413507 & 10:00:24.2 & +02:27:41.3 & 22.52 & 2.4794 & 1.07 & 111 & 10.3 & 100 & 1.1 & 0.57 \\
ZC413597 & 09:59:36.4 & +02:27:59.1 & 22.58 & 2.4498 & 0.92 & 84 & 9.1 & 110 & 1.0 & 0.51 \\
ZC415876 & 10:00:09.4 & +02:36:58.4 & 22.38 & 2.4362 & 1.15 & 94 & 8.2 & 100 & 1.0 & 0.68 \\
\enddata
\tablecomments{The stellar population properties are taken from \citet{mancini11} and \citet{forster-schreiber11a}. We use the results computed for the \citet{bruzual03} model and constant star formation from \citet{mancini11}, assuming a \citet{chabrier03} IMF, solar metallicity, the \citet{calzetti00} reddening law, and either constant or exponentially declining SFRs. The uncertainties on the stellar properties are dominated by systematics from the model assumption and are up to a factor of $\sim2-3$ for $M_{\star}$ and at least $\sim3$ for SFRs.}
\tablenotetext{a}{Typical uncertainties on the $K$-band magnitudes range from $\approx 0.05$ for the brightest sources and up to $\approx 0.15$ for the faintest. }
\tablenotetext{b}{Spectroscopic redshifts are based on the source-integrated H$\alpha$ emission from the AO SINFONI data. Taken from FS15.}
\tablenotetext{c}{Rest frame $U-V$ colors have uncertainties of order 0.1 mag. Taken from FS15. }
\tablenotetext{d}{For Q2346-BX482, no $K$-band magnitude is available, and listed here is the $H$-band magnitude measured from HST/NICMOS imaging with the NIC2 camera.}

\end{deluxetable}
\capstarttrue

\clearpage
\capstartfalse 
\begin{deluxetable}{lcccccc}
\tabletypesize{\scriptsize}
\tablecolumns{7}
\tablewidth{0pc}
\tablecaption{Overview of the available HST imaging data. \label{tbl:hst_data_overview}}
\tablehead{
\colhead{Source} & \colhead{$V_{606}$} & \colhead{$I_{814}$} & \colhead{$Y_{105}$} & \colhead{$J_{110}$} & \colhead{$J_{125}$} & \colhead{$H_{160}$} 
}
\startdata

D3A-15504 & --- & --- & --- & A & --- & A  \\
D3A-6004 & --- & --- & --- & A & --- & A  \\
GK-2303 & --- & C & --- & --- & C & C  \\
GK-2363 & --- & C & --- & --- & C & C  \\
GK-2540 & --- & C & --- & --- & C & C  \\
K20-ID6 & --- & C & C & --- & C & C  \\
K20-ID7 & --- & C & C & --- & C & C  \\
Q1623-BX502 & --- & --- & --- & A & --- & D  \\
Q1623-BX599 & --- & --- & --- & A & --- & A  \\
Q2343-BX389 & --- & --- & --- & A & --- & B  \\
Q2343-BX610 & --- & --- & --- & A & --- & B  \\
Q2346-BX482 & --- & --- & --- & A & --- & B  \\
ZC400528 & --- & E & --- & A & --- & A  \\
ZC400569 & --- & E & --- & A & --- & A  \\
ZC401925 & --- & E & --- & A & --- & A  \\
ZC404221 & --- & E & --- & A & --- & A  \\
ZC405226 & --- & E & --- & A & --- & A  \\
ZC405501 & --- & E & --- & A & --- & A  \\
ZC406690 & --- & E & --- & A & --- & A  \\
ZC407302 & --- & E & --- & A & --- & A  \\
ZC407376 & --- & E & --- & A & --- & A  \\
ZC409985 & --- & E & --- & A & --- & A  \\
ZC410041 & C & E & --- & --- & C & C  \\
ZC410123 & --- & E & --- & A & --- & A  \\
ZC411737 & C & E & --- & --- & C & C  \\
ZC412369 & --- & E & --- & A & --- & A  \\
ZC413507 & --- & E & --- & --- & C & C  \\
ZC413597 & --- & E & --- & A & --- & A  \\
ZC415876 & --- & E & --- & A & --- & A  \\
\enddata
\tablecomments{Labels: \\
A: \# GO12578 (this work): exposure time $t_{H}=5223.5$ s and $t_{J}=2611.8$ s\\
B: HST/NICMOS NIC2 $H$ \citep{forster-schreiber11a}, $t_{H}=10239.5$ s \\
C: CANDELS Data, see \citet{grogin11, koekemoer11}: $t_{H}=3200.0$ s and $t_{J}=1900.0$ s\\
D: \# GO11694 (PI Law): $t_{H}=5395.4$ s \\
E: COSMOS Data, see \citet{scoville07} \\
F: HST/WFPC2 F702W \# GO6557 (PI Steidel)}
\end{deluxetable}
\capstarttrue

\clearpage
\capstartfalse 
\begin{landscape}
\begin{deluxetable}{lccccccccccccc}
\tabletypesize{\scriptsize}
\tablecolumns{14}
\tablewidth{0pc}
\tablecaption{Overview measurements done on the $H$-band images. \label{tbl:measurements_overview}}
\tablehead{
\colhead{}&\colhead{}&\colhead{}&\multicolumn{4}{c}{ \texttt{GALFIT}: Single Component}&\colhead{}&\multicolumn{6}{c}{ \texttt{GALFIT}: Double Component (Disk+Bulge)}\\
\cline{4-7}\cline{9-14}\\
\colhead{Source} & \colhead{Classi.$^{\dagger}$} & \colhead{} & \colhead{$r_{e}$ [kpc]} & \colhead{$n$} & \colhead{$b/a$} & \colhead{P.A. [deg]} & \colhead{} & \colhead{$r_{e, disk}$ [kpc]} & \colhead{$(b/a)_{disk}$} & \colhead{P.A.$_{\mathrm{disk}}$ [deg]} & \colhead{$r_{e, bulge}$ [kpc]} & \colhead{$n_{bulge}$} & \colhead{$B/T$} }
\startdata

D3A15504 & RD &  & 5.0$\pm$0.4 & 1.3$\pm$0.3 & 0.68$\pm$0.13 & -28$\pm$2 &  & 4.9$\pm$0.3 & 0.73$\pm$0.14 & -24$\pm$2 & 0.5$\pm$0.1 & 8.0$\pm$2.1 & 0.06$^{+0.05}_{-0.05}$ \\
D3A6004 & RD &  & 4.6$\pm$0.5 & 2.6$\pm$0.5 & 0.81$\pm$0.11 & 55$\pm$6 &  & 4.7$\pm$0.5 & 0.84$\pm$0.12 & 55$\pm$6 & 0.6$\pm$0.1 & 1.0$\pm$0.2 & 0.07$^{+0.04}_{-0.04}$ \\
GK2303 & RD &  & 1.3$\pm$0.4 & 0.9$\pm$0.3 & 0.64$\pm$0.16 & 19$\pm$13 &  & 1.3$\pm$0.4 & 0.64$\pm$0.15 & 17$\pm$11 & 0.9$\pm$0.3 & 1.0$\pm$0.4 & 0.03$^{+0.10}_{-0.03}$ \\
GK2363 & RD &  & 1.7$\pm$0.5 & 1.3$\pm$0.4 & 0.52$\pm$0.13 & 61$\pm$2 &  & 1.8$\pm$0.5 & 0.52$\pm$0.13 & 60$\pm$2 & 0.1$\pm$0.1 & 5.3$\pm$1.7 & 0.07$^{+0.05}_{-0.06}$ \\
GK2540 & PD &  & 7.9$\pm$0.9 & 1.4$\pm$0.4 & 0.86$\pm$0.15 & 86$\pm$17 &  & 7.5$\pm$0.8 & 0.89$\pm$0.16 & 88$\pm$17 & 0.3$\pm$0.1 & 2.4$\pm$0.7 & 0.03$^{+0.04}_{-0.03}$ \\
K20ID6 & PD &  & 3.6$\pm$0.3 & 0.6$\pm$0.3 & 0.86$\pm$0.15 & 37$\pm$5 &  & 3.8$\pm$0.3 & 0.85$\pm$0.14 & 40$\pm$5 & 3.8$\pm$0.3 & 1.0$\pm$0.4 & 0.16$^{+0.17}_{-0.10}$ \\
K20ID7$^{\ddagger}$ & PD &  & 5.7$\pm$0.7 & 0.2$\pm$0.2 & 0.46$\pm$0.21 & 18$\pm$1 &  & --- & --- & --- & --- & --- & --- \\
Q1623-BX502 & RD &  & 0.9$\pm$0.6 & 1.0$\pm$0.6 & 0.66$\pm$0.17 & 1$\pm$6 &  & --- & --- & --- & --- & --- & --- \\
Q1623-BX599 & M &  & 2.0$\pm$0.5 & 2.0$\pm$0.6 & 0.69$\pm$0.14 & -57$\pm$13 &  & 4.9$\pm$1.2 & 0.68$\pm$0.14 & -3$\pm$1 & 1.3$\pm$0.3 & 1.0$\pm$0.3 & 0.57$^{+0.09}_{-0.09}$ \\
Q2343-BX389$^{\ddagger}$ & PD &  & 3.3$\pm$0.7 & 0.3$\pm$0.3 & 0.28$\pm$0.13 & -48$\pm$2 &  & 3.8$\pm$0.8 & 0.32$\pm$0.15 & -45$\pm$2 & 3.3$\pm$0.7 & 1.0$\pm$1.0 & 0.02$^{+0.10}_{-0.02}$ \\
Q2343-BX610$^{\ddagger}$ & RD &  & 3.3$\pm$0.8 & 0.4$\pm$0.2 & 0.54$\pm$0.15 & 17$\pm$1 &  & 3.3$\pm$0.8 & 0.56$\pm$0.15 & -15$\pm$1 & 0.5$\pm$0.1 & 8.0$\pm$4.3 & 0.08$^{+0.06}_{-0.06}$ \\
Q2346-BX482$^{\ddagger}$ & PD &  & 4.2$\pm$0.6 & 0.1$\pm$0.1 & 0.50$\pm$0.20 & -62$\pm$9 &  & --- & --- & --- & --- & --- & --- \\
ZC400528 & PD &  & 2.1$\pm$0.6 & 3.8$\pm$0.7 & 0.78$\pm$0.13 & -65$\pm$9 &  & 3.0$\pm$0.9 & 0.88$\pm$0.15 & -85$\pm$11 & 1.7$\pm$0.5 & 1.0$\pm$0.2 & 0.3$^{+0.12}_{-0.12}$ \\
ZC400569 & PD &  & 6.7$\pm$1.3 & 4.8$\pm$0.7 & 0.89$\pm$0.15 & 38$\pm$9 &  & 5.6$\pm$1.1 & 0.8$\pm$0.13 & 22$\pm$5 & 0.9$\pm$0.2 & 1.0$\pm$0.2 & 0.19$^{+0.04}_{-0.03}$ \\
ZC401925 & DD &  & 1.6$\pm$0.4 & 1.5$\pm$0.6 & 0.38$\pm$0.11 & -60$\pm$6 &  & 1.8$\pm$0.4 & 0.36$\pm$0.11 & -58$\pm$6 & 0.5$\pm$0.1 & 8.0$\pm$3.2 & 0.19$^{+0.10}_{-0.10}$ \\
ZC404221 & DD &  & 0.5$\pm$0.4 & 5.0$\pm$1.8 & 0.45$\pm$0.09 & -13$\pm$1 &  & --- & --- & --- & --- & --- & --- \\
ZC405226 & RD &  & 4.2$\pm$0.7 & 0.7$\pm$0.3 & 0.60$\pm$0.15 & -66$\pm$10 &  & 4.6$\pm$0.7 & 0.59$\pm$0.15 & -67$\pm$10 & 2.6$\pm$0.4 & 8.0$\pm$3.3 & 0.01$^{+0.04}_{-0.01}$ \\
ZC405501$^{\ddagger}$ & RD &  & 3.2$\pm$0.5 & 0.1$\pm$0.1 & 0.31$\pm$0.13 & 11$\pm$1 &  & 3.4$\pm$0.5 & 0.43$\pm$0.19 & 15$\pm$1 & 1.9$\pm$0.3 & 1.0$\pm$4.0 & 0.37$^{+0.09}_{-0.08}$ \\
ZC406690$^{\ddagger}$ & PD &  & 5.5$\pm$0.9 & 0.1$\pm$0.1 & 0.62$\pm$0.17 & -77$\pm$7 &  & --- & --- & --- & --- & --- & --- \\
ZC407302$^{\star}$ & PD &  & 2.4$\pm$0.8 & 1.1$\pm$0.4 & 0.45$\pm$0.14 & 47$\pm$3 &  & 2.0$\pm$0.7 & 0.37$\pm$0.12 & 52$\pm$4 & 3.5$\pm$1.2 & 1.0$\pm$0.3 & 0.20$^{+0.21}_{-0.12}$ \\
ZC407376 & M &  & 1.5$\pm$0.5 & 3.6$\pm$0.7 & 0.81$\pm$0.14 & -68$\pm$9 &  & 2.6$\pm$0.8 & 0.7$\pm$0.12 & -66$\pm$9 & 0.5$\pm$0.2 & 1.0$\pm$0.2 & 0.42$^{+0.08}_{-0.08}$ \\
ZC409985 & DD &  & 1.6$\pm$0.2 & 1.6$\pm$0.3 & 0.67$\pm$0.16 & -14$\pm$3 &  & 1.6$\pm$0.2 & 0.67$\pm$0.16 & -22$\pm$5 & 0.2$\pm$0.1 & 4.3$\pm$0.8 & 0.09$^{+0.05}_{-0.06}$ \\
ZC410041$^{\star}$ & RD &  & 2.2$\pm$0.6 & 0.5$\pm$0.5 & 0.22$\pm$0.08 & -64$\pm$1 &  & 1.8$\pm$0.5 & 0.18$\pm$0.06 & -69$\pm$1 & 3.2$\pm$0.8 & 1.0$\pm$1.0 & 0.11$^{+0.16}_{-0.08}$ \\
ZC410123 & PD &  & 2.0$\pm$0.4 & 1.5$\pm$0.6 & 0.38$\pm$0.13 & 16$\pm$5 &  & 2.3$\pm$0.5 & 0.36$\pm$0.12 & 19$\pm$6 & 0.9$\pm$0.2 & 5.1$\pm$2.1 & 0.21$^{+0.10}_{-0.11}$ \\
ZC411737 & RD &  & 1.7$\pm$0.2 & 0.9$\pm$0.2 & 0.83$\pm$0.13 & -19$\pm$7 &  & 1.8$\pm$0.2 & 0.75$\pm$0.11 & -5$\pm$2 & 1.4$\pm$0.2 & 1.0$\pm$0.3 & 0.20$^{+0.16}_{-0.10}$ \\
ZC412369 & M &  & 2.1$\pm$0.8 & 4.9$\pm$1.7 & 0.47$\pm$0.12 & -54$\pm$10 &  & 3.1$\pm$1.1 & 0.40$\pm$0.10 & -58$\pm$10 & 0.2$\pm$0.1 & 8.0$\pm$2.7 & 0.37$^{+0.20}_{-0.21}$ \\
ZC413507$^{\star}$ & RD &  & 2.1$\pm$0.4 & 0.8$\pm$0.4 & 0.66$\pm$0.21 & -22$\pm$7 &  & 2.0$\pm$0.4 & 0.62$\pm$0.19 & -19$\pm$6 & 2.9$\pm$0.6 & 1.0$\pm$0.5 & 0.11$^{+0.15}_{-0.09}$ \\
ZC413597$^{\star}$ & DD &  & 1.1$\pm$0.4 & 1.7$\pm$0.6 & 0.46$\pm$0.14 & 15$\pm$9 &  & 0.5$\pm$0.2 & 0.22$\pm$0.06 & 8$\pm$5 & 2.2$\pm$0.7 & 1.0$\pm$0.3 & 0.44$^{+0.17}_{-0.15}$ \\
ZC415876$^{\star}$ & PD &  & 2.1$\pm$0.9 & 1.5$\pm$0.6 & 0.78$\pm$0.15 & -80$\pm$38 &  & 1.7$\pm$0.7 & 0.65$\pm$0.12 & 87$\pm$41 & 3.2$\pm$1.3 & 1.3$\pm$0.5 & 0.29$^{+0.14}_{-0.13}$ \\

\enddata
\tablecomments{
$\dagger$ Classification based on rest frame optical light distribution (morphology) and kinematic data: rotating disks (RD), irregular disks (ID), mergers (M), and unresolved systems (UNR). \\ 
$\ddagger$ Single-component fits are less reliable because light distribution is not centrally concentrated (assumed models do not represent the galaxy well). \\
$^{\star}$ Double-component fits gives non-physical solution, i.e. bulge component is larger than disk component. }
\end{deluxetable}
\clearpage
\end{landscape}
\capstarttrue

\capstartfalse 
\begin{landscape}
\begin{deluxetable}{lcccccccccccc}
\tabletypesize{\scriptsize}
\tablecolumns{13}
\tablewidth{0pc}
\tablecaption{Overview measurements done on the $J$-band images. \label{tbl:measurements_overview2}}
\tablehead{
\colhead{}&\colhead{}&\multicolumn{4}{c}{ \texttt{GALFIT}: Single Component}&\colhead{}&\multicolumn{6}{c}{ \texttt{GALFIT}: Double Component (Disk+Bulge)}\\
\cline{3-6}\cline{8-13}\\
\colhead{Source} & \colhead{} & \colhead{$r_{e}$ [kpc]} & \colhead{$n$} & \colhead{$b/a$} & \colhead{P.A. [deg]} & \colhead{} & \colhead{$r_{e, disk}$ [kpc]} & \colhead{$(b/a)_{disk}$} & \colhead{P.A.$_{\mathrm{disk}}$ [deg]} & \colhead{$r_{e, bulge}$ [kpc]} & \colhead{$n_{bulge}$} & \colhead{$B/T$} }
\startdata

D3A15504 &  & 5.0$\pm$0.8 & 0.9$\pm$0.4 & 0.68$\pm$0.13 & -28$\pm$2 &  & 4.9$\pm$0.7 & 0.73$\pm$0.14 & -24$\pm$2 & 0.5$\pm$0.1 & 8.0$\pm$3.4 & 0.00$^{+0.07}_{-0.00}$ \\
D3A6004 &  & 6.8$\pm$2.9 & 2.9$\pm$0.8 & 0.81$\pm$0.13 & 55$\pm$10 &  & 4.7$\pm$2.1 & 0.84$\pm$0.14 & 55$\pm$10 & 0.6$\pm$0.3 & 1.0$\pm$0.3 & 0.03$^{+0.09}_{-0.03}$ \\
GK2303 &  & 1.3$\pm$0.4 & 1.0$\pm$0.4 & 0.64$\pm$0.16 & 19$\pm$17 &  & 1.3$\pm$0.4 & 0.64$\pm$0.16 & 17$\pm$15 & 0.9$\pm$0.3 & 1.0$\pm$0.4 & 0.03$^{+0.1}_{-0.03}$ \\
GK2363 &  & 1.7$\pm$0.4 & 1.0$\pm$0.3 & 0.52$\pm$0.14 & 61$\pm$2 &  & 1.8$\pm$0.4 & 0.52$\pm$0.14 & 60$\pm$2 & 0.1$\pm$0.1 & 5.3$\pm$1.7 & 0.02$^{+0.06}_{-0.02}$ \\
GK2540 &  & 8.8$\pm$1.4 & 1.3$\pm$0.4 & 0.86$\pm$0.17 & 86$\pm$10 &  & 7.5$\pm$1.2 & 0.89$\pm$0.18 & 88$\pm$11 & 0.3$\pm$0.1 & 2.4$\pm$0.6 & 0.03$^{+0.04}_{-0.03}$ \\
K20ID6 &  & 3.7$\pm$0.4 & 0.5$\pm$0.2 & 0.86$\pm$0.13 & 37$\pm$5 &  & 3.8$\pm$0.4 & 0.85$\pm$0.13 & 40$\pm$5 & 3.8$\pm$0.4 & 1.0$\pm$0.5 & 0.30$^{+0.20}_{-0.21}$ \\
K20ID7$^{\ddagger}$ &  & 5.9$\pm$1.4 & 0.1$\pm$0.1 & 0.46$\pm$0.15 & 18$\pm$1 &  & --- & --- & --- & --- & --- & --- \\
Q1623-BX502 &  & 1.0$\pm$0.4 & 0.5$\pm$0.5 & 0.66$\pm$0.18 & 1$\pm$4 &  & --- & --- & --- & --- & --- & --- \\
Q1623-BX599 &  & 2.8$\pm$0.6 & 2.6$\pm$0.7 & 0.69$\pm$0.14 & -57$\pm$23 &  & 4.9$\pm$1.1 & 0.68$\pm$0.14 & -3$\pm$1 & 1.3$\pm$0.3 & 1.0$\pm$0.3 & 0.43$^{+0.11}_{-0.11}$ \\
Q2343-BX389$^{\ddagger}$ &  & 3.1$\pm$0.4 & 0.2$\pm$0.2 & 0.28$\pm$0.14 & -48$\pm$3 &  & 3.8$\pm$0.6 & 0.32$\pm$0.16 & -45$\pm$3 & 3.3$\pm$0.5 & 1.0$\pm$2.5 & 0.20$^{+0.15}_{-0.10}$ \\
Q2343-BX610$^{\ddagger}$ &  & 3.5$\pm$0.5 & 0.1$\pm$0.1 & 0.54$\pm$0.14 & 17$\pm$1 &  & 3.3$\pm$0.5 & 0.56$\pm$0.15 & -15$\pm$0 & 0.5$\pm$0.1 & 8.0$\pm$16.0 & 0.15$^{+0.15}_{-0.12}$ \\
Q2346-BX482$^{\ddagger}$ &  & 4.3$\pm$0.6 & 0.1$\pm$0.1 & 0.50$\pm$0.20 & -62$\pm$10 &  & --- & --- & --- & --- & --- & --- \\
ZC400528 &  & 3.0$\pm$1.2 & 6.1$\pm$1.6 & 0.78$\pm$0.12 & -65$\pm$9 &  & 3.0$\pm$1.2 & 0.88$\pm$0.13 & -85$\pm$12 & 1.7$\pm$0.7 & 1.0$\pm$0.3 & 0.35$^{+0.11}_{-0.09}$ \\
ZC400569 &  & 8.1$\pm$2.0 & 3.6$\pm$0.6 & 0.89$\pm$0.17 & 38$\pm$5 &  & 5.6$\pm$1.4 & 0.8$\pm$0.15 & 22$\pm$3 & 0.9$\pm$0.2 & 1.0$\pm$0.2 & 0.13$^{+0.05}_{-0.05}$ \\
ZC401925 &  & 1.7$\pm$0.6 & 1.6$\pm$0.6 & 0.38$\pm$0.12 & -60$\pm$6 &  & 1.8$\pm$0.6 & 0.36$\pm$0.12 & -58$\pm$5 & 0.5$\pm$0.2 & 8.0$\pm$3.1 & 0.11$^{+0.11}_{-0.09}$ \\
ZC404221 &  & 0.4$\pm$0.2 & 7.2$\pm$1.5 & 0.45$\pm$0.15 & -13$\pm$3 &  & --- & --- & --- & --- & --- & --- \\
ZC405226 &  & 4.1$\pm$0.8 & 0.5$\pm$0.3 & 0.60$\pm$0.16 & -66$\pm$18 &  & 4.6$\pm$0.8 & 0.59$\pm$0.16 & -67$\pm$18 & 2.6$\pm$0.5 & 8.0$\pm$4.1 & 0.04$^{+0.12}_{-0.04}$ \\
ZC405501$^{\ddagger}$ &  & 3.3$\pm$0.5 & 0.1$\pm$0.1 & 0.31$\pm$0.13 & 11$\pm$1 &  & 3.4$\pm$0.5 & 0.43$\pm$0.18 & 15$\pm$1 & 1.9$\pm$0.3 & 1.0$\pm$4.0 & 0.33$^{+0.16}_{-0.13}$ \\
ZC406690$^{\ddagger}$ &  & 5.4$\pm$1.0 & 0.1$\pm$0.1 & 0.62$\pm$0.18 & -77$\pm$5 &  & --- & --- & --- & --- & --- & --- \\
ZC407302$^{\star}$ &  & 2.3$\pm$0.4 & 1.2$\pm$0.3 & 0.45$\pm$0.15 & 47$\pm$3 &  & 2.0$\pm$0.3 & 0.37$\pm$0.12 & 52$\pm$3 & 3.5$\pm$0.6 & 1.0$\pm$0.3 & 0.24$^{+0.13}_{-0.12}$ \\
ZC407376 &  & 1.8$\pm$1.0 & 5.0$\pm$1.8 & 0.81$\pm$0.16 & -68$\pm$39 &  & 2.6$\pm$1.4 & 0.7$\pm$0.13 & -66$\pm$38 & 0.5$\pm$0.3 & 1.0$\pm$0.4 & 0.49$^{+0.13}_{-0.14}$ \\
ZC409985 &  & 1.6$\pm$0.2 & 2.2$\pm$0.5 & 0.67$\pm$0.15 & -14$\pm$4 &  & 1.6$\pm$0.3 & 0.67$\pm$0.15 & -22$\pm$6 & 0.2$\pm$0.1 & 4.3$\pm$1.0 & 0.14$^{+0.06}_{-0.07}$ \\
ZC410041$^{\star}$ &  & 2.2$\pm$0.5 & 0.2$\pm$0.2 & 0.22$\pm$0.10 & -64$\pm$1 &  & 1.8$\pm$0.4 & 0.18$\pm$0.08 & -69$\pm$1 & 3.2$\pm$0.8 & 1.0$\pm$1.9 & 0.11$^{+0.16}_{-0.07}$ \\
ZC410123 &  & 2.2$\pm$0.8 & 1.8$\pm$0.5 & 0.38$\pm$0.12 & 16$\pm$5 &  & 2.3$\pm$0.8 & 0.36$\pm$0.12 & 19$\pm$6 & 0.9$\pm$0.3 & 5.1$\pm$1.6 & 0.22$^{+0.13}_{-0.12}$ \\
ZC411737 &  & 1.7$\pm$0.2 & 0.8$\pm$0.2 & 0.83$\pm$0.13 & -19$\pm$17 &  & 1.8$\pm$0.2 & 0.75$\pm$0.11 & -5$\pm$4 & 1.4$\pm$0.2 & 1.0$\pm$0.3 & 0.23$^{+0.16}_{-0.12}$ \\
ZC412369 &  & 2.7$\pm$1.7 & 5.2$\pm$2.3 & 0.47$\pm$0.13 & -54$\pm$7 &  & 3.1$\pm$2.1 & 0.4$\pm$0.11 & -58$\pm$8 & 0.2$\pm$0.1 & 8.0$\pm$3.5 & 0.39$^{+0.21}_{-0.22}$ \\
ZC413507$^{\star}$ &  & 1.9$\pm$0.5 & 0.6$\pm$0.5 & 0.66$\pm$0.19 & -22$\pm$10 &  & 2.0$\pm$0.5 & 0.62$\pm$0.18 & -19$\pm$9 & 2.9$\pm$0.8 & 1.0$\pm$0.8 & 0.04$^{+0.05}_{-0.04}$ \\
ZC413597$^{\star}$ &  & 1.1$\pm$0.4 & 2.2$\pm$1.0 & 0.46$\pm$0.14 & 15$\pm$10 &  & 0.5$\pm$0.2 & 0.22$\pm$0.07 & 8$\pm$5 & 2.2$\pm$0.8 & 1.0$\pm$0.5 & 0.45$^{+0.12}_{-0.12}$ \\
ZC415876$^{\star}$ &  & 1.7$\pm$0.7 & 1.4$\pm$0.6 & 0.78$\pm$0.15 & -80$\pm$42 &  & 1.7$\pm$0.7 & 0.65$\pm$0.13 & 87$\pm$46 & 3.2$\pm$1.3 & 1.3$\pm$0.5 & 0.23$^{+0.10}_{-0.09}$ \\

\enddata
\tablecomments{
$\ddagger$ Single-component fits are less reliable because light distribution is not centrally concentrated (assumed models does not represent the galaxy well). \\
$^{\star}$ Double-component fits gives non-physical solution, i.e. bulge component is larger than disk component. }
\end{deluxetable}
\clearpage
\end{landscape}
\capstarttrue

\end{document}